\documentclass{article}
\usepackage[letterpaper,left=1in,right=1in,top=1in,bottom=1in]{geometry}
\usepackage[T1]{fontenc}
\usepackage{microtype}
\usepackage{amsmath,amssymb,amsthm}
\usepackage{mathtools,mathrsfs,bm,bbm,accents}
\usepackage{graphicx,xcolor}
\graphicspath{{media/}}
\usepackage{wrapfig,subcaption,float,flafter}
\usepackage{booktabs,enumitem}
\usepackage{nicefrac,xfrac,cancel}
\usepackage{comment}
\usepackage[ruled,noend]{algorithm2e}
\usepackage{tikz}
\usetikzlibrary{positioning,chains,fit,shapes,calc,arrows,arrows.meta}
\usepackage[numbers,square]{natbib}
\usepackage{url}
\usepackage{hyperref}
\usepackage[capitalise]{cleveref}
\usepackage{appendix}
\allowdisplaybreaks
\usepackage{etoolbox}
\patchcmd{\section}{\normalfont\Large\bfseries}
  {\normalfont\Large\bfseries\raggedright}{}
  {\PackageError{arxiv-layout}{Could not configure section wrapping}{}}

\newtheorem{theorem}{Theorem}[section]
\newtheorem{lemma}[theorem]{Lemma}
\newtheorem{proposition}[theorem]{Proposition}

\newtheorem{assumption}{Assumption}
\makeatletter

\AtBeginDocument{\let\c@proposition\c@theorem}
\makeatother

\usepackage[suppress]{color-edits}

\title{Sparse Predict and Demote:\\
Optimal Scheduling with Minimal Size Advice}

\author{Chengyi Liu\thanks{California Institute of Technology, \texttt{c.liu@caltech.edu}} \and Thodoris Lykouris\thanks{Massachusetts Institute of Technology, \texttt{lykouris@mit.edu}} \and
Junxuan Shen\thanks{ Massachusetts Institute of Technology, \texttt{junxuan@mit.edu}}
\and 
Adam Wierman\thanks{California Institute of Technology, \texttt{adamw@caltech.edu}}}

\date{}

\begin{document}
\hypersetup{pageanchor=false}
\maketitle
\begin{abstract}
Knowledge of job sizes is crucial for optimizing the mean response time in single-server scheduling.  However, predictions of job sizes are typically inexact and may be costly to obtain. Thus, a long-standing research question is: what is the minimal information about job sizes required to obtain near-optimal mean response time?  We introduce a novel policy, \emph{Sparse Predict and Demote (SPD)}, and prove that it achieves order-optimal mean response time in heavy traffic while requiring only one-bit job-size predictions for a vanishing fraction of arrivals. SPD achieves heavy-traffic order-optimality even in situations when predictions are (i) inexact up to any fixed multiplicative error bound, and (ii) costly, i.e., the server is used to compute predictions and is thus unavailable to serve jobs while processing prediction queries.  Finally, we use numerical experiments to illustrate that SPD maintains near-optimal performance beyond the heavy-traffic regime. 
\end{abstract}
\thispagestyle{empty}
\clearpage
\setcounter{page}{1}
\hypersetup{pageanchor=true}

\section{Introduction}
A foundational result in scheduling theory is the optimality of Shortest Remaining Processing Time (SRPT) scheduling for a single-server queue \cite{schrage1966queue,schrage1968proof}. SRPT is optimal for the mean response time (i.e., sojourn time or flow time) in any single-server system, regardless of the arrival process or job size distribution (i.e., service distribution).  Not only is SRPT optimal, but the reduction in mean response time that results from using size information to prioritize small jobs is an order of magnitude \cite{lin_heavy-traffic_2011,gromoll2011diffusion}.  For example, consider a single-server system with arrival rate $\lambda$, service rate $\mu$, and load $\rho=\lambda/\mu$. In \emph{heavy traffic} ($\rho\rightarrow 1$), SRPT has mean response time $\Theta(1/((1-\rho)\log(1/(1-\rho)))$, which improves over the mean response time $\Theta\left(1/(1-\rho)\right)$ of optimal size-oblivious algorithms. This contrast is not unique to exponential service distributions; it extends to a general class of distributions (see Section~\ref{subsec:service-time-classes} for detailed discussion).

While SRPT is optimal, it is rarely used `as is' in applications.  In particular, schedulers typically do not have access to the exact sizes of jobs.  Job sizes need to be estimated and, often, determining a job size estimate requires use of the server.  For example, in the case of scheduling in compute systems, a server may use an AI model to estimate job sizes, which requires allocating the server to an inference task for some period of time during which it is unavailable to serve arriving jobs. Further, these estimates are not perfect; they arrive with prediction errors. These challenges are not unique to computer systems. Another example comes from emergency room resource allocation.  Patient prioritization is informed by quick but error-prone initial screening that is performed by the same clinicians who treat patients. An automobile service center faces the same tradeoff. A mechanic must perform an initial shallow check to diagnose the problem and better schedule an incoming job, but the time spent on these checks comes at the expense of serving existing jobs. 

Given the challenges associated with using SRPT in practice, it is not surprising that a long-standing direction for the field has been to understand how to (nearly) match the performance of SRPT while using less information about jobs.  For example, previous work has sought to understand the impact of noisy predictions \cite{wierman_scheduling_2008, dellamico_scheduling_2019,dellamico_psbs_2016, mitzenmacher_scheduling_2019, scully_uniform_2021, mitzenmacher2021queues},
has studied the role of using a finite (small) set of priority classes instead of exact sizes for prioritization (e.g., classifying jobs as either small or large) \cite{harchol-balter_performance_2013,reiman1983some,mitzenmacher2021queues,chen2021scheduling}; and recently the impact of using predictions of only a subset of the jobs has begun to receive attention \cite{shahout_skippredict_2024}. However, while these lines of work provide specific policies that reduce the size information required by SRPT, the broad question of what is the minimal information that is needed to match the performance of SRPT remains open. In this work, we ask:    

\begin{center}\emph{What is the minimal size information that enables order-optimal mean response time in heavy traffic? \\
How should scheduling policies adapt to predictions that are costly and/or inexact?}
\end{center}
More specifically, our work both (i) introduces and analyzes policies that effectively use the minimal number of costly, inexact job-size estimates to attain order-optimal mean response time in heavy traffic, when achievable, and (ii) characterizes what size information is necessary to achieve order-optimal mean response time in heavy traffic.  Further, we show how these results are impacted by the cost and error of size predictions.

\subsection{Our results}
We begin by focusing on the case where size predictions are both free and exact.  In this setting, our first set of results (Section \ref{sec:one-free-bit}) establishes that \emph{binary size advice for each job} suffices for order-optimal mean response time in heavy traffic in a general class of lognormal-or-lighter service distributions that includes the exponential service distribution. In particular, we introduce the \emph{Predict and Demote} policy that operates a high-priority and a low-priority queue as follows. Upon arrival of a job, the advice communicates whether the job size is above a carefully selected size threshold; this size advice is binary (and thus minimal) for each job. If jobs are present in both queues, the high-priority queue is always prioritized. Within each queue, we prioritize jobs in increasing order of their arrival, i.e., based on \emph{First Come First Served (FCFS)}. Surprisingly, this simple algorithm attains order-optimal mean response time in heavy traffic (Theorem~\ref{thm:free-bit-sufficient}). We complement this result by showing that for a class of distributions outside this family, no assignment of jobs to two FCFS priority queues using binary size advice attains the SRPT order (Theorem~\ref{thm:free-bit-impossible}). The latter result shows that binary size advice does not always recover the improvements of full size information. 

Our second set of results (Section \ref{sec:sparse-predict}) builds on the first to show that binary predictions are not needed for every job; rather, the aforementioned improvements only require advice for a \emph{vanishing fraction of incoming jobs}. In particular, we introduce the \emph{Sparse Predict and Demote (SPD)} policy, which obtains binary size advice for a sparse selection of jobs and demotes to the low-priority queue any queried job that has remaining size greater than a carefully chosen size threshold that we refer to as the \emph{residual demotion barrier}. The simplest way to attain the sparse set is by querying any arriving job with a vanishingly small probability.  We show that SPD achieves an order-optimal mean response time in heavy traffic for any lognormal-or-lighter service distribution with such sparse predictions at arrival (Theorem~\ref{thm:sparse-arrival-optimal}). That said, the same guarantee can also be attained when the algorithm queries jobs only if it has not finished after a particular amount of service that is carefully selected such that a vanishing fraction of jobs end up being queried (Theorem~\ref{thm:sparse-uniform-optimality}).  

We next address the question of what happens outside of the lognormal-or-lighter family of distributions (Section \ref{sec:sparse-srpt}).  We show that SPD can attain heavy-traffic order-optimality for a class of service distributions outside this family by obtaining exact sizes for a vanishing fraction of jobs and using SRPT in the low-priority queue (Theorem~\ref{thm:sparse-srpt-intermediate}). In particular, knowing the remaining sizes of demoted jobs lets shorter jobs bypass larger ones in this queue. This controls the delay of a larger demoted population, allowing us to remove enough work from high priority to attain the SRPT order while still querying a vanishing fraction of arrivals.

Our final analytic result is for the setting where size advice comes at a cost and is inexact (Section~\ref{sec:sparse-costly}).  Here, we show how the design of SPD must adapt to these challenges.  In particular, for a given error parameter $\varepsilon\in(0,1)$, we allow the size advice to be any number within a $(1\pm \varepsilon)$ multiplicative interval of the true size (possibly adversarially selected). We also assume that a size query for a specific job requires some inference duration during which the server is occupied and cannot serve any job. As long as the inference duration asymptotically decays as a power function of the system slack $1-\rho$, we establish that SPD remains heavy-traffic order-optimal for lognormal-or-lighter service distributions under such costly and inexact size advice (Theorem~\ref{thm:sparse-constant-error}).  Interestingly, in the presence of costs, the two sparsity mechanisms (query with some probability at arrival and query upon receiving some amount of service) are not interchangeable: heavy-traffic optimality requires carefully tuning these two mechanisms to the service distribution (Proposition~\ref{prop:sparse-cost-distributions}). 

Our analytic results focus exclusively on heavy-traffic performance.  We conclude the paper with numerical evaluations that establish that SPD performs well outside of the heavy-traffic regime too (Section \ref{sec:experiments}).  These experiments highlight that the design insights that lead to SPD are useful beyond heavy-traffic.

\subsection{Our techniques}
To demonstrate our technical contributions, we use exponential service distributions as an illustrative example. Recall that the optimal scheduling algorithm (SRPT) has mean response time that scales as $\Theta(1/((1-\rho)\log(1/(1-\rho)))$ in this setting, where $1-\rho$ is the system slack. In contrast, FCFS enjoys the optimal size-oblivious mean response time of $\Theta\left(1/(1-\rho)\right)$. 

The key to attaining the SRPT order with sparse binary advice is to balance the system slack of the high-priority queue with the fraction of jobs that are demoted to the low-priority queue. By demoting jobs with sufficiently large remaining size (via an appropriate residual demotion barrier), we remove sufficient load from the high-priority queue. This increases its system slack to $1-\rho^{\textsc{HIGH}}=\Theta((1-\rho)\log\left(1/(1-\rho)\right))$, which enables FCFS to attain the optimal size-aware mean response time on this queue. That said, if we demote too many jobs, the low-priority queue may end up suffering. To circumvent this, the selection of the residual barrier together with the sparsity mechanism is done in a way that only a $O(1-\rho)$ fraction of jobs is demoted; this controls the contribution of demoted jobs to the overall mean response time. Interestingly, we can show that a queried job has remaining size greater than the selected residual demotion barrier with probability $\omega(1-\rho)$ (see Lemma~\ref{lem:sparse-candidate-abundance}). Hence, querying a vanishingly small fraction of jobs suffices to ensure that $1-\rho$ fraction of jobs are demoted.

Moving to costly and inexact predictions, we need a more refined approach towards the sparsity mechanism. In the presence of costs, predictions introduce additional load to the system. To maintain the optimality guarantees, the sparsity mechanism should be selected in a way that ensures that this additional load is at most a constant fraction of the original system slack. This is only achievable if the cost from the minimum number of queries that are needed to ensure sufficient demotion of jobs to the low-priority queue is controllable. As a result, for a given cost per query, the optimal sparsity mechanism needs to minimize the \emph{self-information}, the negative logarithm of the probability that a queried job qualifies for demotion under exact predictions. We show that, for our results to extend to costly predictions, this self-information must be small enough to preserve a fixed fraction of the slack after accounting for inference costs (Lemma~\ref{cor:sparse-gamma-cost}). This result provides a toolbox towards handling inexact predictions. In particular, we create a finite collection of candidate sparsity mechanisms whose size depends only on the inference-cost exponent and show that their self-information values sum to at most $-\log(1-\rho)$, so at least one mechanism has inference load negligible relative to the original slack (Lemma~\ref{lem:sparse-error-slack-preserving-pair}), thereby asymptotically preserving the original system slack  despite the additional inference load.

\subsection{Related work}
Our work contributes to two distinct literatures: (i) the study of size-aware scheduling in queueing systems and (ii) the study of online algorithms with advice (a.k.a. learning-augmented online algorithms).  We discuss each in turn. 

\textit{Size-aware scheduling.} 
The field of scheduling in single-server queueing systems is classical.  Priority-based policies, including size-based and age-based policies, have well-known closed-form characterizations; e.g., see \cite{harchol-balter_performance_2013, kleinrock_queueing_1974}. In fact, recent work has provided an analysis of a very broad class of priority-based policies, beyond the classically studied policies, via the SOAP classification (see Section \ref{subsec:soap-overview}) \cite{scully_soap_2018}.  SRPT has been known to be optimal for mean response time for decades, since \cite{schrage1966queue,schrage1968proof} and has received considerable attention as a result.  Most relevant to the current work is analysis of its heavy traffic behavior  \cite{lin_heavy-traffic_2011,gromoll2011diffusion,banerjee2022heavy}.  However, the task of understanding the minimal amount of size information necessary to be nearly optimal remains unanswered and has received considerable attention in recent years. In particular, while classic results focus on situations where policies are either size-aware or size-oblivious, e.g., \cite{scully2020simple,nuyens_foregroundbackground_2008,scully2021gittins}, recently work has begun to focus on situations where exact sizes are not known, but approximate sizes are used for prioritization \cite{wierman_scheduling_2008, dellamico_scheduling_2019,dellamico_psbs_2016, mitzenmacher_scheduling_2019, scully_uniform_2021, mitzenmacher2021queues}. A traditional approach for system design using prioritization is the multilevel queue (MLQ) policy, which adjusts priorities as jobs receive service without requiring advance knowledge of their processing times \cite{Corbató2001, becchetti2006average}. Recent work has increasingly studied policies that place jobs into a finite set of priority classes, typically based on size, as a way to minimize the amount of size information that needs to be predicted.  Such policies have a long history, e.g., see \cite{kleinrock_queueing_1974,harchol-balter_performance_2013,reiman1983some} and the references therein, and have received renewed attention in recent years motivated by the goal of scheduling with minimal size information \cite{mitzenmacher2021queues,chen2021scheduling,scully2021schedule}. 
Additionally, motivated by predictions via costly AI inferences, very recent work has focused on the goal of reducing the number of jobs for which size predictions are requested \cite{shahout_skippredict_2024, mishra2026scheduling}.

From the literature reviewed above, the most related papers to our work are those that focus on the design of algorithms that use limited size information to provide performance that is robust to costly and/or inexact predictions.  In particular, \cite{scully_uniform_2021} shows that a simple scheduling policy can achieve near-optimal performance with accurate estimates while remaining robust to prediction errors. Building on this, more recent work in \cite{shahout_skippredict_2024} introduces algorithms that selectively request predictions, outperforming those that predict every job.  The DelayPredict policy presented in this work serves as a baseline in our experimental results and is a special case of the SPD policy we introduce in this paper. Finally, the most related work to the current paper is \cite{mitzenmacher2021queues, chen2021scheduling}, which both study the performance of policies that use one-bit predictions in order to place jobs into two priority queues.  \cite{mitzenmacher2021queues} has a different focus than the current paper.  It uses mean-field analysis to study a multi-server system.  On the other hand, \cite{chen2021scheduling} studies the heavy-traffic behavior of a single server with two priority queues facing general arrival and service rates.  The results of \cite{chen2021scheduling} are similar in spirit to our warmup results in Section \ref{sec:free-bit-sufficient}.  They show that, by categorizing jobs as either large or small, it is possible to asymptotically recover the mean response time of SRPT in heavy traffic.  Although this result is qualitatively similar to Theorem~\ref{thm:free-bit-sufficient} of our paper, the results are not directly comparable. \cite{chen2021scheduling} considers more general arrival distribution patterns, which prevents the derivation and use of explicit characterizations of the mean response time as a function of load. Moreover, their result is restricted to a special case of service distributions.  In contrast, we provide a full characterization across all service distributions, showing when a two-class policy (a.k.a. one-bit predictions) is, and is not, enough. Further, there is no parallel in \cite{chen2021scheduling} for our results on the use of a vanishing fraction of predictions or the consideration of costly or inexact predictions. 

\textit{Online algorithms with advice.}
The field of online algorithms with (machine-learned) advice is a relatively new area, which emerged following the work of \cite{lykouris_competitive_2021}. Since then, a wide range of online problems have been explored wthin this framework; a recent survey of this field can be found in \cite{mitzenmacher_algorithms_2022} and an online repository of related work is available at \cite{lindermayr_alps_nodate}.
Within the broad area of scheduling and resource allocation, a number of problems have been studied, including quantifying prediction error costs \cite{mitzenmacher_scheduling_2019}, designing robust policies under misprediction \cite{scully_uniform_2021}, and load-balancing of large databases \cite{ahmadian_robust_2023}. Notably, the key assumption in nearly all of this work is that predictions are provided at no cost. An exception to this is the literature on online learning with hints \cite{bhaskara_online_2020, cutkosky_artificial_2019} and parsimonious learning-augmented online algorithms \cite{im_parsimonious_2022, bampis_parsimonious_2024}, where the economic cost associated with predictions is considered. However, the question of how to manage the interference of computational resources in algorithms with machine-learned advice has not been explored. Our work presents results addressing this interference in the context of online scheduling.

\section{Model and Preliminaries}
\label{sec: prelims}
We study the M/GI/1 queueing model in heavy traffic. In this model, jobs arrive at a unit-speed single server as a Poisson process of rate \(\lambda_\rho\). Their sizes \(S\sim F\) are independent and identically distributed (i.i.d.) random variables drawn from a service distribution \(F\) and independent of the arrival process. With mean job size \(\mathbb E[S]\), the unit-speed server has service rate $ \mu:=1 / {\mathbb E[S]}$. We assume that (in the absence of inference work) the system is stable, that is, the arrival rate is smaller than the service rate. As a result, $ \lambda_\rho:=\rho\mu$ for some $\rho\in(0,1)$, where \(\rho=\lambda_\rho\mathbb E[S]\) is the ordinary-service load, and \(1-\rho\) is the \emph{ordinary-service slack}, the fraction of server capacity left after ordinary service, before inference work is added. 
A scheduling policy \(\pi\) decides which job receives service and when to request a prediction; service may be preempted and resumed without repeating previously completed service. In a stable system, a policy \(\pi\) induces a steady-state response time~\(T^\pi\), the time from arrival to completion of a typical job. Our objective is to minimize its mean \(\mathbb E[T^\pi]\). 

We study the heavy-traffic limit \(\rho\uparrow1\), equivalently \(\lambda_\rho\uparrow\mu\), with the service distribution fixed. As the ordinary-service slack \(1-\rho\) vanishes, we compare the rates at which mean response times diverge under different scheduling policies. When the load is fixed, we suppress its subscript and write \(\lambda\).
To state the relevant heavy-traffic benchmarks, we define the workload distribution
\begin{equation}
    G(x)
    :=
    \frac{\mathbb E[S\mathbf 1\{S\le x\}]}
         {\mathbb E[S]},
\end{equation}
which is the fraction of the  workload contributed by jobs of size at most \(x\). Hence, \(G^{-1}(\rho)\) is the size cutoff above which jobs contribute exactly a \(1-\rho\) fraction of the total workload: \(\mathbb E[S\mathbf1\{S>G^{-1}(\rho)\}]=(1-\rho)\mathbb E[S]\).

Shortest Remaining Processing Time (SRPT) always serves the job with the smallest true remaining size. Under Assumption~\ref{asm:heavy-traffic-regularity} on the job size distribution (see Section~\ref{subsec:service-time-classes}), the heavy-traffic characterization of \cite{lin_heavy-traffic_2011} gives that the optimal size-aware mean response time, attained by SRPT:
\begin{equation}
\label{eq:srpt-heavy-traffic-benchmark}
    \mathbb E\left[T^{\mathrm{SRPT}}\right]
    =
    \Theta\!\left(
        \frac{1}
        {(1-\rho)G^{-1}(\rho)}
    \right).
\end{equation}
Since the service distribution has unbounded support, \(G^{-1}(\rho)\to\infty\) as \(\rho\uparrow1\).
We call a policy \emph{heavy-traffic order-optimal} if its mean response time is within a constant factor of this benchmark as \(\rho\uparrow1\). Our goal is to design heavy-traffic order-optimal policies when only given access to inexact and costly size predictions for only a small fraction of jobs. We first assume predictions are exact and free (Sections~\ref{sec:one-free-bit} and~\ref{sec:sparse-predict}). We then allow predictions to require inference work on the same server competing with job service (Section~\ref{sec:sparse-costly-exact}), and be inexact (Section~\ref{sec:sparse-predict-errors}).

\subsection{Classes of Service Distributions}
\label{subsec:service-time-classes}
The relevance of job-size information depends strongly on the service-time tail. Let \(f\) denote the density of \(F\), and write $\overline F(x):=\mathbb P(S>x)$. The hazard rate $h(x):={f(x)}/{\overline F(x)}$ is the instantaneous completion rate at size \(x\), conditional on the job requiring at least \(x\) units of service. For ease of exposition, we restrict attention to distributions with unbounded support and with continuous, strictly positive hazard rate for \(x>0\). 
We organize service distributions according to the asymptotic growth of their cumulative hazard 
\[
H(x):=\int_0^x h(t)\,dt=-\log \overline F(x);
\]
faster growth of \(H\) means faster decay of the service-time tail.

Our main focus is the \emph{lognormal-or-lighter regime}, which begins at \(H(x)=\Theta((\log x)^2)\) and contains all faster-growing cumulative hazards. This regime contains many classical service distributions. Lognormal-type tails lie at the boundary, with \(H(x)\sim A(\log x)^2\) for \(A>0\). Weibull tails have \(H(x)=(x/c)^\nu\) for \(c,\nu>0\), which grows faster than \((\log x)^2\); the exponential distribution is the classical Weibull case \(\nu=1\). Doubly exponential tails provide another example: \(H(x)=e^{cx}-1\) with \(c>0\) growing faster than every power of~\(x\).
To place them in context, letting \(f(x)\ll g(x)\) denote \(f(x)=o(g(x))\), we distinguish three tail regimes:\footnote{Since \(\overline F(x)=e^{-H(x)}\), larger values of \(H(x)\) mean smaller tail probabilities. Faster growth of \(H\) therefore corresponds to a lighter service-time tail.}
\[
\underbrace{H(x)=\Omega\!\left((\log x)^2\right)}_{\text{lognormal-or-lighter}}
    \qquad\Big|\qquad
    \underbrace{\log x\ll H(x)\ll(\log x)^2}_{\text{intermediate}}
    \qquad\Big|\qquad
    \underbrace{H(x)=O(\log x)}_{\text{Pareto-or-heavier}},
\]
The intermediate regime consists of tails that decay faster than every power law but remain heavier than the boundary of the lognormal-or-lighter regime; for example, \(H(x)\sim(\log x)^r\) with \(1<r<2\). 

The Pareto-or-heavier regime contains Pareto and other power-law-type tails: \(H(x)=O(\log x)\) implies that \(\overline F(x)\ge x^{-C}\) for some \(C<\infty\) and all sufficiently large \(x\). For such distributions, it is known that size-oblivious policies can maintain heavy-traffic order-optimality.  For example, for Pareto service distributions with finite variance, the size-unaware Foreground--Background (FB) policy, which prioritizes jobs with the least attained service, is known to achieve the same heavy-traffic order as size-aware SRPT \cite{nuyens_foregroundbackground_2008,bansal2006handling}. This motivates our focus on the intermediate and lognormal-or-lighter regimes.

Assumption~\ref{asm:heavy-traffic-regularity} states the conditions on the service distribution that are assumed throughout the paper. Note that we omit it from individual result statements to avoid repetition.
\begin{assumption}
\label{asm:heavy-traffic-regularity} The service distributions are assumed to satisfy the following conditions.
\begin{enumerate}
    \item The service distribution satisfies \(\frac{H(x)}{\log x}\longrightarrow\infty\).
    \item \( \alpha(\overline F)<-2\), where \(\alpha(\overline F)\) denotes the upper Matuszewska index \footnote{
    For a positive function \(g\), by a slight modification of Definition 1 in \cite{lin_heavy-traffic_2011}, its upper Matuszewska index can be defined by
    \[
        \alpha(g)
        :=
        \inf\left\{
            a\in\mathbb R:
            \exists C,x_0>0
            \text{ such that }
            g(y)
            \le
            C\left(\frac{y}{x}\right)^a g(x)
            \text{ for all }y\ge x\ge x_0
        \right\}.
    \]} of \(\overline F\).
    \item The extended limit $\lim_{x\to\infty}\frac{x h(x)}{\log x}\in[0,\infty]$ exists.
\end{enumerate}
\end{assumption}
The first condition restricts attention to the intermediate and lognormal-or-lighter regimes, and excludes Pareto-or-heavier tails. The second is a regularity condition used in the SRPT heavy-traffic characterization of \cite{lin_heavy-traffic_2011}. The third rules out persistent oscillations across the boundary of the lognormal-or-lighter regime and distinguishes the two regimes. Under Assumption~\ref{asm:heavy-traffic-regularity}, \(\mathbb E[S]\) and \(\mathbb E[S^2]\) are finite as \(\rho\uparrow1\) (Appendix~\ref{app:service-finite-moments}), the extended limit classifies the tail regimes (Appendix~\ref{subsec:log-quadratic-boundary}), and the tail workload relates to the tail probability (Appendix~\ref{app:free-bit-tail-work-proofs}).

\subsection{Brief Overview of the SOAP Framework}
\label{subsec:soap-overview}

In our analysis, we repeatedly apply the SOAP framework \cite{scully_soap_2018} to derive response times of SPD and its variations. In this framework, a job is described by its \emph{age}~\(a\), the total service it has received, including inference work when a costly prediction is requested, together with a \emph{descriptor} \(d\) recording job-specific information such as a prediction or whether the job is selected for a query. SOAP then assigns the job a rank \(r(d,a)\), with smaller rank corresponding to higher priority, and always serves the job with smallest rank, breaking ties in favor of the job that arrived first (first-come, first-served, or FCFS). For delayed predictions, the rank function simply does not use the prediction until the age at which it becomes available.\footnote{Equivalently, for the purpose of the SOAP representation, the descriptor may encode the eventual prediction outcome, while the rank function ignores that component before the prediction is activated.} 
In the SOAP formulas, \(x\) denotes a tagged job's total server work; it equals its ordinary size when predictions are free and includes inference work otherwise. For a tagged job with descriptor \(d\) and total work \(x\), let \(T_{d,x}\) denote its response time. This is the sum of its \emph{waiting time}, from arrival until it first receives service, and its \emph{residence time}, from first service until completion, including any interruptions.

We first consider the simpler case in which $a\longmapsto r(d,a)$  is non-increasing for every descriptor \(d\). In this case, a job's priority can only improve as it receives service; this is indeed the case for  several canonical policies including SRPT and fixed-rank priority policies. To describe the response time of a tagged job, fix a rank \(r\). With FCFS tie-breaking, an older job wins an equal-rank tie. Service is \emph{relevant to rank \(r\)} if it can be served ahead of a tagged job of rank \(r\). Thus old-job service is relevant to
ranks at most \(r\), while new-job service is relevant only to 
ranks strictly below \(r\). Let \(X^{\mathrm{new}}[r]\) denote the total server work of a newly arriving job if its initial rank is strictly below \(r\), and zero otherwise; a non-increasing rank remains below \(r\) once it starts there. For jobs that arrived before the tagged job, SOAP similarly distinguishes relevant service according to whether the job is relevant from its arrival or becomes relevant only after receiving some service. We denote the corresponding amounts of service by \(X_0^{\mathrm{old}}[r]\) and \(X_1^{\mathrm{old}}[r]\), respectively. When ranks are non-increasing, a job can cross the threshold \(r\) at most once, so there are no further relevant intervals. Define
$\rho^{\mathrm{new}}[r] := \lambda \mathbb E[X^{\mathrm{new}}[r]]$ and $\rho_0^{\mathrm{old}}[r] := \lambda \mathbb E[X_0^{\mathrm{old}}[r]]$.

\begin{proposition}[Non-increasing-rank case of SOAP {\citep[Theorem~5.5]{scully_soap_2018}}]
\label{prop:soap-nonincreasing}
Suppose that \(a\mapsto r(d,a)\) is non-increasing for every descriptor \(d\). Then the conditional mean response time of a job with descriptor \(d\) and total work \(x\) is
\[
    \mathbb E[T_{d,x}]
    =
    \underbrace{
        \frac{
            \lambda\left(
                \mathbb E\!\left[(X_0^{\mathrm{old}}[r(d,0)])^2\right]
                +
                \mathbb E\!\left[(X_1^{\mathrm{old}}[r(d,0)])^2\right]
            \right)
        }{
            2\left(1-\rho_0^{\mathrm{old}}[r(d,0)]\right)
            \left(1-\rho^{\mathrm{new}}[r(d,0)]\right)
        }
    }_{\text{expected waiting time}}
    +
    \underbrace{
        \int_0^x
        \frac{da}{1-\rho^{\mathrm{new}}[r(d,a)]}
    }_{\text{expected residence time}} .
\]
\end{proposition}
\stepcounter{proposition}
To interpret this formula, note that in the waiting-time denominator, \(1-\rho_0^{\mathrm{old}}[r]\) measures the slack against relevant work already present when the tagged job arrives, while \(1-\rho^{\mathrm{new}}[r]\) measures the slack against relevant work generated by future arrivals. The residence term accumulates the reciprocal of this latter slack as the tagged job progresses through its rank trajectory. 
For example, with no inference work, FCFS has \(\rho_0^{\mathrm{old}}=\rho\) and \(\rho^{\mathrm{new}}=0\), so the vanishing slack in the waiting-time denominator gives \(\mathbb E[T^{\mathrm{FCFS}}]=\Theta(1/(1-\rho))\), a factor of order \(G^{-1}(\rho)\) above SRPT. Details of this calculation are given in Appendix~\ref{app:soap-fcfs}.

A similar response-time formula holds even when ranks increase at some ages; Appendix~\ref{app:soap-general-intuition} gives this more general result in Proposition~\ref{prop:soap-response-time} and its intuition. The proof of Proposition~\ref{prop:soap-nonincreasing} follows directly from this more general proposition by specializing to non-increasing ranks.

\section{The Power and Limits of Binary-Level Queues}
\label{sec:one-free-bit}

A \emph{multilevel queue} (MLQ) scheduler organizes jobs into a finite set of priority queues. Scheduling then involves two decisions: which queue receives service and which job is served within that queue. In the strict preemptive-priority version considered here, the server always works on the highest-priority nonempty queue; a job in a lower-priority queue is interrupted whenever higher-priority work becomes available. A separate rule, such as FCFS, determines the service order within each queue. Assigning jobs to queues therefore allows the scheduler to favor some jobs without maintaining a complete ordering of all jobs by size.

In this section, we study the simplest such system: \emph{binary-level queues} (BLQ), with just a high-priority queue and a low-priority queue, both served according to FCFS. For each load \(\rho\), every arriving job receives a size prediction, which the scheduler uses to assign a binary rank \(R_\rho\in\{0,1\}\). Ranks \(0\) and \(1\) designate the high-priority and low-priority queues, respectively. Each job keeps this rank until completion. Thus, the prediction is used only for the initial queue assignment: afterward, the scheduler retains one bit per job (its rank) and uses FCFS within each queue.

Formally, a BLQ policy is specified by a measurable assignment map \(\varphi_\rho:\mathbb R_+\to[0,1]\), where $\mathbb P(R_\rho=1\mid S=s)=\varphi_\rho(s)$. Thus, \(\varphi_\rho(s)\) is the probability of assigning low priority to a job of size \(s\). Values in \(\{0,1\}\) give deterministic size partitions; values in \((0,1)\) allow randomization, independent across arrivals and of the queue state conditional on job sizes. We also use \(\varphi_\rho\) to denote the induced BLQ policy. The high-priority load and slack are
$\rho^{\mathrm{HIGH}}
    :=\lambda_\rho\mathbb E[S\mathbf 1\{R_\rho=0\}]$ and $
    1-\rho^{\mathrm{HIGH}}
    =(1-\rho)+\lambda_\rho\mathbb E[S\mathbf 1\{R_\rho=1\}]$ respectively.
Note that assigning work to the low-priority queue
creates additional slack for the high-priority queue.
Since ranks remain fixed (and are thus non-increasing), Proposition~\ref{prop:soap-nonincreasing} applies. The next two lemmas give the response-time contribution of each queue and follow from standard arguments \cite{kleinrock_queueing_1974,harchol-balter_performance_2013}. We provide their proof for completeness in Appendix~\ref{app:free-bit-response-identities} and a proof sketch below. 

\begin{lemma}[Response time of the high-priority queue]
\label{lem:free-bit-high-priority-response}
Under any BLQ policy $\varphi_\rho$,
the contribution of the high-priority queue to the overall mean response time is
\begin{equation*}
    \mathbb E\!\left[
        T^{\varphi_\rho
        }\mathbf 1\{R_\rho=0\}
    \right]
    =
    \mathbb P(R_\rho=0)
    \frac{
        \lambda_\rho
        \mathbb E\!\left[
            S^2\mathbf 1\{R_\rho=0\}
        \right]
    }{
        2(1-\rho^{\mathrm{HIGH}})
    }
    +
    \mathbb E\!\left[
        S\mathbf 1\{R_\rho=0\}
    \right].
\end{equation*}
\end{lemma}

\begin{lemma}[Response time of the low-priority queue]
\label{lem:free-bit-low-priority-response}
Under any BLQ policy $\varphi_\rho$, the contribution of the low-priority queue to the overall mean response time is
\begin{equation*}
    \mathbb E\!\left[
        T^{\varphi_\rho}\mathbf 1\{R_\rho=1\}
    \right]
    =
    \mathbb P(R_\rho=1)
    \frac{
        \lambda_\rho\mathbb E[S^2]
    }{
        2(1-\rho)(1-\rho^{\mathrm{HIGH}})
    }
    +
    \frac{
        \mathbb E\!\left[
            S\mathbf 1\{R_\rho=1\}
        \right]
    }{
        1-\rho^{\mathrm{HIGH}}
    }.
\end{equation*}
\end{lemma}
\begin{proof}[Proof sketch of Lemmas~\ref{lem:free-bit-high-priority-response} and~\ref{lem:free-bit-low-priority-response}] 
Both identities follow from Proposition~\ref{prop:soap-nonincreasing}. A high-priority tagged job waits only for older high-priority work; FCFS keeps future arrivals behind it. A low-priority tagged job waits for all older work and can be overtaken only by future high-priority arrivals. High-priority residence is uninterrupted, whereas high-priority interruptions stretch low-priority residence by the reciprocal high-priority slack. Substituting these quantities and averaging over each rank gives the identities.
\end{proof}
These two lemmas show how the queue assignment affects mean response time. Routing a job to the low-priority queue removes its size from the load that can delay high-priority jobs. However, the demoted job must now wait for high-priority work as well. Its longer delay is weighted by the probability of demotion in the overall mean. Selecting large jobs for demotion can therefore reduce the high-priority load substantially while making only a small fraction of arrivals incur low-priority delay. We next choose a size cutoff that balances the work removed from high priority against the fraction of arrivals sent to low priority.

\subsection{When Does One Free Bit Suffice?}
\label{sec:free-bit-sufficient}

We begin by showing that a simple BLQ policy suffices to obtain heavy-traffic order-optimal mean response time.  In particular, our warm-up policy, \emph{Predict and Demote}, PD($b$), uses a residual demotion barrier \(b\ge0\), predicting at the arrival of every job whether its \emph{remaining size} exceeds \(b\). Since the job has received no service at arrival, its remaining   size  is exactly \(S\), giving $
    R_\rho=\mathbf 1\{S>b\}$.
The comparison is made once, so a demoted job stays in the low-priority queue through completion.

To choose \(b\), we first consider how often jobs should be demoted. In Lemma~\ref{lem:free-bit-low-priority-response}, the low-priority waiting contribution contains the factor \(\mathbb P(R_\rho=1)/(1-\rho)\). Setting the demotion probability to \(1-\rho\) makes this factor equal to one: the small fraction of demoted jobs offsets the additional waiting factor \(1/(1-\rho)\) they experience. Both queues' waiting contributions can then be bounded on the scale \(1/(1-\rho^{\mathrm{HIGH}})\). Among all assignments with demotion probability \(1-\rho\), choosing the largest jobs removes the most expected service per arrival from the high-priority queue. This motivates the \emph{slack-matched size cutoff}
\begin{equation}
\label{eq:free-bit-threshold}
    \bar{s}_\rho=\overline F^{-1}(1-\rho).
\end{equation}
Under \(\mathrm{PD}(\bar{s}_\rho)\), this cutoff selects the largest \(1-\rho\) fraction of arrivals for the low-priority queue. The next lemma states this demotion probability, which we use to bound the low-priority queue's contribution to the mean response time. Its proof follows directly as \(\mathbb P(R_\rho=1)=\overline F(\bar{s}_\rho)=1-\rho\).

\begin{lemma}[Low-priority fraction at $\bar{s}_\rho$]
\label{lem:free-bit-low-priority-fraction}
Under \(\mathrm{PD}(\bar{s}_\rho)\),
$\mathbb P(R_\rho=1)=1-\rho$. 
\end{lemma}
This demotion probability determines the fraction of arrivals that incur low-priority delay. We must also show that their demotion removes enough work from the high-priority queue. Every demoted job has more than \(\bar{s}_\rho\) service remaining, so together they remove at least \((1-\rho)\bar{s}_\rho\) service per arrival from that queue. In the  lognormal-or-lighter   regime, the next lemma (proof in Appendix~\ref{app:free-bit-sufficient}) shows that \(\bar{s}_\rho\) is on the \(G^{-1}(\rho)\) scale. The resulting high-priority slack is therefore at least of order \((1-\rho)G^{-1}(\rho)\), which makes the waiting-time bound match the SRPT benchmark.

\begin{lemma}[Asymptotic scale of $\bar{s}_\rho$]
\label{lem:free-bit-threshold-location}
For distributions in the lognormal-or-lighter   regime, there exists a constant \(\bar{\theta}\in(0,1]\) such that the slack-matched size cutoff satisfies $\frac{\bar{s}_\rho}{G^{-1}(\rho)}
    \longrightarrow
    \bar{\theta}$. 
\end{lemma}

\begin{theorem}[One free bit suffices in the lognormal-or-lighter  regime]
  \label{thm:free-bit-sufficient}
For distributions in the     lognormal-or-lighter   regime, \(\mathrm{PD}(\bar{s}_\rho)\) is heavy-traffic order-optimal, i.e., $\mathbb E[T^{\mathrm{PD}(\bar{s}_\rho)}] =
    \Theta\!\left(
        \mathbb E[T^{\mathrm{SRPT}}]
    \right)$.
\end{theorem}

\begin{proof}
By Lemmas~\ref{lem:free-bit-low-priority-fraction} and~\ref{lem:free-bit-threshold-location}, \(\mathbb P(R_\rho=1)=1-\rho\) and \(\bar{s}_\rho=\Theta(G^{-1}(\rho))\). Since each demoted job has size greater than \(\bar{s}_\rho\), the high-priority slack satisfies
\begin{equation}
\label{eq:free-bit-high-slack}
    1-\rho^{\mathrm{HIGH}}
    \ge \lambda_\rho\bar{s}_\rho(1-\rho)
    =\Omega\!\left((1-\rho)G^{-1}(\rho)\right).
\end{equation}
For the waiting contributions, Lemmas~\ref{lem:free-bit-high-priority-response} and~\ref{lem:free-bit-low-priority-response} give the following bounds for the high-priority and low-priority queues, respectively:
\[
\begin{aligned}
    \mathbb P(R_\rho=0)
    \frac{\lambda_\rho\mathbb E[S^2\mathbf 1\{R_\rho=0\}]}
         {2(1-\rho^{\mathrm{HIGH}})}
    &\le
    \frac{\lambda_\rho\mathbb E[S^2]}
         {2(1-\rho^{\mathrm{HIGH}})},\\[1mm]
    \mathbb P(R_\rho=1)
    \frac{\lambda_\rho\mathbb E[S^2]}
         {2(1-\rho)(1-\rho^{\mathrm{HIGH}})}
    &=
    \frac{(1-\rho)\lambda_\rho\mathbb E[S^2]}
         {2(1-\rho)(1-\rho^{\mathrm{HIGH}})}=
    \frac{\lambda_\rho\mathbb E[S^2]}
         {2(1-\rho^{\mathrm{HIGH}})}.
\end{aligned}
\]

In the second calculation, Lemma~\ref{lem:free-bit-low-priority-fraction} lets us replace \(\mathbb P(R_\rho=1)\) by \(1-\rho\), which cancels the same factor in the denominator. The residence contributions are at most \(\mathbb E[S]\) and \(1/\lambda_\rho\), respectively; the latter follows from
$
    1-\rho^{\mathrm{HIGH}}
    =(1-\rho)+\lambda_\rho\mathbb E[S\mathbf 1\{R_\rho=1\}]
    \ge\lambda_\rho\mathbb E[S\mathbf 1\{R_\rho=1\}]$.
Summing the waiting and residence contributions from Lemmas~\ref{lem:free-bit-high-priority-response} and~\ref{lem:free-bit-low-priority-response} gives
\[
\begin{aligned}
    \mathbb E[T^{\mathrm{PD}(\bar{s}_\rho)}]
    &\le
    \frac{\lambda_\rho\mathbb E[S^2]}{1-\rho^{\mathrm{HIGH}}}
    +\mathbb E[S]+\frac{1}{\lambda_\rho}=O\!\left(\frac{1}{(1-\rho)G^{-1}(\rho)}\right),
\end{aligned}
\]
where we used \eqref{eq:free-bit-high-slack}, \(\mathbb E[S^2]<\infty\), and \(\lambda_\rho=\Theta(1)\). Finally, SRPT optimality gives $\mathbb E[T^{\mathrm{PD}(\bar{s}_\rho)}]\ge\mathbb E[T^{\mathrm{SRPT}}]$; the SRPT mean response time in~\eqref{eq:srpt-heavy-traffic-benchmark} supplies the matching lower bound.
\end{proof}

\subsection{When Does One Free Bit Not Suffice?}
\label{sec:free-bit-impossible}
\label{app:free-bit-impossible}
\label{app:free-bit-impossibility-proof}

In the previous section, we determined a class of distributions for which one free bit suffices under PD($b$); we now focus on the opposite.  In particular, in this section we show that for service distributions in the intermediate regime of Section~\ref{subsec:service-time-classes}, no BLQ policy is heavy-traffic order-optimal. This statement holds for every measurable assignment map \(\varphi_\rho:\mathbb R_+\to[0,1]\), even when the map is chosen based on the load~\(\rho\).

An order-optimal policy must satisfy two requirements, formalized in Lemma~\ref{lem:free-bit-necessary-structure} (proved in Appendix~\ref{app:free-bit-necessary-structure-proof}). First, it must demote enough work to give the high-priority queue sufficient slack. Second, it must demote sufficiently few jobs to control the low-priority contribution to the overall mean response time. Together, these requirements force the average size of demoted jobs to be at least of order \(G^{-1}(\rho)\).

\begin{lemma}[Necessary size scale of the low-priority queue]
\label{lem:free-bit-necessary-structure}
Suppose a BLQ policy \(\varphi_\rho\) is heavy-traffic order-optimal along some sequence \(\rho\uparrow1\). Along that sequence,
\[
    \mathbb E[S\mathbf 1\{R_\rho=1\}]=\Omega\!\left((1-\rho)G^{-1}(\rho)\right)     
    \qquad \text{and} \qquad
    \mathbb E[S\mid R_\rho=1]=\Omega\!\left(G^{-1}(\rho)\right)
    .
\]
\end{lemma}
In the intermediate regime, even all jobs above any fixed positive multiple of \(G^{-1}(\rho)\) carry $o\!\left((1-\rho)G^{-1}(\rho)\right)$ work per arrival. The next lemma establishes this scarcity bound (proof in Appendix~\ref{app:free-bit-tail-work-scarcity-proof}); combined with the average-size requirement, this prevents a BLQ policy from demoting enough work to obtain optimality.
\begin{lemma}[Insufficient tail work in the intermediate regime]
\label{lem:free-bit-tail-work-scarcity}
For distributions in the intermediate regime, under Assumption~\ref{asm:heavy-traffic-regularity}, for every fixed \(d>0\),
$\mathbb E\!\left[
        S\mathbf 1\{
            S>dG^{-1}(\rho)
        \}
    \right]
    =
    o\!\left(
        (1-\rho)G^{-1}(\rho)
    \right)$.
\end{lemma}
Combining Lemmas~\ref{lem:free-bit-necessary-structure} and~\ref{lem:free-bit-tail-work-scarcity} gives the following impossibility result.

\begin{theorem}[BLQ impossibility in the intermediate regime]
\label{thm:free-bit-impossible}
For distributions in the intermediate regime, every BLQ policy \(\varphi_\rho\) is not heavy-traffic order-optimal, i.e.,
    \[
\mathbb E[T^{\varphi_\rho}]
    =
    \omega\!\left(
        \frac{1}{(1-\rho)G^{-1}(\rho)}
    \right).
\] 
\end{theorem}

\begin{proof}[Proof of Theorem~\ref{thm:free-bit-impossible}]
If the claimed \(\omega\) bound failed, there would be a sequence \(\rho\uparrow1\) along which \(\mathbb E[T^{\varphi_\rho}]=O(1/((1-\rho)G^{-1}(\rho)))\). Lemma~\ref{lem:free-bit-necessary-structure} gives constants \(c_0,c_1>0\) such that, eventually along the same sequence,
\[
    \mathbb E[S\mid R_\rho=1]\ge c_0G^{-1}(\rho) \qquad \text{and} \qquad
    \mathbb E[S\mathbf 1\{R_\rho=1\}]
    \ge c_1(1-\rho)G^{-1}(\rho).
\]

First consider the demoted jobs whose size is at most \((c_0/2)G^{-1}(\rho)\). Their work per arrival is at most this size threshold times the probability of demotion. Hence
\[
\begin{aligned}
    \mathbb E[S\mathbf 1\{R_\rho=1,\ S\le(c_0/2)G^{-1}(\rho)\}]
    &\le
    \frac{c_0}{2}G^{-1}(\rho)\mathbb P(R_\rho=1)\le
    \frac12\mathbb E[S\mathbf 1\{R_\rho=1\}].
\end{aligned}
\]
The second inequality follows from the average-size bound: multiplying \(\mathbb E[S\mid R_\rho=1]\ge c_0G^{-1}(\rho)\) by \(\mathbb P(R_\rho=1)\) gives \(\mathbb E[S\mathbf 1\{R_\rho=1\}]\ge c_0G^{-1}(\rho)\mathbb P(R_\rho=1)\). Thus demoted jobs at or below the threshold account for at most half of the demoted work, so demoted jobs above it account for at least half. The first expectation below counts the work of all jobs above the threshold, in either queue. The second counts only the demoted jobs above it. Since these are a subset of all jobs above the threshold and \(S\ge0\), we obtain
\[
\begin{aligned}
    \mathbb E[S\mathbf 1\{S>(c_0/2)G^{-1}(\rho)\}]
    \ge
    \mathbb E[S\mathbf 1\{R_\rho=1,\ S>(c_0/2)G^{-1}(\rho)\}]\ge
    \frac12\mathbb E[S\mathbf 1\{R_\rho=1\}]\ge
    \frac{c_1}{2}(1-\rho)G^{-1}(\rho).
\end{aligned}
\]
The last inequality uses the lower bound on total demoted work.
This contradicts Lemma~\ref{lem:free-bit-tail-work-scarcity} with \(d=c_0/2\).
\end{proof}

\section[Sparse Predict \&
Demote: A Vanishing Prediction Rate Suffices]{Sparse Predict \& Demote: A Vanishing Prediction Rate Suffices}
\label{sec:sparse-predict}

\begin{wrapfigure}{R}{0.5\textwidth}
\includegraphics[width=0.5\textwidth]{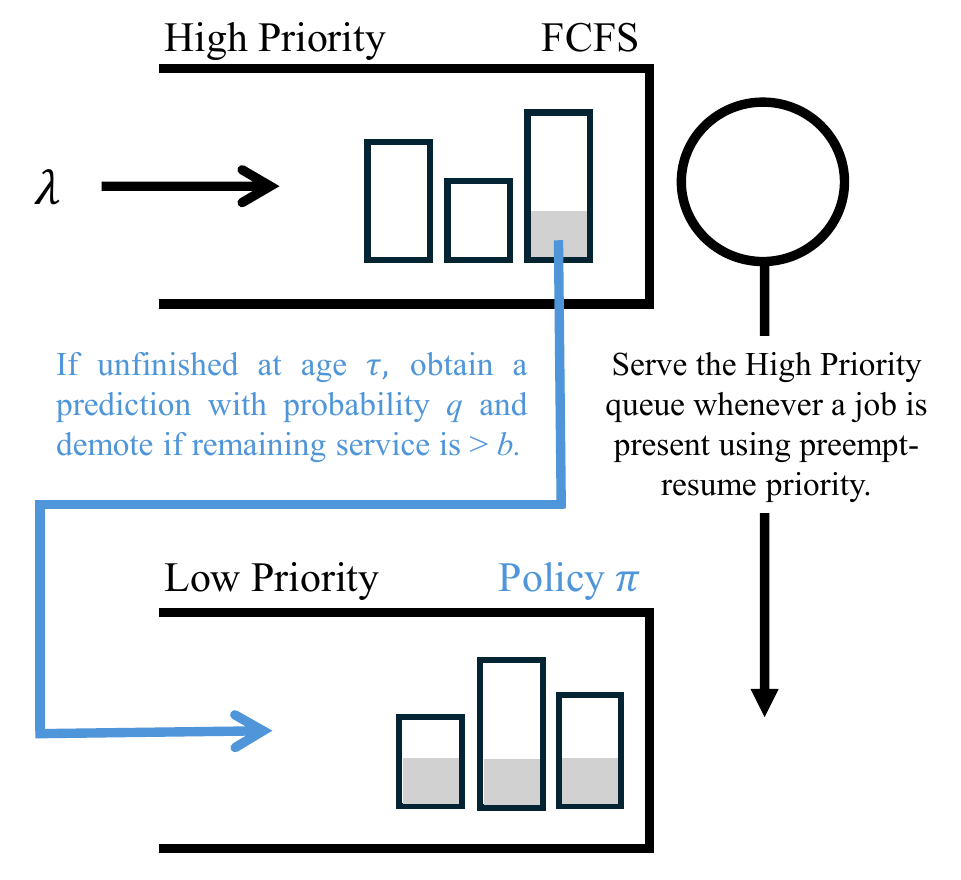}
\caption{Overview of the operation of Sparse Predict and Demote \(\mathrm{SPD}(b,q,\tau,\pi)\)  } \label{fig:spdoverview}
\end{wrapfigure}

The previous section highlights that a scheduler can be heavy-traffic order-optimal using only one bit of information per job when the job size distribution is lognormal-or-lighter. Here, we show that one can use even less job size information while maintaining heavy-traffic order-optimal mean response times. Specifically, we ask if one bit is needed for every job or if a policy can still maintain heavy-traffic order-optimality without querying size predictions from every job.   

To answer this question, we introduce a new algorithm, \emph{Sparse Predict and Demote} (SPD, a.k.a., ``speedy''), and show that it achieves heavy-traffic order-optimality using only a vanishing fraction of one-bit job size queries.  It achieves this by lowering the residual demotion barrier---the threshold on remaining service---to create more eligible jobs, then querying jobs at random to keep the demoted fraction at \(1-\rho\). At arrival, the eligible fraction grows relative to this target, allowing the fraction of arriving jobs that are queried to vanish. 

In the remainder of this section, we first establish this guarantee when deciding if a job will have a prediction queried for it at arrival, and then consider the case where queries may be delayed until a job has received a chosen amount of service, called the \emph{delay trigger}.  This additional lever can be consequential when predictions are costly (as we establish in Section~\ref{sec:sparse-costly}).
In both cases, our results characterize how to choose policy parameters (e.g., sampling probabilities and delay triggers) that preserve order-optimality while querying only a vanishing fraction of arrivals. We then use full-size predictions and SRPT scheduling in the low-priority queue to extend the guarantee to the intermediate regime.

Before moving to our main results, we begin by introducing SPD scheduling formally (see Figure \ref{fig:spdoverview} for an overview). 
\(\mathrm{SPD}(b,q,\tau,\pi)\) has four parameters: a residual demotion barrier \(b\ge0\) on remaining service, a query probability \(q\in[0,1]\), a delay trigger \(\tau\ge0\) measured in attained service, and a background scheduling policy \(\pi\).
Under SPD, every job begins at high priority. A job still unfinished at age \(\tau\) is queried independently with probability \(q\) and is permanently demoted if its remaining service exceeds \(b\).\footnote{Note that each binary query returns only whether the predicted remaining service exceeds the chosen barrier; the full-size variant in Section~\ref{sec:sparse-srpt} explicitly requests the size itself.} The high priority queue uses FCFS and strictly preempts the background queue, which uses \(\pi\).
The defaults are \(q=1\), \(\tau=0\), and \(\pi=\mathrm{FCFS}\); we omit trailing default arguments, so \(\mathrm{PD}(b)=\mathrm{SPD}(b)\).   An interesting special case of SPD is the DelayPredict algorithm introduced in \cite{shahout_skippredict_2024}.  DelayPredict corresponds to $\mathrm{SPD}(1,1,\tau,\mathrm{SRPT})$ and was proposed in the context of costly predictions, rather than with the goal of using minimal size information.  In Section \ref{sec:sparse-srpt} we provide results for the case when SPD uses SRPT scheduling. We compare variations of SPD to DelayPredict outside of heavy traffic in Section \ref{sec:experiments}. 

\subsection[Predictions upon Arrival]{Sparse Predictions upon Arrival}
\label{sec:sparse-predict-arrival}
We first consider sparsity through random sampling at arrival (delay trigger \(\tau=0\)). The goal is to query a vanishing fraction of arrivals while retaining the two properties used in the previous section's proof of order-optimality: a \(1-\rho\) fraction of arrivals is demoted, and each demoted job has size at least a fixed positive fraction of \(G^{-1}(\rho)\). The idea is to lower the demotion barrier to make more jobs eligible, then sample arrivals at random. To make the intuition above formal, recall that \(\overline F(\bar s_\rho)=1-\rho\) and \(\bar s_\rho/G^{-1}(\rho)\to\bar\theta\in(0,1]\) by Lemma~\ref{lem:free-bit-threshold-location}. 
For a fixed parameter \(\theta\in(0,\bar{\theta})\), we define the \emph{\(\theta\)-tuned residual barrier} and the \emph{\(\theta\)-tuned query probability}:
\[
 b_\rho(\theta):=\theta G^{-1}(\rho) \quad and \quad q_\rho(\theta):=(1-\rho)/\overline F(b_\rho(\theta)).\quad 
\]

Lowering \(\theta\) makes more jobs eligible for demotion and reduces the \(\theta\)-tuned query probability. It also lowers the guaranteed size of a demoted job, but every fixed positive \(\theta\) retains a lower bound that is a fixed multiple of \(G^{-1}(\rho)\), so the demotions still remove enough work from high priority. This probability is valid, with \(q_\rho(\theta)\in[0,1]\), exactly when \(b_\rho(\theta)\le\bar{s}_\rho\). Since \(\theta<\bar\theta\), eventually \(b_\rho(\theta)<\bar s_\rho\), and hence \(q_\rho(\theta)<1\).

For the at-arrival policy   \(\mathrm{SPD}(b_\rho(\theta),q_\rho(\theta))\) , we draw   \(Q_\rho(\theta)\sim\operatorname{Bernoulli}(q_\rho(\theta))\)  independently of each job's size. We query the job if   \(Q_\rho(\theta)=1\) , and assign rank \(1\) exactly when the queried job has size above   \(b_\rho(\theta)\) ; otherwise we assign rank \(0\). The assigned rank remains fixed, so this policy belongs to the binary-level queue (BLQ) class of Section~\ref{sec:one-free-bit}.

\begin{lemma}[Demotion for \(\theta\)-tuned parameters]
\label{lem:sparse-arrival-demotion}
If \(q_\rho(\theta)\le1\) , the policy   \(\mathrm{SPD}(b_\rho(\theta),q_\rho(\theta))\)  assigns a \(1-\rho\) fraction of arrivals to low priority. Every such job has size greater than   \(b_\rho(\theta)=\theta G^{-1}(\rho)\). 
\end{lemma}
\begin{proof}
Recall that jobs in the low-priority queue have rank \(1\). The assigned rank is   \(R_\rho(\theta)=Q_\rho(\theta)\mathbf1\{S>b_\rho(\theta)\}\), which gives the size bound. Independence of   \(Q_\rho(\theta)\)  and \(S\) also gives   \(\mathbb P(R_\rho(\theta)=1)=q_\rho(\theta)\overline F(b_\rho(\theta))=1-\rho\).
\end{proof}
The lemma is the analogue of Lemma~\ref{lem:free-bit-low-priority-fraction} for the sampled policy.
The remaining question is whether the fraction of jobs with remaining size above the barrier grows sufficiently relative to the target \(1-\rho\). The next lemma states that their ratio, which equals   \(1/q_\rho(\theta)\) , diverges; taking its reciprocal yields that   \(q_\rho(\theta)\to0\) .
\begin{lemma}[Abundance below slack-matched threshold]
\label{lem:sparse-candidate-abundance}
Suppose the service distribution is in the   lognormal-or-lighter  regime. For every fixed \(\theta\in(0,\bar{\theta})\), it holds that 
$\frac{\overline F(\theta G^{-1}(\rho))}{1-\rho}\longrightarrow\infty$.
\end{lemma}

\begin{proof}[Proof sketch.]
Since \(\theta<\bar\theta\), the residual barrier \(\theta G^{-1}(\rho)\) stays a fixed fraction below the slack-matched cutoff~\(\bar{s}_\rho\). In this regime, the hazard accumulated across that gap grows without bound as the cutoffs increase. Tail probabilities depend exponentially on the cumulative hazard rate, so lowering the cutoff increases the fraction of jobs above the residual barrier by an unbounded factor relative to the target demotion fraction \(1-\rho\). The \(\theta\)-tuned query probability is the reciprocal of this ratio and, therefore, vanishes. Appendix~\ref{app:sparse-candidate-abundance-proof} gives the full proof.
\end{proof}
Equipped with these lemmas, the following theorem shows that predicting every job is unnecessary: the same heavy-traffic order is obtained by querying only a vanishing fraction of arrivals.

\begin{theorem}[Querying a vanishing fraction of arrivals suffices]
\label{thm:sparse-arrival-optimal}
Suppose the service distribution is in the   lognormal-or-lighter regime. For every fixed \(\theta\in(0,\bar{\theta})\), the policy   \(\mathrm{SPD}(b_\rho(\theta),q_\rho(\theta))\)  is well defined for all sufficiently large \(\rho\), has   \(q_\rho(\theta)\to0\) , and is heavy-traffic order-optimal.
\end{theorem}

\begin{proof}
Lemma~\ref{lem:sparse-candidate-abundance} gives   \(1/q_\rho(\theta)\to\infty\) , so the\(\theta\)-tuned query probability$q_\rho(\theta)$  is eventually valid and tends to zero. By Lemma~\ref{lem:sparse-arrival-demotion}, the high-priority slack satisfies
  \[
    1-\lambda_\rho\mathbb E[S\mathbf1\{R_\rho(\theta)=0\}]
    \ge\lambda_\rho(1-\rho)b_\rho(\theta)
    =\Omega\!\left((1-\rho)G^{-1}(\rho)\right).
\]
Applying Lemmas~\ref{lem:free-bit-high-priority-response} and~\ref{lem:free-bit-low-priority-response} as in the proof of Theorem~\ref{thm:free-bit-sufficient} completes the upper bound. The service moments in these formulas are finite and independent of \(\rho\) as discussed in Section~\ref{subsec:service-time-classes}. Lemma~\ref{lem:sparse-arrival-demotion} supplies the required demotion fraction and size scale: both waiting contributions are \(O(1/((1-\rho)G^{-1}(\rho)))\), and the residence contributions are \(O(1)\). Finally, SRPT optimality gives \(\mathbb E[T^{\mathrm{SPD}(b_\rho(\theta),q_\rho(\theta))}]\ge\mathbb E[T^{\mathrm{SRPT}}]\). The SRPT mean response time in~\eqref{eq:srpt-heavy-traffic-benchmark} supplies the matching lower bound.
\end{proof}

\subsection[Delayed Prediction Queries]{Delayed Prediction Queries}
\label{sec:sparse-predict-delay}
We have established that binary size information for a vanishing fraction of jobs suffices to attain heavy-traffic order-optimality. Until now, this sparsity is achieved by querying jobs at arrival with a \(\theta\)-tuned query probability $q_{\rho}(\theta)$ that vanishes as $\rho\rightarrow 1$. A similar sparsity can also be achieved by delaying the query until only a small fraction of jobs remains unfinished. We thus now consider \(\mathrm{SPD}(b, q, \tau)\) that also includes a delay trigger~$\tau$ and requests a prediction with probability $q$ for any job that reaches age $\tau$, and demotes the job to the low-priority queue if its remaing size exceeds the residual demotion barrier $b$.

For each residual barrier \(b_\rho(\theta)\), we seek a range of triggers and associated  probabilities that preserve heavy-traffic order-optimality while only querying a vanishing fraction of jobs. To this end, we keep \(\theta\in(0,\bar{\theta})\) fixed and use the residual barrier \(b_\rho(\theta)\)  defined above. Then, a fraction \(\overline F(\tau)\) of arrivals remains unfinished at age \(\tau\). Among these jobs, the conditional probability of having more than   \(b_\rho(\theta)\)  units of service left is   \(\overline F(\tau+b_\rho(\theta))/\overline F(\tau)\) . Multiplying this with the probability of reaching the trigger gives   \(\overline F(\tau+b_\rho(\theta))\), fraction of jobs that are large enough for demotion if queried. The sampling rule independently selects a fraction \(q\) of these eligible jobs, so the probability of eventual demotion is  \(q\overline F(\tau+b_\rho(\theta))\). To make the eventually demoted fraction equal to \(1-\rho\), we define the \emph{$(\theta,\tau)$-tuned
query probability}
\begin{equation}
\label{eq:sparse-trigger-query-probability} q_\rho(\theta,\tau):=\frac{1-\rho}{\overline F(\tau+b_\rho(\theta))}.
\end{equation}
This recovers   \(q_\rho(\theta,0)=q_\rho(\theta)\)  at arrival. 
For \(0<\theta<\bar\theta\) and \(\tau\ge0\), we call the pair \((\theta,\tau)\) \emph{feasible} at load \(\rho\) if \(q_\rho(\theta,\tau)\le1\), or equivalently \(\tau+b_\rho(\theta)\le\bar s_\rho\). Beyond this point, too few arrivals are eligible to attain the demotion target even with query probability one.
For fixed \(\theta\in(0,\bar\theta)\), the \emph{feasible triggers}, i.e., those \(\tau\) for which \((\theta,\tau)\) is feasible, form the \emph{feasible trigger interval} \([0,\bar\tau_\rho(\theta)]\), where \(\bar\tau_\rho(\theta):=\bar s_\rho-b_\rho(\theta)\). When \(\bar\tau_\rho(\theta)\ge0\), its upper endpoint is the \emph{latest feasible trigger}. This interval is nonempty for all sufficiently large \(\rho\), since Lemma~\ref{lem:free-bit-threshold-location} and the definition of \(b_\rho(\theta)\) give \(\bar\tau_\rho(\theta)/G^{-1}(\rho)=\bar s_\rho/G^{-1}(\rho)-\theta\to\bar\theta-\theta>0\). In this interval we use   \(\mathrm{SPD}(b_\rho(\theta),q_\rho(\theta,\tau),\tau)\), whose \emph{ex-ante query probability} \footnote{The query probability \(q\) is conditional on the job remaining unfinished at age \(\tau\), that is, on \(S>\tau\). The ex-ante query probability averages over all arrivals, including jobs that finish before the trigger. It therefore captures both the filtering effect of delay and the random selection of jobs that reach the trigger.} is
\[ 
p_\rho^{\mathrm{qry}}(\theta,\tau) :=\text{\(q_\rho(\theta,\tau)\overline F(\tau)\)} =\frac{1-\rho}{\overline F(\tau+b_\rho(\theta))}\,\overline F(\tau).
\]
This factorization separates the two sources of sparsity. At \(\tau=0\), sparsity comes entirely from random sampling. At the latest feasible trigger  \(\tau=\bar\tau_\rho(\theta)\), the \((\theta,\tau)\)-tuned query probability is one, so sparsity comes entirely from the small fraction of arrivals that reach the trigger. Intermediate triggers combine the two mechanisms. The next lemma extends Lemma~\ref{lem:sparse-candidate-abundance} from queries at arrival to every trigger in the feasible trigger interval.
\begin{lemma}[Vanishing ex-ante query probability]
\label{lem:sparse-delayed-query-vanishing}
Suppose the service distribution is in the   lognormal-or-lighter  regime. For each fixed \(\theta\in(0,\bar{\theta})\), the ex-ante query probability vanishes uniformly over the feasible trigger interval:
\[
\sup_{0\le\tau\le\bar\tau_\rho(\theta)} p_\rho^{\mathrm{qry}}(\theta,\tau)\longrightarrow0
\quad\text{as }\rho\to 1.
\]
\end{lemma}

\begin{proof}
A job is queried only if it reaches the trigger and is then selected, so   \(p_\rho^{\mathrm{qry}}(\theta,\tau)=\overline F(\tau)q_\rho(\theta,\tau)\). We split the feasible triggers into two cases. For early triggers, \(0\le\tau<(\bar{\theta}-\theta)G^{-1}(\rho)/2\), we have \(\tau+b_\rho(\theta)<(\bar{\theta}+\theta)G^{-1}(\rho)/2\). Using \(\overline F(\tau)\le1\), we bound the ex-ante query probability by the \((\theta,\tau)\)-tuned query probability. For late triggers, \((\bar{\theta}-\theta)G^{-1}(\rho)/2\le\tau\le\bar\tau_\rho(\theta)\), feasibility gives \(q_\rho(\theta,\tau)\le1\), so we instead bound the ex-ante query probability by the fraction reaching the trigger. Together these bounds give
  \[    \sup_{0\le\tau\le\bar\tau_\rho(\theta)}
    p_\rho^{\mathrm{qry}}(\theta,\tau)
    \le
    \max\left\{
        \frac{1-\rho}{\overline F\!\left(\frac{\bar{\theta}+\theta}{2}G^{-1}(\rho)\right)},
        \overline F\!\left(\frac{\bar{\theta}-\theta}{2}G^{-1}(\rho)\right)
    \right\}.
\]
The first bound tends to zero by Lemma~\ref{lem:sparse-candidate-abundance}, because \((\bar{\theta}+\theta)/2<\bar{\theta}\). The second tends to zero since \((\bar{\theta}-\theta)G^{-1}(\rho)/2\to\infty\). Both bounds are independent of the particular trigger, giving uniform convergence.
\end{proof}

To bound response time, we next quantify the work removed from high priority.
As above, we define   \(Q_\rho(\theta,\tau)\sim\operatorname{Bernoulli}(q_\rho(\theta,\tau))\)  as the policy's size-independent sampling coin, which may equivalently be drawn at arrival and used only if \(S>\tau\). The \emph{actual-query indicator} is
\[
 Q_\rho^{\mathrm{qry}}(\theta,\tau)
 :=Q_\rho(\theta,\tau)\mathbf1\{S>\tau\}.
\]
It equals one exactly when the job reaches the trigger unfinished and is selected for a query, so
\(\mathbb E[Q_\rho^{\mathrm{qry}}(\theta,\tau)]=p_\rho^{\mathrm{qry}}(\theta,\tau)\). The eventual demotion indicator is
$R_\rho(\theta,\tau):=Q_\rho(\theta,\tau)\mathbf1\{S>\tau+b_\rho(\theta)\}.$ Here   \(R_\rho(\theta,\tau)=0\)  means the job is never demoted. In particular, if \(S\le\tau\), it finishes without a query and has   \(R_\rho(\theta,\tau)=0\), even when its unused coin has \(Q_\rho(\theta,\tau)=1\). A job with   \(R_\rho(\theta,\tau)=1\)  changes from rank zero to rank one upon reaching attained service age \(\tau\), while still unfinished.
The low-priority service contributed by one arrival is   \(\max\{0,S-\tau\}R_\rho(\theta,\tau)\). Its first \(\min\{S,\tau\}\) units always receive high priority; the remaining service does so only if the job is never demoted. Accordingly, the high-priority load is
  \[  \rho_\rho^{\mathrm{HIGH}}(\theta,\tau)
:=\lambda_\rho\mathbb E\!\left[
        \min\{S,\tau\}+\max\{0,S-\tau\}\mathbf1\{R_\rho(\theta,\tau)=0\}
    \right].
\]

\begin{lemma}[High-priority slack from demotion]
\label{lem:sparse-delayed-high-slack}
Suppose the service distribution is in the   lognormal-or-lighter  regime, and fix \(\theta\in(0,\bar{\theta})\). For every feasible trigger \(\tau\), the probability of eventual demotion is   \(\mathbb P(R_\rho(\theta,\tau)=1)=1-\rho\) . Uniformly over the feasible trigger interval,   \(1-\rho_\rho^{\mathrm{HIGH}}(\theta,\tau)=\Omega\!\left((1-\rho)G^{-1}(\rho)\right)\) .
\end{lemma}
\begin{proof}
A fraction   \(\overline F(\tau+b_\rho(\theta))\)  of arrivals has enough remaining service to qualify for demotion at age \(\tau\). The policy samples each such job with probability   \(q_\rho(\theta,\tau)\) , independently of size, so \begin{equation*}
    \mathbb P(R_\rho(\theta,\tau)=1)
        =q_\rho(\theta,\tau)\overline F(\tau+b_\rho(\theta)) =\frac{1-\rho}{\overline F(\tau+b_\rho(\theta))}\,\overline F(\tau+b_\rho(\theta))=1-\rho.
\end{equation*}
Subtracting the expected low-priority work from the total load therefore yields \begin{equation*}
\begin{aligned}
    1-\rho_\rho^{\mathrm{HIGH}}(\theta,\tau)
    &=(1-\rho)+\lambda_\rho\mathbb E[\max\{0,S-\tau\}R_\rho(\theta,\tau)]
    \ge (1-\rho)+\lambda_\rho(1-\rho)b_\rho(\theta),
\end{aligned}
\end{equation*}
because each eventually demoted job has more than   \(b_\rho(\theta)\)  service remaining at its demotion time. Since   \(b_\rho(\theta)=\theta G^{-1}(\rho)\)  and \(\lambda_\rho\to1/\mathbb E[S]>0\) with a fixed finite service mean as noted in Section~\ref{subsec:service-time-classes}, this gives an \(\Omega((1-\rho)G^{-1}(\rho))\) lower bound, uniformly over the feasible trigger interval.
\end{proof}

These lemmas give the bounds needed to extend Theorem~\ref{thm:sparse-arrival-optimal} to delayed predictions.
\begin{theorem}[Heavy-traffic order-optimality with delayed predictions]
\label{thm:sparse-uniform-optimality}
Suppose the service distribution is in the   lognormal-or-lighter  regime, and fix \(\theta\in(0,\bar{\theta})\). Uniformly over \(\tau\) in the feasible trigger interval,   \(\mathrm{SPD}(b_\rho(\theta),q_\rho(\theta,\tau),\tau)\)  is heavy-traffic order-optimal and achieves vanishing ex-ante query probability as \(\rho\uparrow1\).
\end{theorem}

\begin{proof}[Proof sketch.]Lemma~\ref{lem:sparse-delayed-query-vanishing} already gives vanishing prediction use. Lemma~\ref{lem:sparse-delayed-high-slack} gives the same lower bound on high-priority slack as in the at-arrival proof, while   \(\mathbb P(R_\rho(\theta,\tau)=1)=1-\rho\) . If ranks were non-increasing in attained service, Proposition~\ref{prop:soap-nonincreasing} and Lemmas~\ref{lem:free-bit-high-priority-response}--\ref{lem:free-bit-low-priority-response} would complete the response-time argument just as in Theorem~\ref{thm:sparse-arrival-optimal}.
For a positive trigger, however, a job that is eventually demoted has its rank increased from zero to one when it reaches the trigger, violating the non-increasing-rank condition of Proposition~\ref{prop:soap-nonincreasing}.

That said, we can still apply the general SOAP formula in Proposition~\ref{prop:soap-response-time}
separately for
jobs that are never demoted and for jobs that are eventually demoted to characterize their corresponding mean response times. The first group includes jobs that finish before the trigger, jobs that finish after the trigger but their residual size then is at most the residual barrier, and jobs that are not sampled for query. The second group includes jobs that are sampled for query at the time of the trigger and have residual size larger than the residual barrier. Weighting the contributions of the  groups
by \(\rho\) and \(1-\rho\), respectively, and using Lemma~\ref{lem:sparse-delayed-high-slack} gives the claimed response-time order. Appendix~\ref{app:sparse-uniform-optimality-proof} states the conditional response-time identities for the two groups and combines them with the demotion probability and high-priority slack bound to complete the theorem proof. Appendix~\ref{app:sparse-delayed-identity-proof} derives those identities from the general SOAP formula.
\end{proof}

\subsection{Vanishing Full-Size Predictions Suffice for the Intermediate Regime
}
\label{sec:sparse-srpt}
So far, our upper bounds considered service distributions that are lognormal-or-lighter. 
In the intermediate regime, Theorem~\ref{thm:free-bit-impossible} rules out heavy-traffic order-optimality for binary-level queues. We now come back to this intermediate regime and show that SPD can obtain heavy-traffic order-optimality in this regime as well if it uses full-size predictions and SRPT scheduling in the low-priority queue.

\begin{theorem}[Order-optimality with a vanishing fraction of full-size predictions]
\label{thm:sparse-srpt-intermediate}
Suppose the service distribution is in the intermediate regime or satisfies \(H(x)\sim A(\log x)^m\), with \(A>0\) and \(m\ge2\). With full-size predictions, fix any \(\beta\in(0,1)\) and set
\[
 \tau_\rho=\overline F^{-1}\!\left((1-\rho)^\beta\right).
\]
Then \(\mathrm{SPD}(\tau_\rho,1,\tau_\rho,\mathrm{SRPT})\) is heavy-traffic order-optimal, with ex-ante query probability \((1-\rho)^\beta\to0\).
\end{theorem}

Although binary-level queues cannot achieve heavy-traffic order-optimality in the intermediate regime, SPD with full-size predictions can do so while querying only a vanishing fraction of arrivals.  Specifically, \(\beta>0\) makes the ex-ante query probability vanish, while \(\beta<1\) ensures, under the stated tail assumptions, that enough work is demoted to make the response-time contribution of never-demoted jobs negligible relative to SRPT. Appendix~\ref{app:sparse-srpt} actually proves a more general result that allows inference costs, which reduces to this theorem when the inference cost is zero.

\section{Extension to Costly and Inexact Predictions}
\label{sec:sparse-costly}
\label{sec:sparse-costly-model}
\label{subsec:prediction-model}
Up to this point, our analysis has focused on the setting of free, exact job size predictions.  In this section, predictions may require server work and may be inexact. 
Specifically, each queried job requires \(\delta_\rho\ge0\) units of work on the server to obtain its prediction. This work prevents the server from processing other jobs, so inference causes \emph{interference}. For a scheduling policy \(\pi\), let \(T^\pi(\delta_\rho)\) denote the response time of a typical job, including inference; \(T^\pi(0)\) recovers the free-prediction setting with the same policy parameters. 

The inference work described above returns a prediction of the queried job's total size, which may be inexact. At each load \(\rho\), we describe this output by a \emph{prediction rule}, a measurable map \(\Phi_\rho:\mathbb R_+\times[0,1]\to\mathbb R_+\). For a job of true size \(S\), its predicted total size is \(\widehat S=\Phi_\rho(S,U)\), where \(U\) is uniform on \([0,1]\). Every job uses the same map with fresh randomness, independent across jobs and of all sizes, arrivals, and sampling coins. A deterministic rule ignores \(U\). Predictions are \emph{\(\varepsilon\)-inexact} if, for a fixed \(\varepsilon\in[0,1)\), every job satisfies
\begin{equation}
\label{eq:sparse-multiplicative-error}
 (1-\varepsilon)S\le\widehat S\le(1+\varepsilon)S.
\end{equation}
Our guarantees are uniform over all prediction rules in this model satisfying the above condition, including biased or randomized rules chosen adversarially at each load. The case \(\varepsilon=0\) recovers exact predictions. 
In the remainder of this section, we first study the case of inference costs and exact predictions \(\varepsilon=0\) in Section~\ref{sec:sparse-costly-exact}.  Section~\ref{sec:sparse-predict-errors} then allows for inexact predictions \(\varepsilon>0\).

\subsection[Costly Exact Predictions]{Costly Exact Predictions}
\label{sec:sparse-costly-exact}
\label{sec:sparse-costly-gamma}
\label{sec:sparse-advice-optimal-trigger}
\label{sec:sparse-cost-distributions}

When predictions are free, Theorem~\ref{thm:sparse-uniform-optimality} gives heavy-traffic order-optimality for every feasible trigger with a fixed barrier scale \(\theta\in(0,\bar\theta)\) in the lognormal-or-lighter regime. 
With costly predictions, SPD keeps a queried job in the high-priority queue until inference finishes, when its prediction determines whether it is demoted.
The barrier scale and delay trigger determine how often the server performs this extra work; a high ex-ante query probability can exhaust the available slack or lead to instability. Our main result in this setting shows how to preserve slack and maintain heavy-traffic order-optimality for SPD.

\vspace{.1in}
\noindent\textbf{Slack Preservation.} To retain the heavy-traffic order-optimality, we need to ensure that a sufficient fraction of the ordinary-service slack remains after adding inference work. To that end, we call a load-dependent pair \((\theta_\rho,\tau_\rho)\) \emph{slack-preserving} if, for some fixed \emph{constant-slack parameter} \(\zeta\in(0,1)\), the pair is feasible and satisfies
\begin{equation}
\label{eq:sparse-costly-slack-condition}
 \lambda_\rho\delta_\rho p_\rho^{\mathrm{qry}}(\theta_\rho,\tau_\rho)
 \le(1-\zeta)(1-\rho)
\end{equation}
at all sufficiently high loads. Since the total load is \(\rho(\delta_\rho):=\rho+\lambda_\rho\delta_\rho p_\rho^{\mathrm{qry}}(\theta_\rho,\tau_\rho)\), this budget reserves at least \(\zeta(1-\rho)\) slack after inference is included. When predictions are free, \(\rho(0)=\rho\) and every pair that is feasible for all sufficiently high loads is slack-preserving.

\begin{lemma}[From free to costly predictions]
\label{lem:sparse-cost-transfer}
Fix a constant-slack parameter \(\zeta\in(0,1)\). At a load \(\rho\in[1/2,1)\), let \(\pi\) be an SPD policy with FCFS in both queues, a prediction rule, and an ex-ante query probability of at least \(1-\rho\). If \(\pi\) satisfies the slack-preservation condition~\eqref{eq:sparse-costly-slack-condition} with this ex-ante query probability, then it is stable and
\[
\mathbb E[T^\pi(\delta_\rho)]\le\frac{6}{\zeta^2}\,\mathbb E[T^\pi(0)].
\]
\end{lemma}
Lemma~\ref{lem:sparse-cost-transfer} transfers heavy-traffic order-optimality from free to costly predictions under slack preservation; Appendix~\ref{app:sparse-costly-extension-proof} gives its proof.
To apply Lemma~\ref{lem:sparse-cost-transfer} at a given inference cost, we now examine how the trigger choice controls the ex-ante query probability and hence the inference load. At arrival, sparsity comes from random sampling; at the latest feasible trigger, the \((\theta,\tau)\)-tuned query probability is one, so sparsity comes entirely from the small fraction of jobs that remain unfinished.
With costly predictions, the ex-ante query probability induced by the pair \((\theta,\tau)\) determines the inference load and can separate heavy-traffic order-optimality from instability. To understand which trigger uses fewer queries, consider how often a queried survivor (a job that has size greater than the inference trigger) is eligible for demotion. For a feasible pair \((\theta,\tau)\), define its \emph{self-information} by
\[
 I_\rho(\theta,\tau)
 :=-\log\mathbb P(S>\tau+b_\rho(\theta)\mid S>\tau)
 =H(\tau+b_\rho(\theta))-H(\tau).
\]
Using the \((\theta,\tau)\)-tuned query probability in~\eqref{eq:sparse-trigger-query-probability}, which makes the eventual demotion probability equal to \(1-\rho\), we obtain
\[
p_\rho^{\mathrm{qry}}(\theta,\tau)=(1-\rho)e^{I_\rho(\theta,\tau)}.
\]
Smaller self-information means that a larger fraction of queried jobs can be demoted, so fewer queries are needed to meet the target demotion fraction \(1-\rho\). We use this identity both to choose a trigger and to determine how costly predictions can be while preserving slack.

\paragraph{Self-Information and Trigger Choice.}
Fix the service distribution and a parameter \(\theta\in(0,\bar\theta)\). The following proposition characterizes a trigger that minimizes the ex-ante query probability. We use natural logarithms, so information is measured in nats.
When the feasible trigger interval is nonempty, let \(\tau_\rho^{\mathrm{qry}}(\theta)\) denote any trigger in it minimizing \(p_\rho^{\mathrm{qry}}(\theta,\tau)\); the proof below establishes existence.

\begin{proposition}[Exact trigger objective]
\label{prop:sparse-exact-trigger-objective}
For \(\bar\tau_\rho(\theta)\ge0\), a trigger minimizing the ex-ante query probability in the feasible trigger interval \([0,\bar\tau_\rho(\theta)]\) satisfies
\[
\tau_\rho^{\mathrm{qry}}(\theta)
    \in
    \operatorname*{arg\,min}_{0\le\tau\le\bar\tau_\rho(\theta)}
    I_\rho(\theta,\tau).
\]
\end{proposition}
\begin{proof}
Every trigger in \([0,\bar\tau_\rho(\theta)]\) paired with the  \((\theta,\tau)\)-tuned query probability gives the same probability of eventual demotion \(1-\rho\). 
A queried survivor is eligible for demotion exactly when \(S>\tau+b_\rho(\theta)\). Since sampling is independent of service size,
\[
 \mathbb P(S>\tau+b_\rho(\theta)\mid S>\tau)
 =\frac{\overline F(\tau+b_\rho(\theta))}{\overline F(\tau)}
 =e^{-I_\rho(\theta,\tau)}.
\]
Thus, the ex-ante query probability equals the target demotion fraction divided by the conditional probability that a queried survivor is eligible for demotion, \(e^{-I_\rho(\theta,\tau)}\). By the definition of the \((\theta,\tau)\)-tuned query probability \(q_\rho(\theta,\tau)\),
\[
   p_\rho^{\mathrm{qry}}(\theta,\tau)
    =
    q_\rho(\theta,\tau)\overline F(\tau)
    =
    (1-\rho)e^{I_\rho(\theta,\tau)}. \]
The exponential map is increasing, so the two minimization problems are equivalent. Since \(H\) is continuous and the feasible trigger interval is nonempty and compact, \(I_\rho(\theta,\cdot)\) attains its minimum there. Hence a query-minimizing trigger exists.
\end{proof}
With free predictions and fixed \(\theta\), every feasible trigger has the same heavy-traffic response-time order. The trigger can therefore be chosen to minimize prediction use. Given that
\(H'=h\),
$\frac{\partial I_\rho(\theta,\tau)}{\partial\tau}=h(\tau+b_\rho(\theta))-h(\tau)$,
decreasing hazard rates favor delay and increasing hazard rates favor earlier queries.

\paragraph{Inference-Cost Criterion.}
Minimizing the ex-ante query probability reduces inference load; yet, for each feasible parameter pair, stability additionally requires
\(\lambda_\rho\delta_\rho p_\rho^{\mathrm{qry}}(\theta_\rho,\tau_\rho)\) to be less than the ordinary-service slack \(1-\rho\). Preserving a fixed fraction of this slack also permits the response-time comparison in Lemma~\ref{lem:sparse-cost-transfer}. To quantify how inference duration and query choices interact, we use a \emph{$\gamma$-exponential cost scale}:
\begin{equation}
\label{eq:sparse-gamma-cost-scale}
 \lambda_\rho\delta_\rho=(1-\rho)^\gamma
 =e^{-\gamma H(\bar s_\rho)},
 \qquad \gamma\ge0.
\end{equation}
We focus on \(0<\gamma<1\), where queries become cheaper with load but slack shrinks faster.

Each feasible pair \((\theta_\rho, \tau_\rho)\) induces an ex-ante query probability \(p_\rho^{\mathrm{qry}}(\theta_\rho, \tau_\rho)\). Multiplying by the arrival rate and the cost per query gives its inference load. Dividing this load by the available slack gives
\begin{equation}
\label{eq:sparse-gamma-relative-load}
 \frac{\lambda_\rho\delta_\rho p_\rho^{\mathrm{qry}}(\theta_\rho,\tau_\rho)}{1-\rho}
 =\exp\!\left(I_\rho(\theta_\rho,\tau_\rho)-\gamma H(\bar s_\rho)\right).
\end{equation}

\begin{lemma}[\(\gamma\)-exponential stability criterion]
\label{cor:sparse-gamma-cost}
Suppose the service distribution is in the lognormal-or-lighter regime and we have access to exact predictions at a $\gamma$-exponential cost scale where \(\gamma \in (0,1)\). For any load-dependent pair \((\theta_\rho,\tau_\rho)\) that is feasible for all sufficiently high loads, we characterize the performance of
$\mathrm{SPD}(b_\rho(\theta_\rho),q_\rho(\theta_\rho,\tau_\rho),\tau_\rho)$ as follows.
\begin{enumerate}
\item \emph{Stability.} If
$ \limsup_{\rho\uparrow1}
 \frac{I_\rho(\theta_\rho,\tau_\rho)}{H(\bar s_\rho)}<\gamma
$
then the policy is eventually slack-preserving and stable.
\item \emph{Instability.} If
$\liminf_{\rho\uparrow1}
 \frac{I_\rho(\theta_\rho,\tau_\rho)}{H(\bar s_\rho)}>\gamma$
then the policy is eventually unstable.
\item \emph{Boundary.} If $I_\rho(\theta_\rho,\tau_\rho)-\gamma H(\bar s_\rho)
 \longrightarrow d\in\mathbb R$
then \(d<0\) gives eventual slack preservation and stability. If \(d>0\), the policy is eventually unstable. For \(d=0\), this limit alone does not determine stability.
\end{enumerate}
\end{lemma}

\begin{proof}[Proof of Lemma~\ref{cor:sparse-gamma-cost}]
The inference load is \(\lambda_\rho\delta_\rho p_\rho^{\mathrm{qry}}(\theta_\rho,\tau_\rho)\).
We compare it with the ordinary-service slack \(1-\rho\) using the relative inference load identity~\eqref{eq:sparse-gamma-relative-load}. Since \(\overline F(\bar s_\rho)=1-\rho\) by \eqref{eq:free-bit-threshold}, we have \(H(\bar s_\rho)=-\log(1-\rho)\to\infty\).

\smallskip\noindent\emph{Stability.} Under the stability condition, some fixed \(\eta>0\) satisfies
\(I_\rho(\theta_\rho,\tau_\rho)-\gamma H(\bar s_\rho)\le-\eta H(\bar s_\rho)\)
at all sufficiently high loads. The relative inference load therefore satisfies
\[
\begin{aligned}
    \frac{\lambda_\rho\delta_\rho p_\rho^{\mathrm{qry}}(\theta_\rho,\tau_\rho)}{1-\rho}
    &=\exp\!\left(I_\rho(\theta_\rho,\tau_\rho)-\gamma H(\bar s_\rho)\right)\le e^{-\eta H(\bar s_\rho)}
    =(1-\rho)^\eta\longrightarrow0.
\end{aligned}
\]  Hence, the inference load is \(o(1-\rho)\), giving slack preservation~\eqref{eq:sparse-costly-slack-condition} eventually with constant-slack parameter \(\zeta=1/2\).

\smallskip\noindent\emph{Instability.} Under the instability condition, the same exponent is at least \(\eta H(\bar s_\rho)\) for some fixed \(\eta>0\) at all sufficiently high loads. The relative inference load is at least \((1-\rho)^{-\eta}\to\infty\). Inference work eventually exceeds the ordinary-service slack, so total load exceeds one and the policy is unstable. 

\smallskip\noindent\emph{Boundary.} Under the boundary condition, the relative inference load tends to \(e^d\). If \(d<0\), slack preservation~\eqref{eq:sparse-costly-slack-condition} holds eventually with constant-slack parameter \(\zeta=(1-e^d)/2\). If \(d>0\), relative inference load eventually exceeds one, giving instability. If \(d=0\), it tends to one, so no fixed \(\zeta\in(0,1)\) gives slack preservation; finer terms are needed to determine stability. In either slack-preserving case, the total load is eventually at most \(1-\zeta(1-\rho)<1\), giving stability. 
\end{proof}
For a fixed \(\theta\in(0,\bar\theta)\), Theorem~\ref{thm:sparse-uniform-optimality} gives a common free-prediction response-time order across feasible triggers. Whenever Lemma~\ref{cor:sparse-gamma-cost} gives slack preservation, Lemma~\ref{lem:sparse-cost-transfer} transfers this order to the costly policy. We now evaluate the normalized self-information for three tail classes to determine which triggers preserve slack and which cause instability. Here, \(H\) is \emph{regularly varying with index \(\nu\)} if \(H(tx)/H(x)\to t^\nu\) as \(x\to\infty\) for every \(t>0\), and \emph{rapidly varying} if \(H(tx)/H(x)\to\infty\) as \(x\to\infty\) for every \(t>1\).

\begin{proposition}[The sparsity mechanism matters]
\label{prop:sparse-cost-distributions}
Fix \(\theta\in(0,\bar\theta)\) and \(\gamma\in(0,1)\). Suppose we have access to exact predictions at a $\gamma$-exponential cost scale. Then, depending on the distribution, the selection of the delay trigger can be the difference between heavy-traffic order-optimality and instability of SPD. In particular:
\begin{enumerate}
\item If \(H(x)\sim A(\log x)^m\), with \(A>0\) and \(m\ge2\), the arrival policy \(\mathrm{SPD}(b_\rho(\theta),q_\rho(\theta),0)\) is eventually unstable, whereas \(\mathrm{SPD}(b_\rho(\theta),1,\bar\tau_\rho(\theta))\), which queries at the latest feasible trigger, is eventually stable and heavy-traffic order-optimal.
\item If \(H\) is rapidly varying, the arrival policy \(\mathrm{SPD}(b_\rho(\theta),q_\rho(\theta),0)\) is eventually stable and heavy-traffic order-optimal, whereas \(\mathrm{SPD}(b_\rho(\theta),1,\bar\tau_\rho(\theta))\) is eventually unstable.
\item If \(H\) is regularly varying with index \(\nu>0\), fix \(a\in[0,1-\theta)\) and set \(\tau_\rho=aG^{-1}(\rho)\). With this trigger, the policy \(\mathrm{SPD}(b_\rho(\theta),q_\rho(\theta,\tau_\rho),\tau_\rho)\) is eventually stable and heavy-traffic order-optimal when \((a+\theta)^\nu-a^\nu<\gamma\), and eventually unstable when \((a+\theta)^\nu-a^\nu>\gamma\). This threshold decreases with delay for \(0<\nu<1\), increases for \(\nu>1\), and equals \(\theta\) for \(\nu=1\); equality with \(\gamma\) requires finer asymptotics.
\end{enumerate}
\end{proposition}

\begin{proof}[Proof sketch]
Apply Lemma~\ref{cor:sparse-gamma-cost} to the normalized self-information \(I_\rho(\theta,\tau_\rho)/H(\bar s_\rho)\). For polylogarithmic hazards, this ratio tends to \(1\) at arrival and \(0\) at the latest trigger. Since \(0<\gamma<1\), the lemma gives instability at arrival and stability at the latest trigger. For rapidly varying hazards, the limits are reversed, so the lemma reverses these conclusions. Under regular variation, the limit at \(aG^{-1}(\rho)\) is \((a+\theta)^\nu-a^\nu\); comparing it with \(\gamma\) in the lemma gives the stated threshold. In each of these stable cases, Lemma~\ref{cor:sparse-gamma-cost} also gives slack preservation; Theorem~\ref{thm:sparse-uniform-optimality} and Lemma~\ref{lem:sparse-cost-transfer} then give heavy-traffic order-optimality. Appendices~\ref{app:sparse-cost-polylog}, \ref{app:sparse-cost-rapid}, and~\ref{app:sparse-cost-regular} establish these limits and provide detailed comparisons and full proofs.
\end{proof}

\subsection{Costly and Inexact Predictions}
\label{sec:sparse-predict-errors}
With free predictions, 
directly applying the SPD parameters we use for exact-prediction from Section~\ref{sec:sparse-costly-exact} to inexact predictions does not yield heavy-traffic order-optimality under systematic underestimation, even for exponential service times (Appendix~\ref{app:sparse-error-unadjusted-proof}). In the lognormal-or-lighter regime, lowering the arrival barrier to \((1-\varepsilon)b_\rho(\theta)\) with fixed \(\theta\in(0,\bar\theta)\) restores the guarantee uniformly over the allowed prediction rules (Appendix~\ref{app:sparse-error-free-proof}). This repair leaves the \(\theta\)-tuned query probability unchanged, so the inference load may still exhaust the ordinary-service slack when predictions are costly. We therefore adjust the barrier at a delayed trigger and choose the delay trigger and \(\theta\) jointly.

\paragraph{Adjusting the residual barrier at a delayed trigger.}
The arrival repair multiplies the true-size cutoff by \(1-\varepsilon\). At age \(\tau\), that cutoff is \(\tau+b_\rho(\theta)\): we apply the same correction to this total-size threshold, then subtract the \(\tau\) units already served. This gives the \emph{error-adjusted residual demotion barrier}
\[
 \widetilde b_\rho(\theta,\tau;\varepsilon)
 :=(1-\varepsilon)(\tau+b_\rho(\theta))-\tau
 =(1-\varepsilon)b_\rho(\theta)-\varepsilon\tau.
\]
The extra \(-\varepsilon\tau\) appears because total size is predicted with error, whereas the completed service \(\tau\) is known exactly. The test \(\widehat S-\tau>\widetilde b_\rho(\theta,\tau;\varepsilon)\) therefore demotes every queried job with \(S>\tau+b_\rho(\theta)\), even under the worst allowed underprediction.

For a feasible pair \((\theta,\tau)\) with \(\widetilde b_\rho(\theta,\tau;\varepsilon)\ge0\), we run \(\mathrm{SPD}(\widetilde b_\rho(\theta,\tau;\varepsilon),q_\rho(\theta,\tau),\tau)\), keeping the $(\theta, \tau)$-tuned query probability as in Section~\ref{sec:sparse-costly-exact}. Lowering the cutoff can admit smaller jobs, so the candidates below must keep the adjusted residual barrier nonnegative and leave enough ordinary service at demotion. Indeed, a positive prediction satisfies
\[
 S-\tau>
 \frac{(1-\varepsilon)b_\rho(\theta)-2\varepsilon\tau}{1+\varepsilon}.
\]
The trigger pair needs to meet two requirements. First, the trigger should be small enough relative to the barrier so that even when the barrier is lowered, we can ensure every demoted job retains at least a fixed fraction of $G^{-1}(\rho)$ units of service. 
Second, the pair must induce sufficiently few queries so that inference is slack-preserving. Our construction achieves this balance by identifying a family of candidate pairs that all satisfy the first requirement (Lemma~\ref{lem:sparse-error-free-candidates}) and at least one satisfies the second requirement (Lemma~\ref{lem:sparse-error-slack-preserving-pair}).

\paragraph{A finite list of candidate pairs.}
Fix an integer \(N\ge2\). For \(1\le j\le N\), set
\[
 \tau_{\rho,j}:=\frac{\bar\theta}{2}
 \left(\frac{1-\varepsilon}{2(1+\varepsilon)}\right)^jG^{-1}(\rho)
 \qquad and \qquad
 \theta_j:=\frac{(1+3\varepsilon)\bar\theta}{2(1-\varepsilon)}
 \left(\frac{1-\varepsilon}{2(1+\varepsilon)}\right)^j.
\]
Candidate \(j\) uses trigger \(\tau_{\rho,j}\) and a correspondingly tuned query probability \(q_\rho(\theta_j,\tau_{\rho,j})\), and error-adjusted residual demotion barrier \(\widetilde b_\rho(\theta_j,\tau_{\rho,j};\varepsilon)\). In particular, 
\begin{equation}
\label{eq:sparse-error-candidate-barriers}
b_\rho(\theta_j)=\frac{1+3\varepsilon}{1-\varepsilon}\tau_{\rho,j} \qquad \text{and}
\qquad
\widetilde b_\rho(\theta_j,\tau_{\rho,j};\varepsilon)
=(1+2\varepsilon)\tau_{\rho,j}.
\end{equation}
The first identity in~\eqref{eq:sparse-error-candidate-barriers} follows by cancelling the common geometric factor in \(\theta_j\) and \(\tau_{\rho,j}\); substitution into \((1-\varepsilon)b_\rho(\theta_j)-\varepsilon\tau_{\rho,j}\) gives the second. A demoted job therefore satisfies
\[
    (1+\varepsilon)S\ge\widehat S
    >\tau_{\rho,j}+\widetilde b_\rho(\theta_j,\tau_{\rho,j};\varepsilon)
    =2(1+\varepsilon)\tau_{\rho,j}.
\]
As a result, it retains \(S-\tau_{\rho,j}>\tau_{\rho,j}\) ordinary service, as required by the next lemma.

\begin{lemma}[Every candidate works with free predictions]
\label{lem:sparse-error-free-candidates}
Suppose the service distribution is in the lognormal-or-lighter regime, and fix \(\varepsilon\in[0,1)\) and \(N\ge2\). All candidate pairs are feasible for all sufficiently high loads. Under free \(\varepsilon\)-inexact predictions, their corresponding policies have vanishing ex-ante query probabilities and are heavy-traffic order-optimal. 
\end{lemma}
\begin{proof}[Proof sketch]
The candidate endpoints \(\tau_{\rho,j}+b_\rho(\theta_j)\) are eventually below \(\bar s_\rho\), giving feasibility of the candidate pairs.
Their adjusted barriers are nonnegative by~\eqref{eq:sparse-error-candidate-barriers}, so the policies are well defined at these loads;
their diverging triggers give vanishing ex-ante query probabilities. For candidate \(j\), write \(\tau=\tau_{\rho,j}\) and let \(p\) be the demoted fraction and
\(\rho^{\mathrm{HIGH}}(0)\) be
this candidate's high-priority load with free predictions. The adjusted test gives \(p\ge1-\rho\) and leaves every demoted job more than \(\tau\) service.

For high priority, demotion removes at least \(p\tau\) service per arrival, giving slack at least \(\lambda_\rho p\tau\ge\lambda_\rho(1-\rho)\tau\). Jobs that stay in high priority thus contribute at most order \(1/((1-\rho)\tau)\) to mean response time. For jobs sent to low priority (as shown in Appendix~\ref{app:sparse-error-demotion-bound-proof}), the leading response-time term is at most a constant times 
\[
\frac{p}{(1-\rho)\,(1-\rho^{\mathrm{HIGH}}(0))}
\le \frac{p}{(1-\rho)\lambda_\rho p\tau}
=\frac{1}{\lambda_\rho(1-\rho)\tau}.
\]
The remaining service term stays bounded: averaged over arrivals, residual work also adds proportionally to high-priority slack, and each demoted job's service before demotion is smaller than its residual work. Finally, \(\tau=\Theta(G^{-1}(\rho))\), so both groups' contributions have the SRPT upper-bound scale. Appendix~\ref{app:sparse-error-candidate-proof} gives the full calculation, matching lower bound, and uniformity argument.
\end{proof}

\begin{lemma}[At least one candidate preserves slack]
\label{lem:sparse-error-slack-preserving-pair}
For the service distribution and error bound of Lemma~\ref{lem:sparse-error-free-candidates}, fix \(\gamma\in(0,1)\) and an integer \(N>1/\gamma\), and use the candidate pairs above with the \(\gamma\)-exponential cost scale~\eqref{eq:sparse-gamma-cost-scale}. At every sufficiently high load, some candidate preserves slack, uniformly over the allowed prediction rules.
\end{lemma}
\begin{proof}
Note that $\tau_{\rho,j+1}+b_{\rho}(\theta_{j+1})=\tau_{\rho,j+1}+\frac{1+3\varepsilon}{1-\varepsilon}\tau_{\rho,j+1}=\frac{2(1+\varepsilon)}{1-\varepsilon}\tau_{\rho,j+1}=\tau_{\rho,j}$. As a result, the intervals \([\tau_{\rho,j},\tau_{\rho,j}+b_\rho(\theta_j)]\) are adjacent, for \(1\le j<N\) and their hazard increments thus telescope:
\[
 \sum_{j=1}^N I_\rho(\theta_j,\tau_{\rho,j}) =H\!\left(\frac{\bar\theta}{2}G^{-1}(\rho)\right)-H(\tau_{\rho,N})
 \le H(\bar s_\rho)=-\log(1-\rho).
\]
The last inequality uses \(H(\cdot)\ge0\) and the largest candidate endpoint being at most \(\bar s_\rho\) at sufficiently high loads. Hence, there exists a candidate~$j$ such that 
$I_\rho(\theta_j,\tau_{\rho,j})\leq \frac{-\log(1-\rho)}{N}$. This implies that $p_\rho^{\mathrm{qry}}(\theta_j,\tau_{\rho,j})
 =(1-\rho)e^{I_\rho(\theta_j,\tau_{\rho,j})}
 \le(1-\rho)^{1-1/N}.$ 
Given that $N>1/\gamma$, candidate $j$ satisfies slack preservation~\eqref{eq:sparse-costly-slack-condition}  as its inference load relative to the available slack eventually vanishes: 
\(\frac{\lambda_\rho\delta_\rho p_\rho^{\mathrm{qry}}(\theta_j,\tau_{\rho,j})}{1-\rho}\le(1-\rho)^{\gamma-1/N}\longrightarrow0.\)
The above construction does not rely on a particular prediction rule, so this candidate preserves slack uniformly over all allowed prediction rules.
\end{proof}

\paragraph[Self-Information and Affordability]{Self-Information and Affordability.}
Finding one true-size candidate among queried survivors takes \(e^{I_\rho(\theta,\tau)}\) queries on average. Spending all available slack on inference would permit \(e^{\gamma H(\bar s_\rho)}\) queries per such candidate. The choice \(N>1/\gamma\) makes the selected candidate's query demand a vanishing fraction of this allowance. Combining this slack preservation with Lemmas~\ref{lem:sparse-cost-transfer} and~\ref{lem:sparse-error-free-candidates} gives the following theorem.

\begin{theorem}[Robustness with costly inference]
\label{thm:sparse-constant-error}
Suppose the service distribution is in the lognormal-or-lighter regime. For every \(\varepsilon\in[0,1)\) and \(\gamma\in(0,1)\), under the \(\gamma\)-exponential cost scale~\eqref{eq:sparse-gamma-cost-scale}, there is a load-dependent pair \((\theta_\rho,\tau_\rho)\), feasible for all sufficiently high loads, for which
$\mathrm{SPD}\bigl(\widetilde b_\rho(\theta_\rho,\tau_\rho;\varepsilon), q_\rho(\theta_\rho,\tau_\rho),\tau_\rho\bigr)$
is eventually stable, has vanishing ex-ante query probability, and is heavy-traffic order-optimal, uniformly over all \(\varepsilon\)-inexact rules.
\end{theorem}
\begin{proof}
Fix \(N=\lfloor1/\gamma\rfloor+1\). At each sufficiently high load, we choose the smallest qualifying index \(j_\rho\) in Lemma~\ref{lem:sparse-error-slack-preserving-pair} and set \((\theta_\rho,\tau_\rho)=(\theta_{j_\rho},\tau_{\rho,j_\rho})\). Lemma~\ref{lem:sparse-error-free-candidates} gives a uniform free-prediction bound for all candidates, and Lemma~\ref{lem:sparse-error-slack-preserving-pair} makes the selected pair slack-preserving. Lemma~\ref{lem:sparse-cost-transfer} establishes eventual stability and transfers the heavy-traffic order to the costly policy. The selected ex-ante query probability also vanishes by the query bound in the proof of Lemma~\ref{lem:sparse-error-slack-preserving-pair}.
\end{proof}

\paragraph{Finding the pair.}
For fixed \(\varepsilon\) and \(\gamma\), the candidate tuning constants are computed once. At each sufficiently high load, evaluate the hazard increments for at most \(\lfloor1/\gamma\rfloor+1\) candidates and select the first qualifying one. Hence, selection uses \(O(1/\gamma)\) distribution evaluations and is independent of the realized prediction errors.

\section{Numerical Experiments}
\label{sec:experiments}

Our analytic results focus on the heavy-traffic regime; however, the resulting insights into policy design also apply at moderate loads. In this section, we show that the design choices that provide heavy-traffic order-optimality for SPD also yield near-optimal performance outside heavy traffic.

\paragraph{Experimental Setup.}
Our numerical experiments use the closed-form expressions for SPD derived in this work. We consider Poisson arrivals and mean-one service times from the following distributions: doubly exponential, exponential, Weibull with \(\nu=0.5,2\), and lognormal with \(\sigma=1\). Unless otherwise stated, predictions are costly and exact; each query consumes server time \(\delta_\rho\), chosen so that \(\lambda_\rho\delta_\rho=\sqrt{1-\rho}\). At each load, we independently tune \((\tau,b,q)\) over the full family of stable SPD policies for each background scheduler. As baselines, we compare against two size-oblivious policies: FCFS and FB, which are optimal for distributions with increasing and decreasing hazard rates, respectively. We also compare with DelayPredict~\cite{shahout_skippredict_2024}, implemented as \(\mathrm{SPD}(0,1,\tau,\mathrm{SRPT})\) which demotes all queried jobs and runs SRPT on the low-priority queue.

\begin{figure}[!htbp]
  \centering
  \includegraphics[width=\linewidth]{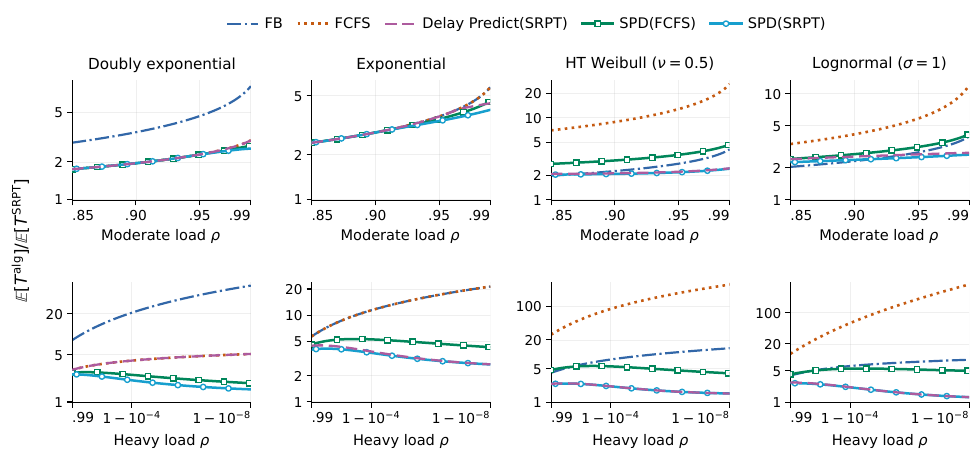}
  \caption{\protect Mean response times with costly, exact predictions, normalized by SRPT with free size information. Top: moderate loads; bottom: high loads with \(\rho\)-labels spaced by \(-\log_{10}(1-\rho)\). Vertical axes are logarithmic. }
 \label{fig:experiments-load}
\end{figure}

\paragraph{\protect Performance Inside and Outside Heavy Traffic.}
Figure~\ref{fig:experiments-load} compares SPD with FCFS, FB, and DelayPredict, with response times normalized by SRPT with free size information. Over moderate loads \(0.85\le\rho\le0.99\), SPD stays within a small multiple of this optimum across all four distributions: the largest observed ratios are 4.72 for SPD(FCFS) and 4.00 for SPD(SRPT). All policies perform comparably for exponential service, but the size-oblivious policies are more sensitive to the service distribution: FB performs poorly for doubly exponential service, whereas FCFS performs poorly for Weibull \(\nu=0.5\) and lognormal service. SPD avoids these pronounced losses across the four examples, although FB can outperform it at some moderate loads.

At high loads, SPD continues to perform well across all four distributions. Over the grid through \(\rho=1-10^{-8}\), its largest observed response ratios are 5.71 for SPD(FCFS) and 4.11 for SPD(SRPT). By comparison, FB's ratio reaches 51.86 for doubly exponential service, while FCFS's reaches 286.17 and 424.75 for Weibull \(\nu=0.5\) and lognormal service. At the highest plotted load, both SPD variants outperform both size-oblivious baselines in all four examples, while each queries at most \(1.42\times10^{-5}\) of arrivals.

DelayPredict is competitive at moderate loads and nearly matches SPD(SRPT) at high loads for Weibull \(\nu=0.5\) and lognormal service. In these cases, tuning SPD(SRPT) selects \(q=1\) and a trigger close to DelayPredict's, so both obtain sparsity by delaying queries. However, for doubly exponential service, both SPD variants improve substantially on DelayPredict: at \(\rho=1-10^{-8}\), the response ratios are 5.10 for DelayPredict, 1.89 for SPD(FCFS), and 1.53 for SPD(SRPT). Here SPD benefits from querying a small fraction of jobs at arrival, while DelayPredict can limit query use only by delaying the trigger. 

\begin{wrapfigure}{r}{0.6\textwidth}
  \centering
  \includegraphics[width=0.6\textwidth]{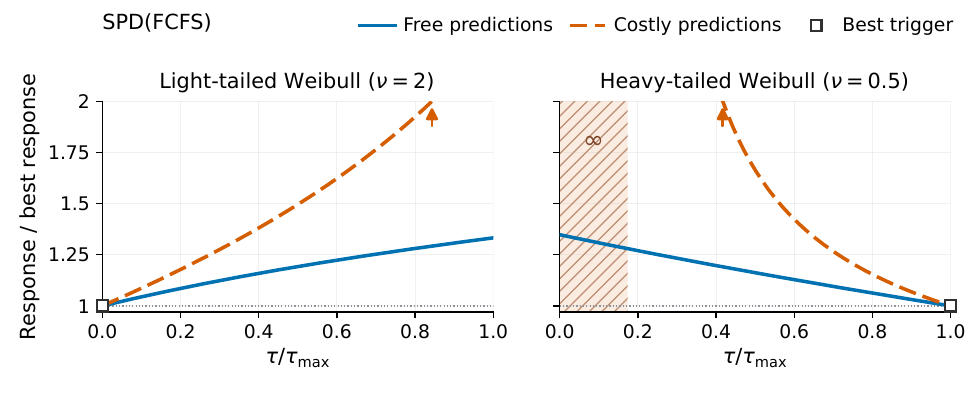}
  \caption{Optimal trigger choice for SPD(FCFS) at \(\rho=0.9\). Each curve is normalized by its best sampled response at the same cost and barrier. Squares mark minima; arrows indicate finite values above two; hatching indicates instability of the costly policy.}
  \label{fig:experiments-trigger-fcfs}
\end{wrapfigure}

\paragraph{Trigger Selection.} 
Figure~\ref{fig:experiments-trigger-fcfs} illustrates how the best trigger depends on the service distribution outside of heavy traffic. In this figure, we set \(\rho=0.9\) and \(b=0.1G^{-1}(\rho)\), and adjust \(q\) as the trigger varies to keep \(q\overline F(\tau+b)=1-\rho\), fixing the fraction of arrivals demoted. We normalize the trigger by \(\tau_{\max}=\bar\tau_\rho(0.1)=F^{-1}(\rho)-b\). The best trigger is at arrival for Weibull \(\nu=2\) and is the latest feasible trigger for \(\nu=0.5\), both with free and with costly predictions.

These curves reflect two effects of trigger choice. Although each demoted job has more than \(b\) units of service remaining, the mean \(\mathbb E[S-\tau\mid S>\tau+b]\) depends on \(\tau\). For Weibull \(\nu=0.5\), an earlier trigger demotes less remaining work per job and requires more queries to demote the same \(1-\rho\) fraction; for \(\nu=2\), a later trigger has both disadvantages. The first effect changes priority allocation even with free predictions; the second adds inference load when predictions are costly. For Weibull \(\nu=0.5\), inference cost makes the first 17.35\% of the trigger range unstable. Thus an incorrect trigger can more than double delay or destroy stability even at the moderate load \(\rho=0.9\). The figure uses FCFS as the background scheduler; SRPT gives qualitatively similar results.

\section{Concluding Remarks}

A fundamental challenge for scheduling is to determine the minimal information about job sizes necessary to achieve near-optimal mean response times in a single-server queue. We introduced a novel policy called Sparse Predict and Demote (SPD), which can achieve heavy-traffic order-optimality, even in situations with costly and/or inexact predictions, using a vanishing fraction of one-bit job-size predictions. Although our analytic guarantees concern the heavy-traffic regime, we use numerical experiments to show that the insights extend beyond heavy traffic. Importantly, our results highlight both when a vanishing fraction of one-bit predictions is enough to achieve heavy-traffic order-optimality and when one-bit predictions are not enough. Our work also points to several avenues of open direction described below.

First, the intermediate regime leaves an unresolved gap in the amount of size information needed to achieve heavy-traffic order-optimality. Our impossibility result applies to binary-level queues, while the positive result requires full-size information to order demoted jobs by remaining size. An open question is whether a fixed number of size classes can replace these full-size predictions while retaining both heavy-traffic order-optimality and a vanishing query fraction. Resolving this gap would clarify the prediction precision required by different service distributions.

Second, it would be desirable to relax the current M/GI/1 assumption; that is, extend sparse prediction to queues with more than one server or non-Poisson arrivals. For renewal arrivals, the diffusion analysis of two-priority-class queues by Chen and Dong~\cite{chen2021scheduling} provides a possible starting point. The challenge is to obtain comparable guarantees for mean response time in these models while accounting for inference work and querying only a vanishing fraction of jobs.

Finally, our robustness analysis gives sufficient conditions for heavy-traffic order-optimality under a fixed multiplicative error bound, leaving open how much error can be tolerated. It also does not exploit improvements in prediction quality over time. As more accurate predictions become available, the policy could adjust when it queries jobs and how it uses the resulting estimates. It remains open when such adaptations reduce inference work or mean response time while preserving heavy-traffic order-optimality.

\subsection*{Acknowledgements}
The authors would like to thank the American Institute of Mathematics (AIM) as this work was initiated during an AIM SQuARE on learning-augmented algorithms as well as the National Science Foundation (NSF) for its support through the grants with award numbers 2512127 and 2326609. 

\bibliographystyle{alpha}
\bibliography{references}

\clearpage
\appendix
\appendixpage

\section[Distributional Regularity]{Consequences of Assumption~\ref{asm:heavy-traffic-regularity} (Section~\ref{subsec:service-time-classes})}
\label{app:distributional-regularity}

We organize the consequences of Assumption~\ref{asm:heavy-traffic-regularity} into three parts: finite service-time moments in Appendix~\ref{app:service-finite-moments}, the boundary of the lognormal-or-lighter regime in Appendix~\ref{subsec:log-quadratic-boundary}, and tail-work estimates in Appendix~\ref{app:free-bit-tail-work-proofs}.

\subsection[Finite Service-Time Moments]{Finite Service-Time Moments}
\label{app:service-finite-moments}

The response-time formulas throughout the paper depend on the first two service-time moments. The following lemma ensures that these moments are finite constants as the load approaches one; in particular, it justifies the FCFS benchmark in Appendix~\ref{app:soap-fcfs}.

\begin{lemma}[Finite service-time moments]
\label{lem:service-finite-moments}
Under Assumption~\ref{asm:heavy-traffic-regularity}, \(0<\mathbb E[S^m]<\infty\) for every \(m>0\). These moments are independent of \(\rho\), since the service distribution is fixed in the heavy-traffic limit.   
\end{lemma}
\begin{proof}[Proof of Lemma~\ref{lem:service-finite-moments}]
Fix \(m>0\). Since \(H(x)/\log x\to\infty\) (by the first condition of Assumption~\ref{asm:heavy-traffic-regularity}), there exists
an \(x_0>1\) such that \(H(x)\ge(m+1)\log x\), and hence \(\overline F(x)\le x^{-(m+1)}\), for every \(x\ge x_0\). The tail-integral identity gives
\[
    \mathbb E[S^m]
    =m\int_0^\infty x^{m-1}\overline F(x)\,dx
    \le x_0^m+m\int_{x_0}^\infty x^{-2}\,dx<\infty.
\]
The moments are positive because the service distribution has unbounded support. Their constancy in \(\rho\) follows from holding this distribution fixed.
\end{proof}

\subsection{Lognormal-or-Lighter Boundary}
\label{subsec:log-quadratic-boundary}
The classification below supplies the hazard lower bound used in Appendix~\ref{app:sparse-candidate-abundance-proof} (Section~\ref{sec:sparse-predict}), so we collect these distributional consequences here.
We define
\[
    L:=\lim_{x\to\infty}\frac{x h(x)}{\log x}\in[0,\infty],
\]
which exists by the third condition of Assumption~\ref{asm:heavy-traffic-regularity}. We now use  \(L\) to classify the regimes in Section~\ref{subsec:service-time-classes}.

\begin{lemma}[Lognormal boundary]
\label{lem:free-bit-log-quadratic}
Under Assumption~\ref{asm:heavy-traffic-regularity}, \(L\) determines the tail regime:
\begin{enumerate}
    \item If \(L=0\), then \(\log x\ll H(x)\ll(\log x)^2\)
    , i.e., the service distribution is in the intermediate regime.
    \item If \(0<L<\infty\), then \(H(x)\sim\frac{L}{2}(\log x)^2\), i.e., the lognormal boundary.
    \item If \(L=\infty\), then \(H(x)/(\log x)^2\to\infty\), i.e., the service distribution is lognormal-or-lighter. 
\end{enumerate}

\end{lemma}

\begin{proof}[Proof of Lemma~\ref{lem:free-bit-log-quadratic}]
By the first condition of Assumption~\ref{asm:heavy-traffic-regularity},
$\frac{H(x)}{\log x}\longrightarrow\infty$,
so in particular \(H(x)\to\infty\). Since \(H'(x)=h(x)\), the rule of de l'H\^{o}pital gives
\[
\begin{aligned}
    \lim_{x\to\infty}
    \frac{H(x)}{(\log x)^2}
    =
    \lim_{x\to\infty}
    \frac{h(x)}{2\log x/x}
    =
    \frac12
    \lim_{x\to\infty}
    \frac{x h(x)}{\log x}
    =
    \frac{L}{2}.
\end{aligned}
\]
This proves the limit in all three cases \(L=0\), \(0<L<\infty\), and \(L=\infty\), and hence the stated classification.
\end{proof}

\subsection{Tail-Work Estimates}
\label{app:free-bit-tail-work-proofs}
The following estimates convert the probability of exceeding a cutoff into the expected work carried by those jobs. The constant-factor bound is used for the tail-work scarcity estimate in Appendix~\ref{app:free-bit-tail-work-scarcity-proof}; the sharper asymptotic is used to compare the size and workload cutoffs in Appendix~\ref{app:free-bit-cutoff-comparison-proof}. For a cutoff \(x>0\), the baseline \(x\overline F(x)\) counts \(x\) units of service for each job exceeding the cutoff. The remaining service is the tail integral:
\[
    \mathbb E[S\mathbf1\{S>x\}]
    =x\overline F(x)+\int_x^\infty\overline F(u)\,du.
\]
The first estimate bounds the excess by a constant multiple of the baseline; the second shows that it is negligible in the lognormal-or-lighter regime.

\begin{lemma}[Tail-work bound]
\label{lem:free-bit-tail-work-bound}
Under Assumption~\ref{asm:heavy-traffic-regularity}, as \(x\to\infty\),
\[
    \mathbb E[S\mathbf 1\{S>x\}]
    =\Theta\!\left(x\overline F(x)\right).
\]
\end{lemma}
\begin{proof}[Proof of Lemma~\ref{lem:free-bit-tail-work-bound}]
The second condition of Assumption~\ref{asm:heavy-traffic-regularity}, \(\alpha(\overline F)<-2\) provides a bound that is uniform over every larger threshold. Choose \(p>2\) with \(\alpha(\overline F)<-p\). By the definition in the footnote to the second condition of Assumption~\ref{asm:heavy-traffic-regularity}, there exist \(C<\infty\) and \(x_0<\infty\) such that
\[
    \frac{\overline F(xy)}{\overline F(x)}
    \le Cy^{-p},
    \qquad x\ge x_0,\quad y\ge1.
\]
Changing variables from \(u\) to \(xy\) therefore gives
\[
    \int_x^\infty\overline F(u)\,du
    =x\int_1^\infty\overline F(xy)\,dy
    \le\frac{C}{p-1}x\overline F(x).
\]
Thus the excess work can enlarge the baseline \(x\overline F(x)\) by at most a fixed factor:
\[
    x\overline F(x)
    \le\mathbb E[S\mathbf 1\{S>x\}]
    \le\left(1+\frac{C}{p-1}\right)x\overline F(x).
\]
This proves the claimed \(\Theta\) estimate.
\end{proof}
In the lognormal-or-lighter regime, the mean excess size above a high cutoff is \(\mathbb E[S-x\mid S>x]=o(x)\). Thus a job exceeding \(x\) has mean size asymptotic to \(x\), and the total work in such jobs is asymptotic to \(x\) times their probability. The explicit factor \(2\) below is sufficient for the cutoff comparison in Appendix~\ref{app:free-bit-cutoff-comparison-proof}.
\begin{lemma}[Tail-work asymptotics]
\label{lem:free-bit-tail-work-asymptotics}
In the lognormal-or-lighter regime, as \(x\to\infty\),
\[
    \mathbb E[S\mathbf 1\{S>x\}]
    =x\overline F(x)(1+o(1)).
\]
In particular, for all sufficiently large \(x\),
\[
    x\overline F(x)
    \le \mathbb E[S\mathbf 1\{S>x\}]
    \le 2x\overline F(x).
\]

\end{lemma}
\begin{proof}[Proof of Lemma~\ref{lem:free-bit-tail-work-asymptotics}]
In the lognormal-or-lighter regime, the bound on relative tail probabilities becomes stronger as the threshold grows. By Lemma~\ref{lem:free-bit-log-quadratic}, \(L\in(0,\infty]\), so there is a \(c>0\) such that \(h(u)\ge c\log u/u\) for all sufficiently large \(u\). For large \(x\) and every \(y\ge1\), this gives
\[
    H(xy)-H(x)
    \ge c\int_x^{xy}\frac{\log u}{u}\,du
    \ge c\log x\log y,
    \qquad
    \frac{\overline F(xy)}{\overline F(x)}
    \le y^{-c\log x}.
\]
Unlike the fixed exponent \(p\) in the previous proof, the exponent \(c\log x\) tends to infinity.

We again start with the integral of \(\overline F\) from \(x\) to infinity and change variables from \(u\) to \(xy\). Once \(c\log x>1\), the tail bound above gives
\[
\begin{aligned}
    \int_x^\infty\overline F(u)\,du
    =x\int_1^\infty\overline F(xy)\,dy
    \le x\overline F(x)\int_1^\infty y^{-c\log x}\,dy=\frac{x\overline F(x)}{c\log x-1}
    =o\!\left(x\overline F(x)\right).
\end{aligned}
\]
The last step follows because \(1/(c\log x-1)\to0\). Thus the expected service beyond the threshold is negligible compared with the baseline work \(x\overline F(x)\). Substituting this estimate into the tail-work decomposition yields
\[
\begin{aligned}
    \mathbb E[S\mathbf 1\{S>x\}]
    =x\overline F(x)+\int_x^\infty\overline F(u)\,du=x\overline F(x)+o\!\left(x\overline F(x)\right)
      =x\overline F(x)(1+o(1)).
\end{aligned}
\]
This proves the asymptotic statement. For all sufficiently large \(x\), the excess integral is at most \(x\overline F(x)\), giving the upper bound with factor \(2\); the lower bound follows from the nonnegative excess integral.

\end{proof}

\section{FCFS Benchmark and SOAP Response-Time (Section~\ref{subsec:soap-overview})}
\label{app:SOAP}

Appendix~\ref{app:soap-fcfs} derives the FCFS mean response time from the SOAP identity and compares its heavy-traffic order with SRPT. Appendix~\ref{app:soap-general-intuition} then states the general SOAP response-time identity and explains how it accounts for changes in priority as jobs receive service. 

\subsection[FCFS Benchmark]{FCFS Benchmark}
\label{app:soap-fcfs}
With no inference work, FCFS has the constant rank \(r(d,a)=0\). An older job contributes all its service to the original relevant interval, while a new arrival cannot pass the tagged job in an FCFS tie. Consequently,
\[
 X_0^{\mathrm{old}}[0]=S,\qquad
 X_i^{\mathrm{old}}[0]=0\ (i\ge1),\qquad
 X^{\mathrm{new}}[0]=0.
\]
The corresponding loads are \(\rho_0^{\mathrm{old}}[0]=\rho\) and \(\rho^{\mathrm{new}}[0]=0\). Substitution in Proposition~\ref{prop:soap-nonincreasing} gives, for a tagged job of size \(x\),
\[
 \mathbb E[T_{d,x}]
 =\frac{\lambda_\rho\mathbb E[S^2]}{2(1-\rho)}+x.
\]
Averaging over \(x=S\) gives
\[
 \mathbb E[T^{\mathrm{FCFS}}]
 =\mathbb E[S]+\frac{\lambda_\rho\mathbb E[S^2]}{2(1-\rho)}.
\] By Lemma~\ref{lem:service-finite-moments}, the service moments are fixed positive constants and \(\lambda_\rho\to1/\mathbb E[S]>0\), so its mean response time has order \((1-\rho)^{-1}\). Comparing with~\eqref{eq:srpt-heavy-traffic-benchmark} gives \(\mathbb E[T^{\mathrm{FCFS}}]/\mathbb E[T^{\mathrm{SRPT}}]=\Theta(G^{-1}(\rho))\).

\subsection[General SOAP Identity and Intuition]{General SOAP Identity and Intuition}
\label{app:soap-general-intuition}

Proposition~\ref{prop:soap-nonincreasing} assumes that a job's rank does not increase as it receives service. This assumption need not hold when a policy requests a size prediction only after a job has received some service: the prediction may cause the job to be assigned lower priority, increasing its rank. We therefore also need a characterization of the mean response time for general rank trajectories. 
For a generic arriving job with descriptor \(D\) and total server work \(X\), let \(X^{\mathrm{new}}[r]\) denote the service received until the job completes or first has rank at least \(r\); this variable is zero if the job arrives with rank at least \(r\). As in Section~\ref{subsec:soap-overview}, define \(\rho^{\mathrm{new}}[r]:=\lambda\mathbb E[X^{\mathrm{new}}[r]]\). The pairs \((D,X)\) are i.i.d. across jobs and independent of the Poisson arrival process.

For general ranks, we need two additions to the notation. First, an old job may enter the region relevant to rank \(r\) multiple times. We let \(X_0^{\mathrm{old}}[r]\) denote its service in the original relevant interval and \(X_i^{\mathrm{old}}[r]\), \(i\geq 1\), its service in the \(i\)-th recycled relevant interval, and define $\rho_i^{\mathrm{old}}[r] := \lambda \mathbb E[X_i^{\mathrm{old}}[r]], \text{ where } i\geq0$.
Second, the tagged job's current rank need not capture all work that can delay it, because its rank may worsen later. SOAP therefore uses its
\emph{worst future rank}
$r_{d,x}^{\mathrm{worst}}(a) := \sup_{a\leq b<x} r(d,b)$.

With these definitions, the general SOAP response-time identity is as follows.

\newcounter{soapSavedProposition}
\setcounter{soapSavedProposition}{\value{proposition}}
\setcounter{proposition}{1}
\begin{proposition}[SOAP for general ranks {\citep[Theorem~5.5]{scully_soap_2018}}]
\label{prop:soap-response-time}
 The conditional mean response time of a tagged job with descriptor \(d\) and total work \(x\) is
\[
 \mathbb E[T_{d,x}]
 =\frac{\lambda\displaystyle\sum_{i=0}^{\infty}
       \mathbb E\!\left[(X_i^{\mathrm{old}}[r])^2\right]}
       {2(1-\rho_0^{\mathrm{old}}[r])(1-\rho^{\mathrm{new}}[r])}
  +\int_0^x\frac{da}{1-\rho^{\mathrm{new}}[r(a)]},
\]
where \(r=r_{d,x}^{\mathrm{worst}}(0)\) and \(r(a)=r_{d,x}^{\mathrm{worst}}(a)\).
\end{proposition}
\setcounter{proposition}{\value{soapSavedProposition}}

We provide some additional intuition for Proposition~\ref{prop:soap-response-time}, focusing on the two features that arise when ranks are not monotone in age.

The first issue is that the tagged job's current rank need not determine all of the work that can delay it. Even if another job has lower priority than the tagged job at its current rank, it may still delay the tagged job if the tagged job's own rank becomes worse later. The Pessimism Principle of \cite{scully_soap_2018} shows that, when determining how much another job can delay a tagged job of descriptor \(d\) and total work \(x\) at age \(a\), it suffices to compare that job against the tagged job's \emph{worst future rank}, defined above.
More strongly, the Worst Future Rank Substitution result of \cite[Appendix~D]{scully_soap_2018} shows that the tagged job may be scheduled as if its rank at every age \(a\) were \(r_{d,x}^{\mathrm{worst}}(a)\), while leaving the arrival sequence and the rank functions of all other jobs unchanged, without changing the tagged job's response time. Since \(a\mapsto r_{d,x}^{\mathrm{worst}}(a)\) is non-increasing, this gives an equivalent non-increasing-rank representation from the perspective of the tagged job. Intuitively, the substitution pessimistically gives the tagged job immediately the worst priority that it will attain in the future; doing so does not change which pieces of competing work ultimately delay its completion.

The second issue is that the other jobs need not have monotone rank trajectories. Relative to a fixed rank~\(r\), another job may therefore move into and out of relevance several times. SOAP decomposes the service received during these periods into relevant intervals. The quantity \(X_0^{\mathrm{old}}[r]\) denotes the service received in the job's original relevant interval, while \(X_i^{\mathrm{old}}[r]\), \(i\geq1\), denotes the service received in its \(i\)-th \emph{recycled} relevant interval. Thus, even though the tagged job has been transformed to have a non-increasing rank trajectory, the non-monotonicity of the other jobs is retained through the collection
$\left\{X_i^{\mathrm{old}}[r]\right\}_{i\geq0}$.
SOAP's Vacation Transformation then represents the delays generated by these recycled relevant intervals in an equivalent transformed queue. Together with the worst-future-rank representation of the tagged job, this yields Proposition~\ref{prop:soap-response-time}.

There is one distinction from the non-increasing-rank case that is useful to keep in mind. When the original rank trajectory is non-increasing, the two terms in the expression of Proposition~\ref{prop:soap-response-time} coincide with the physical waiting- and residence-time contributions defined in Section~\ref{subsec:soap-overview}. For a general rank trajectory, however, the expression is obtained after the Worst Future Rank Substitution and Vacation Transformation, so we do not separately identify these two terms with the original job's physical waiting and residence times. What these transformations preserve is their sum: the response time of the tagged job.

Finally, when \(a\mapsto r(d,a)\) is itself non-increasing for every descriptor \(d\), the worst rank is always equal to the current rank, so no worst-future-rank substitution is needed:
\[
    r_{d,x}^{\mathrm{worst}}(a)=r(d,a).
\]
Moreover, another job can have at most one recycled relevant interval, so \(X_i^{\mathrm{old}}[r]=0\) for all \(i\geq2\). The general SOAP identity therefore reduces exactly to Proposition~\ref{prop:soap-nonincreasing}.

\section[Proofs for Binary-Level Queues (Section~\ref{sec:one-free-bit})]{Proofs for Binary-Level Queues (Section~\ref{sec:one-free-bit})}
\label{app:one-free-bit}

This appendix supplies the deferred proofs from Section~\ref{sec:one-free-bit}. Appendix~\ref{app:free-bit-response-identities} proves Lemmas~\ref{lem:free-bit-high-priority-response} and~\ref{lem:free-bit-low-priority-response}, and Appendix~\ref{app:free-bit-sufficient} proves Lemma~\ref{lem:free-bit-threshold-location}. Appendix~\ref{app:free-bit-cutoff-comparison-proof} proves the auxiliary cutoff comparison in Lemma~\ref{lem:free-bit-cutoff-comparison}. Appendix~\ref{app:free-bit-necessary-structure-proof} proves the necessary workload and average-size bounds in Lemma~\ref{lem:free-bit-necessary-structure}; Appendix~\ref{app:free-bit-tail-work-scarcity-proof} proves the tail-work estimate in Lemma~\ref{lem:free-bit-tail-work-scarcity}. Throughout, Assumption~\ref{asm:heavy-traffic-regularity} is in force.

\subsection[Response-Time Identities (Lemmas~\ref{lem:free-bit-high-priority-response} and~\ref{lem:free-bit-low-priority-response})]{Response-Time Identities (Lemmas~\ref{lem:free-bit-high-priority-response} and~\ref{lem:free-bit-low-priority-response})}
\label{app:free-bit-response-identities}

Both identities follow by identifying the relevant old and new work in Proposition~\ref{prop:soap-nonincreasing}. Fixed ranks imply that there is no recycled old work. We give the full arguments for completeness.

\begin{proof}[Proof of Lemma~\ref{lem:free-bit-high-priority-response}]
Consider a tagged job with rank \(0\). We first analyze its waiting time. Since ranks are fixed, an old job cannot become irrelevant and later become relevant again. Hence there is no recycled old work, so \(X_i^{\mathrm{old}}[0]=0\) for all \(i\ge1\), and in particular \(X_1^{\mathrm{old}}[0]=0\). An old job contributes relevant work precisely when it also has rank \(0\). Thus \(X_0^{\mathrm{old}}[0]=S\mathbf 1\{R_\rho=0\}\) and \(\rho_0^{\mathrm{old}}[0]=\rho^{\mathrm{HIGH}}\). Moreover, no future arrival can delay the tagged job: a future rank-\(0\) job has the same rank but loses the FCFS tie, while a future rank-\(1\) job has lower priority. Hence \(X^{\mathrm{new}}[0]=0\) and \(\rho^{\mathrm{new}}[0]=0\). Finally, the tagged job's residence time is its own   size  as it cannot be preempted by future arrivals. Substituting these quantities into Proposition~\ref{prop:soap-nonincreasing} and averaging over rank-\(0\) jobs yields the lemma.
\end{proof}

\begin{proof}[Proof of Lemma~\ref{lem:free-bit-low-priority-response}]
Consider a tagged job with rank \(1\). We first analyze its waiting time. Since ranks are fixed, an old job cannot become irrelevant and later become relevant again, so \(X_i^{\mathrm{old}}[1]=0\) for all \(i\ge1\).
Every old job contributes relevant work: an old rank-\(0\) job has higher priority, while an old rank-\(1\) job has the same rank and wins the FCFS tie. Hence \(X_0^{\mathrm{old}}[1]=S\) and \(\rho_0^{\mathrm{old}}[1]=\rho\). Among future arrivals, only rank-\(0\) jobs can move ahead of the tagged job, so \(X^{\mathrm{new}}[1]=S\mathbf 1\{R_\rho=0\}\) and \(\rho^{\mathrm{new}}[1]=\rho^{\mathrm{HIGH}}\).
During residence, the tagged job is interrupted only by future rank-\(0\) work. Thus, for a rank-\(1\) job of size~\(x\), Proposition~\ref{prop:soap-nonincreasing} gives the residence contribution
$\int_0^{x}\frac{da}{1-\rho^{\mathrm{new}}[1]}
    =
    \frac{x}{1-\rho^{\mathrm{HIGH}}}$.
Substituting these quantities
into Proposition~\ref{prop:soap-nonincreasing} and averaging over rank-\(1\) jobs yields the lemma.
\end{proof}

\subsection[Slack-Matched Size Cutoff (Lemma~\ref{lem:free-bit-threshold-location})]{Slack-Matched Size Cutoff (Lemma~\ref{lem:free-bit-threshold-location})}
\label{app:free-bit-sufficient}

The   lognormal-or-lighter  regime means that \(H(x)=\Omega((\log x)^2)\): for some \(\kappa>0\), large jobs have tail probability at most \(e^{-\kappa(\log x)^2}\). To locate the slack-matched size cutoff within this regime, we use two auxiliary lemmas. The first compares it with the workload cutoff \(G^{-1}(\rho)\); the second translates an integrated hazard bound into a bound on their logarithmic separation.
Recall from Section~\ref{sec: prelims} that \(G(x)=\mathbb E[S\mathbf1\{S\le x\}]/\mathbb E[S]\). Thus \(G^{-1}(\rho)\) leaves a \(1-\rho\) fraction of total work above the cutoff, whereas \(\bar{s}_\rho\) leaves a \(1-\rho\) fraction of jobs.

\begin{lemma}[Cutoff comparison]
\label{lem:free-bit-cutoff-comparison}
In the   lognormal-or-lighter regime,  as \(\rho\uparrow1\)  ,

\begin{equation*}
\label{eq:free-bit-normalized-hazard-gap}
    \frac{H(G^{-1}(\rho))-H(\bar{s}_\rho)}{\log G^{-1}(\rho)}
    \longrightarrow1.
\end{equation*}

\end{lemma}

We prove the lemma in Appendix~\ref{app:free-bit-cutoff-comparison-proof}.

\begin{lemma}[Logarithmic integral]
\label{lem:free-bit-logarithmic-integral}
For \(1\le a\le b\) with \(b>1\), let \(\Delta=\log(b/a)\). Then
\[
    \frac{1}{\log b}\int_a^b\frac{\log u}{u}\,du
    =\frac{1}{\log b} \cdot \frac{\Delta}{2}(\log b+\log a)
    =\Delta-\frac{\Delta^2}{2\log b},
\]
and hence
\[
    \frac{\Delta}{2}
    \le\frac{1}{\log b}\int_a^b\frac{\log u}{u}\,du
    \le\Delta.
\]
\end{lemma}

\begin{proof}[Proof of Lemma~\ref{lem:free-bit-logarithmic-integral}]
Expanding the integral, factoring, and substituting  \(\log a=\log b-\Delta\)  gives
\[
    \int_a^b\frac{\log u}{u}\,du=\frac{(\log b)^2-(\log a)^2}{2}
    =\frac{(\log b-\log a)(\log b+\log a)}{2}
    =\frac{\Delta(2\log b-\Delta)}{2}.
\]
The first part of the lemma then follows directly, dividing by $\log b$. The upper bound in the second part of the lemma holds because $\log b>0$. 
The lower bound in the second part of the lemma holds because
\(1\le a\le b\) implies \(0\le\Delta=\log b-\log a\le\log b\).
Thus \(2\log b-\Delta\ge\log b\), and
\[
    \frac{1}{\log b}\int_a^b\frac{\log u}{u}\,du
    =\frac{\Delta(2\log b-\Delta)}{2\log b}
    \ge\frac{\Delta\log b}{2\log b}
    =\frac{\Delta}{2}.
\]
\end{proof}

\begin{proof}[Proof of Lemma~\ref{lem:free-bit-threshold-location}]
  \(\overline F(\bar{s}_\rho)=1-\rho\) and unbounded support implies
  \(\bar{s}_\rho\to\infty\); likewise, \(G^{-1}(\rho)\to\infty\). The positive limiting hazard gap in  Lemma~\ref{lem:free-bit-cutoff-comparison}   then gives \(\bar{s}_\rho<G^{-1}(\rho)\) eventually. Thus we compare  hazards on \([\bar{s}_\rho,G^{-1}(\rho)]\). Set
\[
    \Delta_\rho:=\log\frac{G^{-1}(\rho)}{\bar{s}_\rho}\ge0 \qquad \Rightarrow
    \qquad
    \frac{\bar{s}_\rho}{G^{-1}(\rho)}=e^{-\Delta_\rho}.
\]

Recall the (extended) limit \(L:=\lim_{u\to\infty}uh(u)/\log u\), which measures the hazard relative to \((\log u)/u\). By Lemma~\ref{lem:free-bit-log-quadratic}, \(L\in(0,\infty]\) in this regime. We bound \(\Delta_\rho\) in the two cases.

\paragraph{Case 1: \(0<L<\infty\).}
Fix \(\varepsilon\in(0,L)\). Since \(\bar{s}_\rho\to\infty\), the definition of \(L\) gives, for all sufficiently large \(\rho\),
\[
    (L-\varepsilon)\frac{\log u}{u}
    \le h(u)\le(L+\varepsilon)\frac{\log u}{u},
    \qquad \bar{s}_\rho\le u\le G^{-1}(\rho).
\]
Integrating from \(a=\bar{s}_\rho\) to \(b=G^{-1}(\rho)\), using Lemma~\ref{lem:free-bit-logarithmic-integral}, and dividing by $\log b$ yields
\[
    (L-\varepsilon)\left(\Delta_\rho-
        \frac{\Delta_\rho^2}{2\log G^{-1}(\rho)}\right)
    \le\frac{H(G^{-1}(\rho))-H(\bar{s}_\rho)}{\log G^{-1}(\rho)}
    \le(L+\varepsilon)\left(\Delta_\rho-
        \frac{\Delta_\rho^2}{2\log G^{-1}(\rho)}\right).
\]

The integral lemma also bounds the parenthesized quantity below by \(\Delta_\rho/2\). By Lemma~\ref{lem:free-bit-cutoff-comparison}, we can choose \(\rho'\in(0,1)\) such that the preceding inequalities hold and the middle expression is at most \(2\) whenever \(\rho'<\rho<1\). Thus,
\[
    \frac{L-\varepsilon}{2}\Delta_\rho
    \le\frac{H(G^{-1}(\rho))-H(\bar{s}_\rho)}{\log G^{-1}(\rho)}
    \le2.
\]
Consequently, \(0\le\Delta_\rho\le4/(L-\varepsilon)\) for all \(\rho\in(\rho',1)\), so \(\Delta_\rho=O(1)\) as \(\rho\uparrow1\). Since \(G^{-1}(\rho)\to\infty\), the correction \(\Delta_\rho^2/(2\log G^{-1}(\rho))\) tends to zero. Returning to the two-sided bound containing the correction term, taking the upper limit in its left inequality and the lower limit in its right inequality gives

\[
    (L-\varepsilon)\limsup_{\rho\uparrow1}\Delta_\rho
    \le 1
    \le (L+\varepsilon)\liminf_{\rho\uparrow1}\Delta_\rho.
\]

Letting \(\varepsilon\downarrow0\) proves \(\Delta_\rho=\log\frac{G^{-1}(\rho)}{\bar{s}_\rho}\to1/L\), and therefore \(\bar{s}_\rho/G^{-1}(\rho)\to e^{-1/L}\).

\paragraph{Case 2: \(L=\infty\).}
For every fixed \(K>0\), eventually \(h(u)\ge K\log u/u\) throughout \([\bar{s}_\rho,G^{-1}(\rho)]\). Integrating and applying the lower bound in Lemma~\ref{lem:free-bit-logarithmic-integral} gives

\[
    \frac{K}{2}\Delta_\rho
    \le\frac{H(G^{-1}(\rho))-H(\bar{s}_\rho)}{\log G^{-1}(\rho)}
    \longrightarrow1.
\]
Consequently \(0\le\limsup_{\rho\uparrow1}\Delta_\rho\le2/K\). Since \(K\) is arbitrary, \(\Delta_\rho\to0\), and \(\bar{s}_\rho/G^{-1}(\rho)\to1\). 

In both cases the cutoff ratio converges to a limit \(\bar{\theta}\in(0,1]\), as claimed.
\end{proof}

\subsection[Cutoff Comparison (Lemma~\ref{lem:free-bit-cutoff-comparison})]{Cutoff Comparison (Lemma~\ref{lem:free-bit-cutoff-comparison})}
\label{app:free-bit-cutoff-comparison-proof}

\begin{proof}[Proof of Lemma~\ref{lem:free-bit-cutoff-comparison}]
We compare two cutoffs associated with the fraction \(1-\rho\):
\(\bar{s}_\rho\) leaves that fraction of jobs above it, whereas \(G^{-1}(\rho)\) leaves that fraction of total work above it.

By definition, \(G(x)=\mathbb E[S\mathbf1\{S\le x\}]/\mathbb E[S]\). Thus \(G(G^{-1}(\rho))=\rho\) implies \(\mathbb E[S\mathbf1\{S\le G^{-1}(\rho)\}]=\rho\mathbb E[S]\).
Subtracting this contribution from the mean work per arrival, \(\mathbb E[S]\), gives the expected work per arrival contributed by jobs above the cutoff. Using \(\overline F(\bar{s}_\rho)=1-\rho\), we obtain
\[
    \mathbb E[S\mathbf1\{S>G^{-1}(\rho)\}]
    =(1-\rho)\mathbb E[S]
    =\overline F(\bar{s}_\rho)\mathbb E[S].
\]

Because the service distribution has unbounded support, its tail probability and tail work are positive at every finite threshold. Its mean is finite and positive by Lemma~\ref{lem:service-finite-moments}, so both tail targets tend to zero as \(\rho\uparrow1\). Both cutoffs therefore tend to infinity.

A job with size exceeding \(G^{-1}(\rho)\) contributes at least \(G^{-1}(\rho)\) units of work. Consequently, the expected work per arrival contributed by these jobs is at least \(G^{-1}(\rho)\) times the probability that an arrival exceeds this cutoff. Lemma~\ref{lem:free-bit-tail-work-asymptotics} bounds this expected work from above by twice the same quantity once \(\rho\) is sufficiently close to one. Thus,
\[
    G^{-1}(\rho)\overline F(G^{-1}(\rho))
    \le (1-\rho)\mathbb E[S]
    \le 2G^{-1}(\rho)\overline F(G^{-1}(\rho)).
\]
Dividing by
\(\mathbb E[S]\overline F(G^{-1}(\rho))\) and substituting \(1-\rho=\overline F(\bar{s}_\rho)\), we obtain
\[
    \frac{G^{-1}(\rho)}{\mathbb E[S]}
    \le
    \frac{\overline F(\bar{s}_\rho)}
         {\overline F(G^{-1}(\rho))}
    \le
    \frac{2G^{-1}(\rho)}{\mathbb E[S]}.
\]

Thus, the ratio of the tail probabilities lies within a factor of two of \(G^{-1}(\rho)/\mathbb E[S]\). Taking logarithms converts this multiplicative comparison into an additive comparison of cumulative hazards. Since \(H=-\log\overline F\),
\[
    \log\frac{\overline F(\bar{s}_\rho)}
                  {\overline F(G^{-1}(\rho))}
    =H(G^{-1}(\rho))-H(\bar{s}_\rho).
\]
Taking logarithms in the preceding bounds therefore gives
\[
    \log G^{-1}(\rho)-\log\mathbb E[S]
    \le H(G^{-1}(\rho))-H(\bar{s}_\rho)
    \le \log G^{-1}(\rho)-\log\mathbb E[S]+\log2.
\]
Since \(G^{-1}(\rho)\to\infty\) and \(\mathbb E[S]\) is a fixed positive constant, the lower bound is eventually positive. Monotonicity of \(H\) then gives \(\bar{s}_\rho<G^{-1}(\rho)\).
Finally, dividing both bounds by \(\log G^{-1}(\rho)\to\infty\) makes the additive constants vanish, yielding
\[
    \frac{H(G^{-1}(\rho))-H(\bar{s}_\rho)}
         {\log G^{-1}(\rho)}
    \longrightarrow 1.
\]
\end{proof}

\subsection[Necessary Workload and Average-Size Bounds]{Necessary Workload and Average-Size Bounds (Lemma~\ref{lem:free-bit-necessary-structure})}
\label{app:free-bit-necessary-structure-proof}

\begin{proof}[Proof of Lemma~\ref{lem:free-bit-necessary-structure}]
 Suppose that a BLQ policy \(\varphi_\rho\) is heavy-traffic order-optimal along a sequence \(\rho\uparrow1\). All limits and order bounds in this proof are taken along that sequence. For some finite \(C>0\),
\[
    \mathbb E[T^{\varphi_\rho}]
    \le \frac{C}{(1-\rho)G^{-1}(\rho)}.
\]
Recall that the high-priority slack equals the original system slack plus the load carried by demoted jobs:
\[
    1-\rho^{\mathrm{HIGH}}
    =(1-\rho)+\lambda_\rho\mathbb E[S\mathbf 1\{R_\rho=1\}].
\]
We first show that only a small fraction of jobs can be demoted, then that these jobs must nevertheless carry substantial work. Together, the two requirements force the average low-priority job size to be large.

\paragraph{Low-priority delay limits the fraction of demoted jobs.}
The waiting term in Lemma~\ref{lem:free-bit-low-priority-response} is already a lower bound on the overall mean response time. Comparing it with the assumed upper bound gives
\[
    \mathbb P(R_\rho=1)
    \frac{\lambda_\rho\mathbb E[S^2]}
        {2(1-\rho)(1-\rho^{\mathrm{HIGH}})}
    \le \mathbb E[T^{\varphi_\rho}]
    \le \frac{C}{(1-\rho)G^{-1}(\rho)}.
\]
Cancelling the common factor \(1/(1-\rho)\), and using \(\lambda_\rho\to1/\mathbb E[S]>0\), yields

\begin{equation}
\label{eq:free-bit-low-prob-slack}
    \mathbb P(R_\rho=1)G^{-1}(\rho)
    =O\!\left(1-\rho^{\mathrm{HIGH}}\right).
\end{equation}
Because \(1-\rho^{\mathrm{HIGH}}\le1\) and \(G^{-1}(\rho)\to\infty\), we obtain
\[
    \mathbb P(R_\rho=1)
    =O\!\left(\frac{1}{G^{-1}(\rho)}\right)=o(1).
\]
Thus low-priority waiting would be too costly if a nonvanishing fraction of arrivals incurred it.

\paragraph{High-priority delay requires substantial work to be demoted.}
A vanishing low-priority fraction leaves essentially all second-moment mass at high priority. This claim needs justification because the demoted jobs may be precisely the largest ones. For any fixed \(K>0\), splitting the demoted jobs' second-moment contribution at size \(K\) gives
\[
    \mathbb E[S^2\mathbf 1\{R_\rho=1\}]
    \le K^2\mathbb P(R_\rho=1)
       +\mathbb E[S^2\mathbf 1\{S>K\}].
\]
The first term vanishes as \(\rho\uparrow1\); the second can be made arbitrarily small by taking \(K\) large, since \(\mathbb E[S^2]<\infty\)  by Lemma~\ref{lem:service-finite-moments} . This argument applies to every measurable, load-dependent assignment map, including randomized maps. It does not require the sets of demoted jobs to converge as \(\rho\) changes.

It follows that \(\mathbb E[S^2\mathbf 1\{R_\rho=0\}]\to\mathbb E[S^2]>0\), while \(\mathbb P(R_\rho=0)\to1\). The high-priority waiting contribution in Lemma~\ref{lem:free-bit-high-priority-response} therefore gives, for some \(c>0\) and all sufficiently large \(\rho\),
\[
    \frac{c}{1-\rho^{\mathrm{HIGH}}}
    \le \mathbb E[T^{\varphi_\rho}]
    \le \frac{C}{(1-\rho)G^{-1}(\rho)}.
\]
Consequently,
\begin{equation}
\label{eq:free-bit-necessary-slack}
    1-\rho^{\mathrm{HIGH}}
    =\Omega\!\left((1-\rho)G^{-1}(\rho)\right).
\end{equation}

This required slack is much larger than the original slack \(1-\rho\), because \(G^{-1}(\rho)\to\infty\). The slack identity at the start of the proof shows that demotion is the only source of the increase. Subtracting the negligible \(1-\rho\) term and using \(\lambda_\rho=\Theta(1)\) gives
\begin{equation}
\label{eq:free-bit-necessary-low-work}
    \mathbb E[S\mathbf 1\{R_\rho=1\}]
    =\Omega\!\left((1-\rho)G^{-1}(\rho)\right).
\end{equation}

\paragraph{A small fraction carrying substantial work must be large on average.}
The preceding work bound makes the original slack \(1-\rho\) negligible compared with the work demoted per arrival. The slack identity and \(\lambda_\rho=\Theta(1)\) therefore give
\[
    1-\rho^{\mathrm{HIGH}}
    =\Theta\!\left(\mathbb E[S\mathbf 1\{R_\rho=1\}]\right).
\]
Substituting into~\eqref{eq:free-bit-low-prob-slack} yields
\[
    \mathbb P(R_\rho=1)G^{-1}(\rho)
    =O\!\left(\mathbb E[S\mathbf 1\{R_\rho=1\}]\right).
\]
Moreover,~\eqref{eq:free-bit-necessary-low-work} ensures that the expected demoted work is positive for all sufficiently large \(\rho\), and hence \(\mathbb P(R_\rho=1)>0\). We may therefore divide by the demotion probability to obtain
\[
    \mathbb E[S\mid R_\rho=1]
    =\frac{\mathbb E[S\mathbf 1\{R_\rho=1\}]}{\mathbb P(R_\rho=1)}
    =\Omega(G^{-1}(\rho)).
\]
Together with~\eqref{eq:free-bit-necessary-low-work}, this proves both requirements in the lemma. The argument explains their joint role: removing enough work protects the high-priority queue, while selecting only a few large jobs limits the low-priority contribution to the overall mean.
\end{proof}

\subsection[Tail-Work Scarcity]{Tail-Work Scarcity (Lemma~\ref{lem:free-bit-tail-work-scarcity})}
\label{app:free-bit-tail-work-scarcity-proof}

Lemma~\ref{lem:free-bit-necessary-structure} gives the work that an order-optimal BLQ policy must demote. We now show that, in the intermediate regime, even all jobs above any fixed positive multiple of \(G^{-1}(\rho)\) cannot supply this amount. We use the tail-work bound from Lemma~\ref{lem:free-bit-tail-work-bound} in Appendix~\ref{app:free-bit-tail-work-proofs}.

\begin{proof}[Proof of Lemma~\ref{lem:free-bit-tail-work-scarcity}]
Fix \(d>0\). At the threshold \(G^{-1}(\rho)\), the total tail work is exactly \((1-\rho)\mathbb E[S]\), a factor of order \(G^{-1}(\rho)\) smaller than the work required by Lemma~\ref{lem:free-bit-necessary-structure}. If \(d\ge1\), raising the threshold can only reduce the available work, so
\[
    \frac{\mathbb E[S\mathbf 1\{S>dG^{-1}(\rho)\}]}
         {(1-\rho)G^{-1}(\rho)}
    \le\frac{\mathbb E[S]}{G^{-1}(\rho)}\longrightarrow0.
\]
The issue is therefore whether lowering the threshold by a fixed factor \(d<1\) could supply enough additional work to close this gap.

Suppose \(0<d<1\). In the intermediate regime, Lemma~\ref{lem:free-bit-log-quadratic} gives \(uh(u)/\log u\to0\). For any \(\varepsilon>0\), we thus have \(h(u)\le\varepsilon\log u/u\) throughout \([dx,x]\) for all sufficiently large \(x\). Integrating gives
\[
    0\le H(x)-H(dx)
    \le\varepsilon\left(
        \log(1/d)\log x-\frac{(\log d)^2}{2}
    \right).
\]
After division by \(\log x\), the upper bound can be made arbitrarily small by choosing \(\varepsilon\). Hence \(H(x)-H(dx)=o(\log x)\), or equivalently,
\begin{equation}
\label{eq:free-bit-local-tail-ratio}
    \frac{\overline F(dx)}{\overline F(x)}
    =e^{H(x)-H(dx)}
    =x^{o(1)}.
\end{equation}
Thus a fixed proportional reduction of the threshold increases the tail probability by less than any fixed positive power of \(x\). It cannot produce the factor of order \(x\) needed to meet the low-priority work requirement. To transfer this probability comparison to work, Lemma~\ref{lem:free-bit-tail-work-bound} gives a constant \(C_d<\infty\) such that, eventually,
\[
\begin{aligned}
    \mathbb E[S\mathbf 1\{S>dG^{-1}(\rho)\}]
    \le C_d G^{-1}(\rho)\overline F(dG^{-1}(\rho))
    &=C_d G^{-1}(\rho)\overline F(G^{-1}(\rho))\bigl(G^{-1}(\rho)\bigr)^{o(1)}\\
    &\le C_d(1-\rho)\mathbb E[S]\bigl(G^{-1}(\rho)\bigr)^{o(1)}.
\end{aligned}
\]
The last inequality uses \(G^{-1}(\rho)\overline F(G^{-1}(\rho))\le\mathbb E[S\mathbf 1\{S>G^{-1}(\rho)\}]=(1-\rho)\mathbb E[S]\). Dividing by the required work scale yields
\[
    \frac{\mathbb E[S\mathbf 1\{S>dG^{-1}(\rho)\}]}
         {(1-\rho)G^{-1}(\rho)}
    \le C_d\mathbb E[S]\bigl(G^{-1}(\rho)\bigr)^{-1+o(1)}
    \longrightarrow0.
\]
This is the desired scarcity bound. Even assigning every job above \(dG^{-1}(\rho)\) to low priority cannot supply the amount of work demanded by an order-optimal BLQ policy.
\end{proof}

\section[Sparse Predict and Demote (Section~\ref{sec:sparse-predict})]{Sparse Predict and Demote (Section~\ref{sec:sparse-predict})}
\label{app:sparse-predict}
Throughout this appendix, Assumption~\ref{asm:heavy-traffic-regularity} holds and the service distribution is in the   lognormal-or-lighter  regime. The exact-prediction arguments fix \(\theta\in(0,\bar\theta)\) and use   \(b_\rho(\theta)=\theta G^{-1}(\rho)\)  and the feasible trigger interval from Section~\ref{sec:sparse-predict-delay}.

\subsection[Abundance of Jobs above Residual Barrier (Lemma~\ref{lem:sparse-candidate-abundance})]{Abundance of Jobs above Residual Barrier (Lemma~\ref{lem:sparse-candidate-abundance})}
\label{app:sparse-candidate-abundance-proof}
\begin{proof}[Proof of Lemma~\ref{lem:sparse-candidate-abundance}]
The fraction of jobs above the residual barrier is \(\overline F(\theta G^{-1}(\rho))\), while the desired demotion fraction is \(1-\rho=\overline F(\bar{s}_\rho)\). Because \(\overline F(u)=e^{-H(u)}\), the logarithm of this ratio is \(H(\bar{s}_\rho)-H(\theta G^{-1}(\rho))\). In words, the hazard accumulated between the cutoffs measures how much the survival probability decreases when the cutoff is raised. If this hazard increment diverges, lowering the cutoff increases the fraction of jobs above residual barrier by an unbounded factor. We show that even a fixed proportional part of this interval contributes a diverging hazard increment.

Choose \(\widetilde\theta\in(\theta,\bar{\theta})\). Since \(\bar{s}_\rho/G^{-1}(\rho)\to\bar{\theta}\), eventually \(\bar{s}_\rho\ge\widetilde\theta G^{-1}(\rho)\).
By Lemma~\ref{lem:free-bit-log-quadratic}, \(uh(u)/\log u\) has a positive or infinite limit. Thus there is a fixed \(c>0\) such that \(h(u)\ge c\log u/u\) for all sufficiently large \(u\). As \(\theta G^{-1}(\rho)\to\infty\), this bound holds throughout the interval below, giving

\[
\begin{aligned}
    \log\frac{\overline F(\theta G^{-1}(\rho))}{1-\rho}
    &=\int_{\theta G^{-1}(\rho)}^{\bar{s}_\rho}h(u)\,du \ge c\int_{\theta G^{-1}(\rho)}^{\widetilde\theta G^{-1}(\rho)}\frac{\log u}{u}\,du \ge c\log(\theta G^{-1}(\rho))
        \log\!\left(\frac{\widetilde\theta}{\theta}\right)
    \longrightarrow\infty.
\end{aligned}
\]

The last inequality uses \(\log u\ge\log(\theta G^{-1}(\rho))\) throughout the integration interval. Since \(\widetilde\theta/\theta>1\) is fixed, the lower bound diverges. Exponentiating proves the lemma. In particular, the \(\theta\)-tuned query probability  \(q_\rho(\theta)=(1-\rho)/\overline F(\theta G^{-1}(\rho))\) , the reciprocal of this ratio, tends to zero.
\end{proof}

\subsection[Response-Time Guarantee (Theorem~\ref{thm:sparse-uniform-optimality})]{Response-Time Guarantee (Theorem~\ref{thm:sparse-uniform-optimality})}
\label{app:sparse-predict-delay}
\label{app:sparse-uniform-optimality-proof}

For a positive trigger, an eventually demoted job's rank rises from zero to one at age \(\tau\). Thus Proposition~\ref{prop:soap-nonincreasing}, which requires ranks to be non-increasing in attained service, does not apply to this policy. We instead use the general SOAP identity in Proposition~\ref{prop:soap-response-time}. For intuition on its worst-future-rank construction, see Appendix~\ref{app:soap-general-intuition}  and \citep[Appendix~D]{scully_soap_2018}.

Recall that   \(R_\rho(\theta,\tau)=Q_\rho(\theta,\tau)\mathbf1\{S>\tau+b_\rho(\theta)\}\)  records eventual demotion. If   \(R_\rho(\theta,\tau)=0\) , the job is never demoted and receives all its service at high priority. If   \(R_\rho(\theta,\tau)=1\) , the job receives its first \(\tau\) units at high priority and is then demoted while still unfinished. The following analogues of Lemmas~\ref{lem:free-bit-high-priority-response} and~\ref{lem:free-bit-low-priority-response} condition on these two eventual outcomes; their derivations are given in Appendix~\ref{app:sparse-delayed-identity-proof}.

\begin{lemma}[Never demoted jobs]
\label{lem:sparse-delayed-high-response}
For   \(\mathrm{SPD}(b_\rho(\theta),q_\rho(\theta,\tau),\tau)\) with \(\tau\in[0,\bar\tau_\rho(\theta)]\) ,
  \[
\begin{aligned}
 &\mathbb E\!\left[
 T^{\mathrm{SPD}(b_\rho(\theta),q_\rho(\theta,\tau),\tau)}
 \mid R_\rho(\theta,\tau)=0\right]\\
 &\quad=\frac{\lambda_\rho\mathbb E\!\left[\left(
    \min\{S,\tau\}+\max\{0,S-\tau\}\mathbf1\{R_\rho(\theta,\tau)=0\}
\right)^2\right]}{2(1-\rho_\rho^{\mathrm{HIGH}}(\theta,\tau))}
 +\mathbb E[S\mid R_\rho(\theta,\tau)=0].
\end{aligned}
\]
\end{lemma}

\begin{lemma}[Eventually demoted jobs]
\label{lem:sparse-delayed-low-response}
For   \(\mathrm{SPD}(b_\rho(\theta),q_\rho(\theta,\tau),\tau)\) with \(\tau\in[0,\bar\tau_\rho(\theta)]\) ,
  \[
\begin{aligned}
 &\mathbb E\!\left[
 T^{\mathrm{SPD}(b_\rho(\theta),q_\rho(\theta,\tau),\tau)}
 \mid R_\rho(\theta,\tau)=1\right]=
 \frac{\lambda_\rho\mathbb E[S^2]}
 {2(1-\rho)(1-\rho_\rho^{\mathrm{HIGH}}(\theta,\tau))}
 +\frac{\mathbb E[S\mid R_\rho(\theta,\tau)=1]}
 {1-\rho_\rho^{\mathrm{HIGH}}(\theta,\tau)}.
\end{aligned}
\]
\end{lemma}
In each lemma, the first summand is a SOAP waiting contribution and the second is a SOAP residence contribution. For a never-demoted tagged job, its worst future rank equals its actual rank zero, so these are its expected physical waiting and service times. For an eventually demoted tagged job, SOAP instead assigns rank one throughout. This preserves response time, but the separate contributions need not equal its original physical waiting and residence times.

\begin{proof}[Proof of Theorem~\ref{thm:sparse-uniform-optimality}]
Lemma~\ref{lem:sparse-delayed-query-vanishing} establishes that the ex-ante query probability vanishes uniformly. It remains to prove the response-time bound. We combine the two conditional formulas in Lemmas~\ref{lem:sparse-delayed-high-response} and \ref{lem:sparse-delayed-low-response}, weighting them by the probabilities of the eventual outcomes   \(\mathbb P(R_\rho(\theta,\tau)=0)=\rho\) and \(\mathbb P(R_\rho(\theta,\tau)=1)=1-\rho\). Lemma~\ref{lem:sparse-delayed-high-slack} gives these probabilities directly. Lemma~\ref{lem:service-finite-moments} in Appendix~\ref{app:service-finite-moments} ensures that \(\mathbb E[S]\) and \(\mathbb E[S^2]\) are finite constants, with \(\mathbb E[S]>0\); hence \(\lambda_\rho=\rho/\mathbb E[S]\to1/\mathbb E[S]\).
We first bound the entire contribution of jobs that are never demoted. Their probability is \(\rho\le1\), and the high-priority service generated by any competing arrival is at most \(S\). Applying Lemma~\ref{lem:sparse-delayed-high-response} therefore gives
  \[
\begin{aligned}
\mathbb E\!\left[
 T^{\mathrm{SPD}(b_\rho(\theta),q_\rho(\theta,\tau),\tau)}
 \mathbf1\{R_\rho(\theta,\tau)=0\}\right]
&\le
 \frac{\lambda_\rho\mathbb E[S^2]}
 {2(1-\rho_\rho^{\mathrm{HIGH}}(\theta,\tau))}
 +\mathbb E[S\mathbf1\{R_\rho(\theta,\tau)=0\}]\\
&=O\!\left(\frac{1}{(1-\rho)G^{-1}(\rho)}\right)+O(1).
\end{aligned}
\]
Here Lemma~\ref{lem:sparse-delayed-high-slack} bounds the waiting term uniformly in \(\tau\), and the residence term is at most \(\mathbb E[S]\).

Next consider eventually demoted jobs. Multiplying Lemma~\ref{lem:sparse-delayed-low-response} by the demotion probability \(1-\rho\) cancels the extra factor \(1-\rho\) in the waiting denominator and gives
  \[
\begin{aligned}
\mathbb E\!\left[
 T^{\mathrm{SPD}(b_\rho(\theta),q_\rho(\theta,\tau),\tau)}
 \mathbf1\{R_\rho(\theta,\tau)=1\}\right]=
 \frac{\lambda_\rho\mathbb E[S^2]}
 {2(1-\rho_\rho^{\mathrm{HIGH}}(\theta,\tau))}
 +\frac{\mathbb E[S\mathbf1\{R_\rho(\theta,\tau)=1\}]}
 {1-\rho_\rho^{\mathrm{HIGH}}(\theta,\tau)}.
\end{aligned}
\]
 
The first term has the same uniform bound as above. Although the conditional mean   \(\mathbb E[S\mid R_\rho(\theta,\tau)=1]\)  may grow, weighting by the demotion probability gives   \(\mathbb E[S\mathbf1\{R_\rho(\theta,\tau)=1\}]\le\mathbb E[S]\). As a result, the residence term is also bounded uniformly over feasible triggers by
  \[
 \frac{\mathbb E[S\mathbf1\{R_\rho(\theta,\tau)=1\}]}{1-\rho_\rho^{\mathrm{HIGH}}(\theta,\tau)}
 \le\frac{\mathbb E[S]}{1-\rho_\rho^{\mathrm{HIGH}}(\theta,\tau)}
 =O\!\left(\frac{1}{(1-\rho)G^{-1}(\rho)}\right),
\]
The remaining \(O(1)\) term is negligible on this scale: by finite second moment,
\[
 (1-\rho)G^{-1}(\rho)
 =\frac{G^{-1}(\rho)\mathbb E[S\mathbf1\{S>G^{-1}(\rho)\}]}{\mathbb E[S]}
 \le\frac{\mathbb E[S^2\mathbf1\{S>G^{-1}(\rho)\}]}{\mathbb E[S]}
 \longrightarrow0.
\]
Summing the weighted waiting and residence contributions gives
  \[
    \mathbb E\!\left[T^{\mathrm{SPD}(b_\rho(\theta),q_\rho(\theta,\tau),\tau)}\right]
    =O\!\left(\frac{1}{(1-\rho)G^{-1}(\rho)}\right)
\]
uniformly over the feasible trigger interval. SRPT optimality and the benchmark~\eqref{eq:srpt-heavy-traffic-benchmark} give the matching lower bound for every trigger. This proves uniform heavy-traffic order-optimality and completes the theorem.
\end{proof}

\subsection[Proofs of Auxiliary Lemmas]{Proofs of Auxiliary Lemmas (Lemmas~\ref{lem:sparse-delayed-high-response} and~\ref{lem:sparse-delayed-low-response})}
\label{app:sparse-delayed-identity-proof}

We now prove Lemmas~\ref{lem:sparse-delayed-high-response} and~\ref{lem:sparse-delayed-low-response}. As explained above, the rank increase at a positive trigger requires Proposition~\ref{prop:soap-response-time}; the following rank representation supplies its inputs. Fix \(\rho\) and a feasible trigger \(\tau\). Recall that \(Q_\rho(\theta,\tau)\)  is the Bernoulli selection coin used by the policy at the trigger. The policy queries an unfinished job at the trigger with the \((\theta,\tau)\)-tuned query probability   \(q_\rho(\theta,\tau)\) without using job size, so this coin has that Bernoulli law independently of \(S\). For analysis, we may generate it at arrival and use it only if the job reaches the trigger; this gives exactly the same policy. Recall that   \(R_\rho(\theta,\tau)=Q_\rho(\theta,\tau)\mathbf1\{S>\tau+b_\rho(\theta)\}\)  is the eventual demotion indicator. Using the job's size and eventual demotion indicator as its SOAP descriptor, its rank at attained service age \(a\) is
  \[
    r((S,R_\rho(\theta,\tau)),a)
    =R_\rho(\theta,\tau)\mathbf1\{a\ge\tau\},
    \qquad 0\le a<S.
\]
 
The descriptor encodes the eventual demotion decision for analysis. Before the trigger, its rank is zero regardless of the hidden size or sampling outcome; the scheduler still queries only selected jobs that remain unfinished at that age.

The FCFS rule uses original arrival time in both queues. Older jobs win equal-rank ties, whereas future arrivals lose them. Thus old-job relevant service includes ranks at most the tagged rank, and new-job relevant service includes only ranks strictly smaller. A competitor's rank can only increase, so it cannot reenter relevance after leaving: \(X_i^{\mathrm{old}}[r]=0\) for \(i\ge1\) and \(r\in\{0,1\}\). The service variables below describe the relevant service generated by a generic arrival, not its remaining service at an observation time. The corresponding load is the arrival rate \(\lambda_\rho\) multiplied by the mean of that relevant-service variable.

\begin{proof}[Proof of Lemma~\ref{lem:sparse-delayed-high-response}]
A tagged job that is never demoted has rank zero until completion, so its worst future rank is also zero. An older job delays it only during that older job's high-priority service: service until completion or age \(\tau\), whichever comes first, together with any remaining service if the older job is never demoted. Therefore 
  \[
\begin{aligned}
    X_0^{\mathrm{old}}[0]
    &=\min\{S,\tau\}+\max\{0,S-\tau\}\mathbf1\{R_\rho(\theta,\tau)=0\},\\
    \rho_0^{\mathrm{old}}[0]&=\lambda_\rho\mathbb E[X_0^{\mathrm{old}}[0]]=\rho_\rho^{\mathrm{HIGH}}(\theta,\tau).
\end{aligned}
\]
The last equality follows from the definition of \(\rho_\rho^{\mathrm{HIGH}}(\theta,\tau)\) in Section~\ref{sec:sparse-predict-delay}: \(X_0^{\mathrm{old}}[0]\) is exactly the high-priority service generated by a generic arrival.

No future arrival can overtake the tagged job: even another rank-zero job loses the FCFS tie. Hence \(X^{\mathrm{new}}[0]=0\) and \(\rho^{\mathrm{new}}[0]=\lambda_\rho\mathbb E[X^{\mathrm{new}}[0]]=0\). The tagged job's residence contribution is consequently its size. In Proposition~\ref{prop:soap-response-time}, the recycled relevant-work variables are zero, so the waiting numerator is \(\lambda_\rho\mathbb E[(X_0^{\mathrm{old}}[0])^2]\). The two slack factors are \(1-\rho_\rho^{\mathrm{HIGH}}(\theta,\tau)\) and \(1-\rho^{\mathrm{new}}[0]=1\). Thus a tagged job of size \(x\) has the waiting term in the lemma plus residence \(x\). Taking expectation with respect to the conditional distribution of the tagged size given   \(R_\rho(\theta,\tau)=0\)  replaces \(x\) by   \(\mathbb E[S\mid R_\rho(\theta,\tau)=0]\); the waiting term is unchanged. This proves the stated conditional response-time formula. Jobs finishing before the trigger are included automatically: they have   \(R_\rho(\theta,\tau)=0\)  and receive their entire service at high priority.
\end{proof}

\begin{proof}[Proof of Lemma~\ref{lem:sparse-delayed-low-response}]
An eventually demoted tagged job has worst future rank one at every age, including during its initial high-priority service. SOAP substitutes this worst future rank for the tagged job's actual rank while preserving its response time. Competitors retain their original rank functions of attained service; see Appendix~\ref{app:soap-general-intuition}  and \citep[Appendix~D]{scully_soap_2018}.

An older competing job has rank zero or one throughout its service and wins ties at rank one. Hence all of its \(S\) units of service are relevant to the rank-one tag, giving \(X_0^{\mathrm{old}}[1]=S\) and \(\rho_0^{\mathrm{old}}[1]=\lambda_\rho\mathbb E[X_0^{\mathrm{old}}[1]]=\lambda_\rho\mathbb E[S]=\rho\). A future arrival loses ties at rank one, so only its rank-zero service is relevant to this tag. This is the same high-priority work as \(X_0^{\mathrm{old}}[0]\), giving 
  \[
\begin{aligned}
    X^{\mathrm{new}}[1]
    &=X_0^{\mathrm{old}}[0]=\min\{S,\tau\}+\max\{0,S-\tau\}\mathbf1\{R_\rho(\theta,\tau)=0\},\\
    \rho^{\mathrm{new}}[1]&=\lambda_\rho\mathbb E[X^{\mathrm{new}}[1]]=\rho_\rho^{\mathrm{HIGH}}(\theta,\tau).
\end{aligned}
\]
 
Since \(X_0^{\mathrm{old}}[1]=S\) and no recycled work contributes, the waiting numerator in Proposition~\ref{prop:soap-response-time} is \(\lambda_\rho\mathbb E[S^2]\). These inputs give the two slack factors in the waiting term of Proposition~\ref{prop:soap-response-time}: \(1-\rho\) for all older work and   \(1-\rho_\rho^{\mathrm{HIGH}}(\theta,\tau)\)  for high-priority work from future arrivals. The same new-arrival load \(\rho^{\mathrm{new}}[1]=\rho_\rho^{\mathrm{HIGH}}(\theta,\tau)\) also enters the residence denominator. Since the substituted rank is constant, the residence integral for a tagged job of size \(x\) is
  \[
    \int_0^x\frac{da}{1-\rho_\rho^{\mathrm{HIGH}}(\theta,\tau)}
    =\frac{x}{1-\rho_\rho^{\mathrm{HIGH}}(\theta,\tau)}.
\]
Taking expectation with respect to the conditional distribution of the tagged size given   \(R_\rho(\theta,\tau)=1\)  replaces \(x\) in the residence term by   \(\mathbb E[S\mid R_\rho(\theta,\tau)=1]\) ; the waiting term is unchanged. This proves the stated conditional response-time formula. The second moment \(\mathbb E[S^2]\) remains unconditional because it concerns competing arrivals, whose sizes are independent of the tagged job's eventual demotion outcome.
\end{proof}

\section{Proofs for the Costly-Prediction Extension (Section~\ref{sec:sparse-costly-exact})}
\label{app:sparse-costly}

Throughout this appendix, Assumption~\ref{asm:heavy-traffic-regularity} holds. Appendix~\ref{app:sparse-costly-extension-proof} proves the free-to-costly comparison in Lemma~\ref{lem:sparse-cost-transfer} by combining response-time identities with bounds on inference work and slack. Appendix~\ref{app:sparse-cost-helper-proofs} proves the supporting lemmas. The response-time identities allow general prediction rules applied independently across jobs.

The remaining subsections study how trigger choice affects query use and inference load under exact predictions. Appendix~\ref{app:sparse-advice-trigger} establishes cutoff estimates and feasibility of proportional triggers. Appendices~\ref{app:sparse-cost-polylog}--\ref{app:sparse-cost-regular} use these tools to give detailed proofs of the results in Proposition~\ref{prop:sparse-cost-distributions} for polylogarithmic, rapidly varying, and regularly varying cumulative hazards, respectively. Appendices~\ref{app:sparse-advice-trigger}--\ref{app:sparse-cost-regular} additionally assume that the service distribution is in the lognormal-or-lighter regime.

\subsection[Reduction from Free to Costly Predictions (Lemma~\ref{lem:sparse-cost-transfer})]{Reduction from Free to Costly Predictions (Lemma~\ref{lem:sparse-cost-transfer})}
\label{app:sparse-costly-extension-proof}
Fix \(\rho<1\), a deterministic trigger \(\tau\ge0\), and an inference duration \(\delta\ge0\), and suppress load subscripts. Let \(\pi\) denote the policy with these fixed parameters and FCFS in both queues. Recall that \(Q^{\mathrm{qry}}\) is the actual-query indicator from Section~\ref{sec:sparse-predict-delay} and \(R\) for the eventual-demotion indicator, with \(R\le Q^{\mathrm{qry}}\le\mathbf1\{S>\tau\}\), so every demoted job has been queried. The triples \((S,Q^{\mathrm{qry}},R)\) are i.i.d. across jobs and independent of Poisson arrivals of rate \(\lambda\). Their components within a job may be dependent. Recall that inference work is done while the job is still in the high-priority queue and consumes server capacity. During this time, ordinary-service age is paused, and both queues break ties by original arrival time.

Total work per arrival is \(S+\delta Q^{\mathrm{qry}}\), and the work sent to low priority is \((S-\tau)R\). We define 
\[
 \rho(\delta):=\lambda\mathbb E[S+\delta Q^{\mathrm{qry}}],
 \qquad
 \rho^{\mathrm{HIGH}}(\delta)
 :=\lambda\mathbb E[S+\delta Q^{\mathrm{qry}}-(S-\tau)R].
\]
as the corresponding loads (including inference) for the total system and the high-priority queue respectively.
The free-policy loads are \(\rho(0)=\rho\) and \(\rho^{\mathrm{HIGH}}(0)\).

We separate the response-time identity into two lemmas, one for each eventual demotion outcome, and then state bounds on inference duration, slack, and second moments. These four lemmas reduce the proof of Lemma~\ref{lem:sparse-cost-transfer} to a comparison of the SOAP terms. Their proofs follow in Appendix~\ref{app:sparse-cost-helper-proofs}.
\begin{lemma}[Response time of never-demoted jobs with inference]
\label{lem:sparse-cost-high-response}
If \(\rho(\delta)<1\), then
\[
 \mathbb E[T^{\pi}(\delta)(1-R)]
 =\frac{\mathbb P(R=0)\lambda\mathbb E[(S+\delta Q^{\mathrm{qry}}-(S-\tau)R)^2]}
 {2(1-\rho^{\mathrm{HIGH}}(\delta))}
 +\mathbb E[(S+\delta Q^{\mathrm{qry}})(1-R)].
\]
\end{lemma}

\begin{lemma}[Response time of eventually demoted jobs with inference]
\label{lem:sparse-cost-low-response}
If \(\rho(\delta)<1\), then
\[
 \mathbb E[T^{\pi}(\delta)R]
 =\frac{\mathbb P(R=1)\lambda\mathbb E[(S+\delta Q^{\mathrm{qry}})^2]}
 {2(1-\rho(\delta))(1-\rho^{\mathrm{HIGH}}(\delta))}
 +\frac{\mathbb E[(S+\delta Q^{\mathrm{qry}})R]}{1-\rho^{\mathrm{HIGH}}(\delta)}.
\]
\end{lemma}

For never-demoted jobs, the summands are probability-weighted mean physical waiting and residence times. For demoted jobs, they are transformed SOAP contributions whose sum gives the group's contribution to mean response time. Setting \(\delta=0\) gives the free-policy formulas in Appendix~\ref{app:sparse-delayed-identity-proof}. The weighted forms also cover an outcome with probability zero.

Lemma~\ref{lem:sparse-cost-slack-bounds} controls inference duration and the slack denominators in the response-time expressions. Lemma~\ref{lem:sparse-cost-moment-bounds} controls their second-moment numerators. Together they let us compare costly and free policies.
\begin{samepage}
\begin{lemma}[Inference duration and slack preservation]
\label{lem:sparse-cost-slack-bounds}
Suppose \(\zeta\in(0,1)\), \(\mathbb E[Q^{\mathrm{qry}}]\ge1-\rho\), and
$\lambda\delta\mathbb E[Q^{\mathrm{qry}}]\le(1-\zeta)(1-\rho)$.
Then the costly policy is stable, \(\delta\le(1-\zeta)/\lambda\), and
\[
 1-\rho(\delta)\ge\zeta(1-\rho),\quad
 1-\rho^{\mathrm{HIGH}}(\delta)\ge\zeta(1-\rho^{\mathrm{HIGH}}(0)),\quad
 1-\rho^{\mathrm{HIGH}}(0)\ge1-\rho.
\]
\end{lemma}
\end{samepage}
\begin{lemma}[Second moments after adding inference work]
\label{lem:sparse-cost-moment-bounds}
Suppose \(\rho\in[1/2,1)\), \(\zeta\in(0,1)\), \(\mathbb E[Q^{\mathrm{qry}}]\ge1-\rho\), and \(\lambda\delta\mathbb E[Q^{\mathrm{qry}}]\le(1-\zeta)(1-\rho)\). Then
\[
 \delta^2\mathbb E[Q^{\mathrm{qry}}]
 \le\frac{(1-\zeta)^2(1-\rho)}{\lambda^2}
 \le2(1-\zeta)^2\mathbb E[S^2],
\]
and
\[
\begin{aligned}
 \mathbb E[(S+\delta Q^{\mathrm{qry}}-(S-\tau)R)^2]
 &\le2\mathbb E[(S-(S-\tau)R)^2]+2\delta^2\mathbb E[Q^{\mathrm{qry}}],\\
 \mathbb E[(S+\delta Q^{\mathrm{qry}})^2]
 &\le2\bigl(1+2(1-\zeta)^2\bigr)\mathbb E[S^2].
\end{aligned}
\]
\end{lemma}

For high-priority work, we keep the added second-moment term separate; this avoids requiring a lower bound on \(\mathbb E[(S-(S-\tau)R)^2]\).

\begin{proof}[Proof of Lemma~\ref{lem:sparse-cost-transfer}] Fix \(\rho\in[1/2,1)\) and a SPD policy \(\pi\) satisfying the slack-preservation condition~\eqref{eq:sparse-costly-slack-condition}. Compare its free and costly versions with the same policy parameters and prediction law, so \((S,Q^{\mathrm{qry}},R)\) has the same distribution in both.

\vspace{.1in}
\noindent\textbf{Query Probability and Slack Preservation.} By the assumptions of the lemma, the ex-ante query probability satisfies \(\mathbb E[Q^{\mathrm{qry}}]\ge1-\rho>0\). The SPD constructions in Section~\ref{sec:sparse-costly-exact} (exact predictions) and Section~\ref{sec:sparse-predict-errors} (inexact predictions) use the same \((\theta,\tau)\)-tuned query probability \(q_\rho(\theta,\tau)\). At a feasible pair \((\theta,\tau)\), both satisfy the required lower bound:
\begin{equation}
\label{eq:sparse-cost-query-lower-bound}
    \mathbb E[Q^{\mathrm{qry}}]
    =q_\rho(\theta,\tau)\overline F(\tau)
    =(1-\rho)\frac{\overline F(\tau)}
    {\overline F(\tau+b_\rho(\theta))}
    \ge 1-\rho>0.
\end{equation}
This lower bound and the slack-preservation condition~\eqref{eq:sparse-costly-slack-condition} imply that we satisfy the conditions of Lemma~\ref{lem:sparse-cost-slack-bounds}. As a result, applying this lemma, we obtain that $\rho(\delta)\leq 1-\zeta(1-\rho)<1$. This gives stability which, in turn allows us to apply Lemmas~\ref{lem:sparse-cost-high-response} and~\ref{lem:sparse-cost-low-response}.

\vspace{0.1in}\noindent\textbf{Comparison of the Four SOAP Terms.}
We apply Lemmas~\ref{lem:sparse-cost-high-response} and~\ref{lem:sparse-cost-low-response} at \(\delta\) and at \(0\). The common law of \((S,Q^{\mathrm{qry}},R)\) keeps each group's probability weight unchanged when inference is added. We compare waiting and residence for each group.

\paragraph{Never-demoted waiting.}
We start by focusing on the first term in Lemma~\ref{lem:sparse-cost-high-response}, the probability-weighted waiting contribution of never-demoted jobs.
Since \(\rho\ge1/2\), the high-priority second-moment bound in Lemma~\ref{lem:sparse-cost-moment-bounds} applies to the numerator. For the denominator, we use \(1-\rho^{\mathrm{HIGH}}(\delta)\ge\zeta(1-\rho^{\mathrm{HIGH}}(0))\) from Lemma~\ref{lem:sparse-cost-slack-bounds}. Substituting both bounds gives
\[
\begin{aligned}
 \frac{\mathbb P(R=0)\lambda
   \mathbb E[(S+\delta Q^{\mathrm{qry}}-(S-\tau)R)^2]}
 {2(1-\rho^{\mathrm{HIGH}}(\delta))}
 &\quad\le
 \frac{\mathbb P(R=0)\lambda\bigl(2\mathbb E[(S-(S-\tau)R)^2]
       +2\delta^2\mathbb E[Q^{\mathrm{qry}}]\bigr)}
 {2\zeta(1-\rho^{\mathrm{HIGH}}(0))}\\
 &\quad\le
 \frac2\zeta\,
 \frac{\mathbb P(R=0)\lambda\mathbb E[(S-(S-\tau)R)^2]}
 {2(1-\rho^{\mathrm{HIGH}}(0))}
 +\frac{\lambda\delta^2\mathbb E[Q^{\mathrm{qry}}]}
 {\zeta(1-\rho^{\mathrm{HIGH}}(0))}.
\end{aligned}
\]

The second inequality uses \(\mathbb P(R=0)\le1\) only in the added term.

\paragraph{Never-demoted residence.} 
We then move to the second term in Lemma~\ref{lem:sparse-cost-high-response}, \(\mathbb E[(S+\delta Q^{\mathrm{qry}})(1-R)]\), the residence contribution of never-demoted jobs.
Every demoted job is queried, so \(Q^{\mathrm{qry}}R=R\) and \(\mathbb E[Q^{\mathrm{qry}}(1-R)]=\mathbb E[Q^{\mathrm{qry}}]-\mathbb P(R=1)\). Expanding the inference work and discarding the nonpositive term \(-\delta\mathbb P(R=1)\) gives
\[
\begin{aligned}
 \mathbb E[(S+\delta Q^{\mathrm{qry}})(1-R)]
 =\mathbb E[S(1-R)]
   +\delta\bigl(\mathbb E[Q^{\mathrm{qry}}]-\mathbb P(R=1)\bigr)
 \le\mathbb E[S(1-R)]+\delta\mathbb E[Q^{\mathrm{qry}}].
\end{aligned}
\]

\begin{samepage}

\paragraph{Eventually demoted waiting.} 
Next, we focus on the first term in Lemma~\ref{lem:sparse-cost-low-response}, the waiting contribution for eventually demoted jobs in the transformed SOAP calculation.
We apply the total-work second-moment bound in Lemma~\ref{lem:sparse-cost-moment-bounds} to the numerator. For the denominator, Lemma~\ref{lem:sparse-cost-slack-bounds} gives \(1-\rho(\delta)\ge\zeta(1-\rho(0))\) and \(1-\rho^{\mathrm{HIGH}}(\delta)\ge\zeta(1-\rho^{\mathrm{HIGH}}(0))\). Multiplying these positive lower bounds produces the factor \(\zeta^2\):
\nopagebreak[4]
\[
\begin{aligned}
 \frac{\mathbb P(R=1)\lambda\mathbb E[(S+\delta Q^{\mathrm{qry}})^2]}
 {2(1-\rho(\delta))(1-\rho^{\mathrm{HIGH}}(\delta))}
 &\quad\le
 \frac{\mathbb P(R=1)\lambda\,2\bigl(1+2(1-\zeta)^2\bigr)\mathbb E[S^2]}
 {2\zeta^2(1-\rho(0))(1-\rho^{\mathrm{HIGH}}(0))}\\
 &\quad=
 \frac{2\bigl(1+2(1-\zeta)^2\bigr)}{\zeta^2}\,
 \frac{\mathbb P(R=1)\lambda\mathbb E[S^2]}
 {2(1-\rho(0))(1-\rho^{\mathrm{HIGH}}(0))}.
\end{aligned}
\]

\end{samepage}

\paragraph{Eventually demoted residence.}
Finally, we consider the second term in Lemma~\ref{lem:sparse-cost-low-response}, the corresponding transformed residence contribution.
As before, every demoted job was queried. If \(R=1\), then \(Q^{\mathrm{qry}}=1\); if \(R=0\), both \(Q^{\mathrm{qry}}R\) and \(R\) are zero. Hence \(Q^{\mathrm{qry}}R=R\), which gives the numerator \(\mathbb E[SR]+\delta\mathbb P(R=1)\). Apply the high-priority slack bound \(1-\rho^{\mathrm{HIGH}}(\delta)\ge\zeta(1-\rho^{\mathrm{HIGH}}(0))\) from Lemma~\ref{lem:sparse-cost-slack-bounds}, then split this numerator:
\nopagebreak[4]
\[
\begin{aligned}
 \frac{\mathbb E[(S+\delta Q^{\mathrm{qry}})R]}
 {1-\rho^{\mathrm{HIGH}}(\delta)}
 =\frac{\mathbb E[SR]+\delta\mathbb P(R=1)}
 {1-\rho^{\mathrm{HIGH}}(\delta)}\le\frac1\zeta\,
 \frac{\mathbb E[SR]}{1-\rho^{\mathrm{HIGH}}(0)}
 +\frac{\delta\mathbb P(R=1)}{\zeta(1-\rho^{\mathrm{HIGH}}(0))}.
\end{aligned}
\]
The two contributions for eventually demoted jobs have the transformed SOAP interpretation given above. No division by \(\mathbb P(R=1)\) or \(\mathbb P(R=0)\) is used, so the bounds also cover either group having probability zero.

\paragraph{Combining the bounds.}
We first add the four preceding bounds, keeping their separate coefficients and the three added inference terms: 
\[
\begin{aligned}
 \mathbb E[T^{\pi}(\delta)]
 \le{}&\frac2\zeta\,
 \frac{\mathbb P(R=0)\lambda\mathbb E[(S-(S-\tau)R)^2]}
      {2(1-\rho^{\mathrm{HIGH}}(0))}
 +\mathbb E[S(1-R)]\\
 &+\frac{2\bigl(1+2(1-\zeta)^2\bigr)}{\zeta^2}\,
 \frac{\mathbb P(R=1)\lambda\mathbb E[S^2]}
      {2(1-\rho(0))(1-\rho^{\mathrm{HIGH}}(0))}
 +\frac1\zeta\,\frac{\mathbb E[SR]}{1-\rho^{\mathrm{HIGH}}(0)}\\
 &+\frac{\lambda\delta^2\mathbb E[Q^{\mathrm{qry}}]}
        {\zeta(1-\rho^{\mathrm{HIGH}}(0))}
 +\delta\mathbb E[Q^{\mathrm{qry}}]
 +\frac{\delta\mathbb P(R=1)}{\zeta(1-\rho^{\mathrm{HIGH}}(0))}.
\end{aligned}
\]

The unscaled terms in the first two lines are the free-policy contributions in Lemmas~\ref{lem:sparse-cost-high-response} and~\ref{lem:sparse-cost-low-response}. Setting \(\delta=0\) in those lemmas gives
\[
\begin{aligned}
 \mathbb E[T^{\pi}(0)(1-R)]
 &=\frac{\mathbb P(R=0)\lambda\mathbb E[(S-(S-\tau)R)^2]}
        {2(1-\rho^{\mathrm{HIGH}}(0))}
   +\mathbb E[S(1-R)],\\
 \mathbb E[T^{\pi}(0)R]
 &=\frac{\mathbb P(R=1)\lambda\mathbb E[S^2]}
        {2(1-\rho(0))(1-\rho^{\mathrm{HIGH}}(0))}
   +\frac{\mathbb E[SR]}{1-\rho^{\mathrm{HIGH}}(0)}.
\end{aligned}
\]
These four nonnegative terms therefore sum to \(\mathbb E[T^{\pi}(0)(1-R)]+\mathbb E[T^{\pi}(0)R]=\mathbb E[T^{\pi}(0)]\). Since \(0<\zeta<1\), their coefficients \(2/\zeta\), \(1\), \(2(1+2(1-\zeta)^2)/\zeta^2\), and \(1/\zeta\) are all at most \(2(1+2(1-\zeta)^2)/\zeta^2\). Replacing each coefficient by this common upper bound yields

\[
\begin{aligned}
 \mathbb E[T^{\pi}(\delta)]
 \le{}&\frac{2\bigl(1+2(1-\zeta)^2\bigr)}{\zeta^2}\,
        \mathbb E[T^{\pi}(0)]+\frac{\lambda\delta^2\mathbb E[Q^{\mathrm{qry}}]}
 {\zeta(1-\rho^{\mathrm{HIGH}}(0))}
 +\delta\mathbb E[Q^{\mathrm{qry}}]
 +\frac{\delta\mathbb P(R=1)}{\zeta(1-\rho^{\mathrm{HIGH}}(0))}\\
 \le{}&\frac{2\bigl(1+2(1-\zeta)^2\bigr)}{\zeta^2}\,
        \mathbb E[T^{\pi}(0)]
 +\frac{(1-\zeta)^2}{\zeta\lambda}
 +\frac{1-\zeta}{\lambda}
 +\frac{1-\zeta}{\zeta\lambda}\\
 ={}&\frac{2\bigl(1+2(1-\zeta)^2\bigr)}{\zeta^2}\,
        \mathbb E[T^{\pi}(0)]
 +\frac{2(1-\zeta)}{\zeta\lambda}\\
 \le{}&\left[\frac{2\bigl(1+2(1-\zeta)^2\bigr)}{\zeta^2}
       +\frac{4(1-\zeta)}{\zeta}\right]\mathbb E[T^{\pi}(0)]
 =\frac{6-4\zeta}{\zeta^2}\mathbb E[T^{\pi}(0)]
 \le\frac6{\zeta^2}\mathbb E[T^{\pi}(0)].
\end{aligned}
\]
For the second inequality, we bound the three added terms in their displayed order. The first uses \(\delta^2\mathbb E[Q^{\mathrm{qry}}]\le(1-\zeta)^2(1-\rho)/\lambda^2\) from Lemma~\ref{lem:sparse-cost-moment-bounds} and \(1-\rho^{\mathrm{HIGH}}(0)\ge1-\rho\) from Lemma~\ref{lem:sparse-cost-slack-bounds}. The second uses the slack-preservation condition~\eqref{eq:sparse-costly-slack-condition} and \(1-\rho\le1\). For the third, we use \(R\le Q^{\mathrm{qry}}\) to obtain \(\mathbb P(R=1)\le\mathbb E[Q^{\mathrm{qry}}]\), we then apply the same budget and high-priority slack bound. To absorb the remaining \(1/\lambda\) term, every job requires at least its own ordinary service, so \(T^{\pi}(0)\ge S\), and \(\rho\ge1/2\) gives
\[
 \frac1\lambda=\frac{\mathbb E[S]}{\rho}
 \le2\mathbb E[S]\le2\mathbb E[T^{\pi}(0)].
\]
The ex-ante query probability is unchanged because the sampling decision and the condition \(S>\tau\) are the same in both systems. The comparison transfers the free-policy upper bound, and SRPT optimality supplies the matching lower bound for the heavy-traffic conclusion.
\end{proof}

\subsection[Proofs of Auxiliary Lemmas]{Proofs of Auxiliary Lemmas (Lemmas~\ref{lem:sparse-cost-high-response}--\ref{lem:sparse-cost-moment-bounds})}
\label{app:sparse-cost-helper-proofs}

\begin{proof}[Proof of Lemma~\ref{lem:sparse-cost-high-response}]
SOAP age includes inference work, and all recycled relevant-work terms vanish because ranks never decrease. As in the proof of Lemma~\ref{lem:sparse-delayed-high-response}, a never-demoted tagged job has worst future rank zero. Older arrivals contribute high-priority work, now including inference, while future arrivals lose the FCFS tie. Thus the SOAP inputs are\footnote{Here \(\rho(0)\) is the total load when \(\delta=0\), whereas \(\rho_0^{\mathrm{old}}[0]=\lambda\mathbb E[X_0^{\mathrm{old}}[0]]\) is the SOAP load from original, non-recycled old work relevant to rank zero. The subscript and bracketed argument in the latter notation do not specify the inference cost.}
\[
\begin{aligned}
 X_0^{\mathrm{old}}[0]&=S+\delta Q^{\mathrm{qry}}-(S-\tau)R,
 &\rho_0^{\mathrm{old}}[0]&=\rho^{\mathrm{HIGH}}(\delta),\\
 X^{\mathrm{new}}[0]&=0,
 &\rho^{\mathrm{new}}[0]&=0.
\end{aligned}
\]
The tagged job's own work is \(S+\delta Q^{\mathrm{qry}}\). Substituting these inputs into Proposition~\ref{prop:soap-response-time}, averaging over the tagged-job distribution conditional on \(R=0\), and multiplying by \(\mathbb P(R=0)\) proves Lemma~\ref{lem:sparse-cost-high-response}.
\end{proof}

\begin{proof}[Proof of Lemma~\ref{lem:sparse-cost-low-response}]
As in the proof of Lemma~\ref{lem:sparse-delayed-low-response}, an eventually demoted tagged job has worst future rank one throughout. Older arrivals contribute total work, while future arrivals contribute only high-priority work. Including inference gives
\[
\begin{aligned}
 X_0^{\mathrm{old}}[1]&=S+\delta Q^{\mathrm{qry}},
 &\rho_0^{\mathrm{old}}[1]&=\rho(\delta),\\
 X^{\mathrm{new}}[1]&=S+\delta Q^{\mathrm{qry}}-(S-\tau)R,
 &\rho^{\mathrm{new}}[1]&=\rho^{\mathrm{HIGH}}(\delta).
\end{aligned}
\]
All older work remains relevant to a rank-one tagged job until completion, so \(X_i^{\mathrm{old}}[1]=0\) for every \(i\ge1\). These loads give the two waiting denominators. Since the tagged job's worst future rank is constant, its residence contribution is \((S+\delta Q^{\mathrm{qry}})/(1-\rho^{\mathrm{HIGH}}(\delta))\). Applying Proposition~\ref{prop:soap-response-time}, average conditional on \(R=1\) and multiplying by \(\mathbb P(R=1)\), weobtain the identity in Lemma~\ref{lem:sparse-cost-low-response}.
\end{proof}

\begin{proof}[Proof of Lemma~\ref{lem:sparse-cost-slack-bounds}]
By the assumptions of the lemma, given that \(\mathbb E[Q^{\mathrm{qry}}]\ge1-\rho>0\), it holds that: 
\[
 \delta
 \le\frac{(1-\zeta)(1-\rho)}
 {\lambda\mathbb E[Q^{\mathrm{qry}}]}
 \le\frac{1-\zeta}{\lambda}.
\]

For the total load, the same budget gives

\[
 1-\rho(\delta)
 =1-\rho-\lambda\delta\mathbb E[Q^{\mathrm{qry}}]
 \ge\zeta(1-\rho)>0,
\]
so the costly system is stable. Setting \(\delta=0\) in the high-priority load defined above yields

\[
 1-\rho^{\mathrm{HIGH}}(0)
 =1-\rho+\lambda\mathbb E[(S-\tau)R]
 \ge1-\rho,
\]
because demotion requires \(S>\tau\). Consequently,
\[
\begin{aligned}
 1-\rho^{\mathrm{HIGH}}(\delta)
 =1-\rho^{\mathrm{HIGH}}(0)
   -\lambda\delta\mathbb E[Q^{\mathrm{qry}}]
 \ge1-\rho^{\mathrm{HIGH}}(0)-(1-\zeta)(1-\rho)\ge\zeta(1-\rho^{\mathrm{HIGH}}(0)).
\end{aligned}
\]
\end{proof}

\begin{proof}[Proof of Lemma~\ref{lem:sparse-cost-moment-bounds}] By the assumptions of the lemma, it holds that
\[
 \delta^2\mathbb E[Q^{\mathrm{qry}}]= \frac{\delta^2\mathbb E[Q^{\mathrm{qry}}]^2}{\mathbb E[Q^{\mathrm{qry}}]}
 \le\frac{(1-\zeta)^2(1-\rho)^2}{\lambda^2 (1-\rho)}
 =\frac{(1-\zeta)^2(1-\rho)}{\lambda^2}.
\]

The service moments are finite under Assumption~\ref{asm:heavy-traffic-regularity}. Since
\(\mathbb E[S^2]\ge\mathbb E[S]^2\) and
\(\lambda\mathbb E[S]=\rho\),  we obtain

\[
 \frac{\delta^2\mathbb E[Q^{\mathrm{qry}}]}{\mathbb E[S^2]}
 \le\frac{(1-\zeta)^2(1-\rho)}
          {\lambda^2\mathbb E[S^2]}
 \le\frac{(1-\zeta)^2(1-\rho)}{\rho^2}
 \le2(1-\zeta)^2,
\]
where the last step uses \(\rho\ge1/2\). Applying
\((x+y)^2\le2x^2+2y^2\) and
\((Q^{\mathrm{qry}})^2=Q^{\mathrm{qry}}\) gives

\[
\begin{aligned}
 \mathbb E[(S+\delta Q^{\mathrm{qry}}-(S-\tau)R)^2]
 &\le2\mathbb E[(S-(S-\tau)R)^2]
       +2\delta^2\mathbb E[Q^{\mathrm{qry}}],\\
 \mathbb E[(S+\delta Q^{\mathrm{qry}})^2]
 &\le2\mathbb E[S^2]+2\delta^2\mathbb E[Q^{\mathrm{qry}}]\le2\bigl(1+2(1-\zeta)^2\bigr)\mathbb E[S^2].
\end{aligned}
\]

\end{proof}

\subsection[Tools for Distribution-Specific Comparisons]{Tools for Distribution-Specific Comparisons}
\label{app:sparse-advice-trigger}
Throughout Appendices~\ref{app:sparse-cost-polylog}--\ref{app:sparse-cost-regular}, we fix \(\theta\in(0,\bar\theta)\), use exact predictions with residual barrier \(b_\rho(\theta)=\theta G^{-1}(\rho)\), and use the \(\gamma\)-exponential cost scale~\eqref{eq:sparse-gamma-cost-scale} with fixed \(\gamma\in(0,1)\). The tail assumptions describe \(H\) at multiples of the workload cutoff \(G^{-1}(\rho)\), whereas the stability criterion in Lemma~\ref{cor:sparse-gamma-cost} uses \(H(\bar s_\rho)\). The first lemma shows that \(H(\bar s_\rho)\) and \(H(G^{-1}(\rho))\) are asymptotically equal, allowing us to apply the tail assumptions in that criterion. The second establishes feasibility of fixed proportional triggers at sufficiently high loads.

\begin{lemma}[Cutoff scales]
\label{lem:sparse-cutoff-scales}
As \(\rho\uparrow1\), the ratio of cumulative hazard for $\bar s_\rho$ and \(G^{-1}(\rho)\) goes to \(1\):
\[
    \frac{H(\bar s_\rho)}{H(G^{-1}(\rho))}\longrightarrow1.
\]
\end{lemma}

\begin{proof}[Proof of Lemma~\ref{lem:sparse-cutoff-scales}]
By the definition of the workload distribution \(G\), the sizes above \(G^{-1}(\rho)\) contribute a \(1-\rho\) fraction of the total workload. This gives the first equality below; Lemma~\ref{lem:free-bit-tail-work-asymptotics} gives the second:
\[
 (1-\rho)\mathbb E[S]
 =\mathbb E[S\mathbf1\{S>G^{-1}(\rho)\}]
 =G^{-1}(\rho)\overline F(G^{-1}(\rho))(1+o(1)).
\]
Taking logarithms and using \(H(\bar s_\rho)=-\log(1-\rho)\) yields
\[
 H(\bar s_\rho)
 =H(G^{-1}(\rho))-\log G^{-1}(\rho)+\log\mathbb E[S]+o(1).
\]
The first condition of Assumption~\ref{asm:heavy-traffic-regularity} gives \(\log G^{-1}(\rho)/H(G^{-1}(\rho))\to0\) and, by Lemma~\ref{lem:service-finite-moments}, \(\mathbb E[S]\) is finite. Dividing by \(H(G^{-1}(\rho))\) proves the hazard-ratio limit.
\end{proof}

\begin{lemma}[Feasible proportional triggers]
\label{lem:sparse-proportional-triggers}
Fix \(\theta\in(0,\bar\theta)\). For every fixed \(a\in[0,\bar\theta-\theta)\), the pair \((\theta,aG^{-1}(\rho))\) is feasible for all sufficiently high loads. If \(H\) is rapidly varying or regularly varying with positive index, then \(\bar\theta=1\).
\end{lemma}

\begin{proof}[Proof of Lemma~\ref{lem:sparse-proportional-triggers}]
Recall that \(\bar\tau_\rho(\theta)=\bar s_\rho-\theta G^{-1}(\rho)\). Lemma~\ref{lem:free-bit-threshold-location} gives
\[
 \frac{\bar\tau_\rho(\theta)}{G^{-1}(\rho)}
 =\frac{\bar s_\rho}{G^{-1}(\rho)}-\theta
 \longrightarrow\bar\theta-\theta.
\]
Thus, for each fixed \(a\in[0,\bar\theta-\theta)\), we have \(0\le aG^{-1}(\rho)\le\bar\tau_\rho(\theta)\) at all sufficiently high loads. This proves feasibility of the stated pair.

For the last claim, suppose \(\bar\theta<1\) and choose a fixed \(c\in(\bar\theta,1)\). Lemma~\ref{lem:free-bit-threshold-location} gives \(\bar s_\rho<cG^{-1}(\rho)\) eventually. By monotonicity,
\[
 \frac{H(\bar s_\rho)}{H(G^{-1}(\rho))}
 \le\frac{H(cG^{-1}(\rho))}{H(G^{-1}(\rho))}.
\]
The left-hand side tends to \(1\) by Lemma~\ref{lem:sparse-cutoff-scales}. The right-hand side tends to \(0\) under rapid variation, or to \(c^\nu<1\) under regular variation with index \(\nu>0\). Both contradict the inequality, so \(\bar\theta=1\).
\end{proof}

\subsection[Polylogarithmic Cumulative Hazards]{Polylogarithmic Cumulative Hazards}
\label{app:sparse-cost-polylog}

\begin{proposition}[Trigger choice for polylogarithmic hazards]
\label{prop:sparse-cost-polylog}
Suppose \(H(x)\sim A(\log x)^m\), with \(A>0\) and \(m\ge2\), including lognormal-type tails at \(m=2\). Arrival queries make SPD eventually unstable, whereas the latest feasible trigger and every fixed proportional trigger \(aG^{-1}(\rho)\), \(0<a<\bar\theta-\theta\), give eventual stability and heavy-traffic order-optimality.
\end{proposition}

\begin{proof}[Proof of Proposition~\ref{prop:sparse-cost-polylog}]
For each stated trigger choice, Lemma~\ref{lem:sparse-proportional-triggers} gives feasibility at all sufficiently high loads. For every fixed \(t>0\), the polylogarithmic asymptotic and Lemma~\ref{lem:sparse-cutoff-scales} give
\[
 \frac{H(tG^{-1}(\rho))}{H(\bar s_\rho)}
 =\frac{H(tG^{-1}(\rho))}{H(G^{-1}(\rho))}
  \frac{H(G^{-1}(\rho))}{H(\bar s_\rho)}
 \longrightarrow1.
\]
Recall from Section~\ref{sec:sparse-costly-exact} that, for a feasible pair \((\theta,\tau)\), the self-information is \(I_\rho(\theta,\tau)=H(\tau+b_\rho(\theta))-H(\tau)\). We use the preceding limit to evaluate \(I_\rho(\theta,\tau)/H(\bar s_\rho)\) at each trigger and apply Lemma~\ref{cor:sparse-gamma-cost}.

\smallskip\noindent\emph{At-arrival queries.} Since \(H(0)=0\), the definition of self-information gives
\[
 \frac{I_\rho(\theta,0)}{H(\bar s_\rho)}
 =\frac{H(\theta G^{-1}(\rho))}{H(\bar s_\rho)}\longrightarrow1>\gamma.
\]
Lemma~\ref{cor:sparse-gamma-cost} therefore gives eventual instability.

\smallskip\noindent\emph{Positive proportional triggers.} For every fixed \(a\in(0,\bar\theta-\theta)\), both \(a\) and \(a+\theta\) are positive. Substituting \(\tau=aG^{-1}(\rho)\) into the definition of self-information yields
\[
 \frac{I_\rho(\theta,aG^{-1}(\rho))}{H(\bar s_\rho)}
 =\frac{H((a+\theta)G^{-1}(\rho))}{H(\bar s_\rho)}
  -\frac{H(aG^{-1}(\rho))}{H(\bar s_\rho)}
 \longrightarrow0<\gamma.
\]
Lemma~\ref{cor:sparse-gamma-cost} gives slack preservation and eventual stability. Theorem~\ref{thm:sparse-uniform-optimality} and Lemma~\ref{lem:sparse-cost-transfer} then give heavy-traffic order-optimality.

\smallskip\noindent\emph{Latest feasible trigger.} Choose a fixed \(c\in(0,\bar\theta-\theta)\). Lemma~\ref{lem:sparse-proportional-triggers} gives \(cG^{-1}(\rho)\le\bar\tau_\rho(\theta)\le\bar s_\rho\) eventually. Monotonicity therefore yields
\[
 \frac{H(cG^{-1}(\rho))}{H(\bar s_\rho)}
 \le\frac{H(\bar\tau_\rho(\theta))}{H(\bar s_\rho)}\le1.
\]
The left-hand side tends to \(1\). Since \(\bar\tau_\rho(\theta)+b_\rho(\theta)=\bar s_\rho\), the definition of self-information gives
\[
 \frac{I_\rho(\theta,\bar\tau_\rho(\theta))}{H(\bar s_\rho)}
 =1-\frac{H(\bar\tau_\rho(\theta))}{H(\bar s_\rho)}\longrightarrow0<\gamma.
\]
The same stability criterion and free-to-costly comparison give eventual stability and heavy-traffic order-optimality.

\end{proof}

\subsection{Rapidly Varying Cumulative Hazards}
\label{app:sparse-cost-rapid}

\begin{proposition}[Trigger choice under rapid variation]
\label{prop:sparse-cost-rapid}
Suppose \(H\) is rapidly varying, as for doubly exponential tails. Every fixed proportional trigger \(aG^{-1}(\rho)\), \(0\le a<1-\theta\), gives eventual stability and heavy-traffic order-optimality, whereas the latest feasible trigger makes SPD eventually unstable.
\end{proposition}

\begin{proof}[Proof of Proposition~\ref{prop:sparse-cost-rapid}]
Lemma~\ref{lem:sparse-proportional-triggers} gives \(\bar\theta=1\) and feasibility of each fixed proportional trigger with \(a\in[0,1-\theta)\) at sufficiently high loads. Since \(\bar\tau_\rho(\theta)=\bar s_\rho-\theta G^{-1}(\rho)\), Lemma~\ref{lem:free-bit-threshold-location} also gives \(\bar\tau_\rho(\theta)/G^{-1}(\rho)\to1-\theta\).

\smallskip\noindent\emph{Fixed proportional triggers, including arrival.} Since \(a+\theta<1\), rapid variation and Lemma~\ref{lem:sparse-cutoff-scales} give
\[
 0\le\frac{I_\rho(\theta,aG^{-1}(\rho))}{H(\bar s_\rho)}
 \le\frac{H((a+\theta)G^{-1}(\rho))}{H(G^{-1}(\rho))}
     \frac{H(G^{-1}(\rho))}{H(\bar s_\rho)}\longrightarrow0<\gamma.
\]
Lemma~\ref{cor:sparse-gamma-cost} gives slack preservation and eventual stability. Theorem~\ref{thm:sparse-uniform-optimality} and Lemma~\ref{lem:sparse-cost-transfer} then give heavy-traffic order-optimality.

\smallskip\noindent\emph{Latest feasible trigger.} Choose a fixed \(c\in(1-\theta,1)\). Since \(\bar\tau_\rho(\theta)/G^{-1}(\rho)\to1-\theta\), monotonicity and rapid variation give
\[
 0\le\frac{H(\bar\tau_\rho(\theta))}{H(\bar s_\rho)}
 \le\frac{H(cG^{-1}(\rho))}{H(\bar s_\rho)}\longrightarrow0.
\]
Since \(\bar\tau_\rho(\theta)+b_\rho(\theta)=\bar s_\rho\), the definition of self-information gives
\[
 \frac{I_\rho(\theta,\bar\tau_\rho(\theta))}{H(\bar s_\rho)}
 =1-\frac{H(\bar\tau_\rho(\theta))}{H(\bar s_\rho)}\longrightarrow1>\gamma.
\]
Lemma~\ref{cor:sparse-gamma-cost} therefore gives eventual instability.

\end{proof}

\subsection{Regularly Varying Cumulative Hazards}
\label{app:sparse-cost-regular}
\label{sec:sparse-costly-trigger-ranges}

\begin{proposition}[Query use under regular variation]
\label{prop:sparse-cost-regular}
\label{prop:sparse-cost-trigger-ranges}
\label{lem:sparse-cost-trigger-ranges}
Suppose \(H\) is regularly varying with index \(\nu>0\), as for Weibull service times. For any fixed \(0\le a_1<a_2<1-\theta\), the later trigger \(a_2G^{-1}(\rho)\) uses fewer queries than \(a_1G^{-1}(\rho)\) when \(0<\nu<1\), and more when \(\nu>1\), in each case by an unbounded factor.
\end{proposition}

\begin{proof}[Proof of Proposition~\ref{prop:sparse-cost-regular}]
Lemma~\ref{lem:sparse-proportional-triggers} gives \(\bar\theta=1\) and feasibility of \((\theta,aG^{-1}(\rho))\) for every fixed \(a\in[0,1-\theta)\) at all sufficiently high loads. Define
\[
 \Psi_{\nu,\theta}(a):=(a+\theta)^\nu-a^\nu.
\]
Substituting \(\tau=aG^{-1}(\rho)\) into the definition of self-information and applying regular variation and Lemma~\ref{lem:sparse-cutoff-scales} gives
\[
\begin{aligned}
 \frac{I_\rho(\theta,aG^{-1}(\rho))}{H(\bar s_\rho)}
 &=\left(
 \frac{H((a+\theta)G^{-1}(\rho))}{H(G^{-1}(\rho))}
 -\frac{H(aG^{-1}(\rho))}{H(G^{-1}(\rho))}
 \right)\frac{H(G^{-1}(\rho))}{H(\bar s_\rho)}\\
 &\longrightarrow(a+\theta)^\nu-a^\nu=\Psi_{\nu,\theta}(a).
\end{aligned}
\]

\smallskip\noindent\emph{Comparison of two fixed triggers.}
For fixed \(0\le a_1<a_2<1-\theta\), the identity \(p_\rho^{\mathrm{qry}}(\theta,\tau)=(1-\rho)e^{I_\rho(\theta,\tau)}\) and the preceding limits give
\begin{equation}
\label{eq:sparse-query-D-ratio}
\begin{aligned}
 \frac{p_\rho^{\mathrm{qry}}(\theta,a_2G^{-1}(\rho))}
      {p_\rho^{\mathrm{qry}}(\theta,a_1G^{-1}(\rho))}
 &=\exp\!\left(I_\rho(\theta,a_2G^{-1}(\rho))-I_\rho(\theta,a_1G^{-1}(\rho))\right)\\
 &=\exp\!\left(\bigl[\Psi_{\nu,\theta}(a_2)-\Psi_{\nu,\theta}(a_1)+o(1)\bigr]H(\bar s_\rho)\right).
\end{aligned}
\end{equation}
For \(0<\nu<1\), strict concavity of \(x^\nu\) makes \(\Psi_{\nu,\theta}\) strictly decreasing, so the exponent tends to \(-\infty\) and the ratio tends to zero. For \(\nu>1\), strict convexity makes \(\Psi_{\nu,\theta}\) strictly increasing, so the exponent tends to \(+\infty\) and the ratio tends to infinity. When \(\nu=1\), \(\Psi_{1,\theta}(a)=\theta\) for every \(a\), so the leading terms cancel in the exponent of the query-ratio expression~\eqref{eq:sparse-query-D-ratio}. Comparing the query probabilities then requires finer information about \(H\). For example, when service times are exponential, \(H\) is linear, so \(I_\rho(\theta,\tau)\) is independent of \(\tau\) and all feasible triggers have the same ex-ante query probability.

\smallskip\noindent\emph{Inference-cost consequences.}
For each fixed \(a\in[0,1-\theta)\), compare the normalized self-information limit \(\Psi_{\nu,\theta}(a)\) with \(\gamma\). If \(\Psi_{\nu,\theta}(a)<\gamma\), Lemma~\ref{cor:sparse-gamma-cost} gives slack preservation and eventual stability; Theorem~\ref{thm:sparse-uniform-optimality} and Lemma~\ref{lem:sparse-cost-transfer} then give heavy-traffic order-optimality. If \(\Psi_{\nu,\theta}(a)>\gamma\), the same stability criterion gives eventual instability.

By the monotonicity just proved, the set \(\{a\in[0,1-\theta):\Psi_{\nu,\theta}(a)<\gamma\}\), when nonempty, is a terminal interval for \(0<\nu<1\) and an initial interval for \(\nu>1\). At \(\nu=1\), all fixed proportional triggers preserve slack if \(\theta<\gamma\), and all are unstable if \(\theta>\gamma\). Whenever \(\Psi_{\nu,\theta}(a)=\gamma\), the finer terms in the relative inference load identity~\eqref{eq:sparse-gamma-relative-load} determine the inference load.
\end{proof}

\section{Proofs for Inexact Predictions}
\label{app:sparse-predict-errors}

We use the prediction rules and multiplicative error bound from Section~\ref{sec:sparse-costly}. Appendix~\ref{app:sparse-error-unadjusted-proof} shows why the parameters used for exact predictions can fail with prediction error. Appendix~\ref{app:sparse-error-free-proof} states a response-time bound in terms of demoted work and applies it to the repair at arrival. Appendix~\ref{app:sparse-error-candidate-proof} then uses the same bound to give the detailed proof of Lemma~\ref{lem:sparse-error-free-candidates}. Finally, Appendix~\ref{app:sparse-error-demotion-bound-proof} proves the response-time bound.

\subsection{Failure of the Unadjusted Demotion Test}
\label{app:sparse-error-unadjusted-proof}

Recall from Section~\ref{sec:sparse-predict-delay} the residual barrier and \((\theta,\tau)\)-tuned query probability
\[
 b_\rho(\theta)=\theta G^{-1}(\rho),
 \qquad
 q_\rho(\theta,\tau)=\frac{1-\rho}{\overline F(\tau+b_\rho(\theta))}.
\]
A trigger is feasible when \(0\le\tau\le\bar\tau_\rho(\theta)=\bar s_\rho-b_\rho(\theta)\).

\begin{proposition}[Failure of the unadjusted demotion test]
\label{prop:sparse-error-unadjusted-cutoff}
For every fixed \(\theta\in(0,1)\) and \(\varepsilon\in(0,1)\), there exists a service distribution in the lognormal-or-lighter regime and an \(\varepsilon\)-inexact prediction rule such that, with free predictions, \(\mathrm{SPD}(b_\rho(\theta),q_\rho(\theta,\tau_\rho),\tau_\rho)\) is not heavy-traffic order-optimal for any feasible deterministic trigger \(\tau_\rho\).
\end{proposition}

\begin{proof}[Proof.]
We consider mean-one exponential service times and predictions that consistently underestimate size as much as the error model allows: \(\widehat S=(1-\varepsilon)S\). With the parameters in the proposition, \(\mathrm{SPD}(b_\rho(\theta),q_\rho(\theta,\tau_\rho),\tau_\rho)\) demotes a queried job only when \(S>(\tau_\rho+b_\rho(\theta))/(1-\varepsilon)\).
Write \(b=b_\rho(\theta)=\theta G^{-1}(\rho)\). For mean-one exponential service, \(\lambda_\rho=\rho\), \(1-\rho=(G^{-1}(\rho)+1)e^{-G^{-1}(\rho)}\), and \(\bar s_\rho=-\log(1-\rho)=G^{-1}(\rho)-\log(G^{-1}(\rho)+1)\). Feasibility gives \(b\le\tau_\rho+b\le\bar s_\rho<G^{-1}(\rho)\), while the \((\theta,\tau_\rho)\)-tuned query probability is \(q_\rho(\theta,\tau_\rho)=(1-\rho)e^{\tau_\rho+b}\). With \(Q\) the independent sampling coin, the demotion indicator is \(R=Q\mathbf1\{S>(\tau_\rho+b)/(1-\varepsilon)\}\). Direct integration of the exponential tail gives
\[
\begin{aligned}
    \mathbb E[(S-\tau_\rho)R]
    &=(1-\rho)\left(\frac{\tau_\rho+b}{1-\varepsilon}-\tau_\rho+1\right)
    e^{-\varepsilon(\tau_\rho+b)/(1-\varepsilon)}\\
    &\le(1-\rho)\left(\frac{G^{-1}(\rho)}{1-\varepsilon}+1\right)
    e^{-\varepsilon b/(1-\varepsilon)}=o(1-\rho).
\end{aligned}
\]
The upper bound does not depend on the feasible trigger.
This higher threshold makes the expected work removed from high priority \(o(1-\rho)\), uniformly over feasible triggers. We therefore have \(1-\rho^{\mathrm{HIGH}}(0)=1-\rho+\rho\mathbb E[(S-\tau_\rho)R]\sim1-\rho\), uniformly over feasible triggers. Moreover,
\[
\begin{aligned}
    \mathbb P(R=1)
    &=(1-\rho)e^{-\varepsilon(\tau_\rho+b)/(1-\varepsilon)}
    \le 1-\rho,\\
    \mathbb E[(S-(S-\tau_\rho)R)^2]
    &\ge \mathbb E\!\left[S^2\mathbf 1\!\left\{S\le \frac{b}{1-\varepsilon}\right\}\right]
    \longrightarrow \mathbb E[S^2]=2.
\end{aligned}
\]
The second inequality holds because a job of size at most \(b/(1-\varepsilon)\) cannot be demoted at any feasible trigger. Applying Lemma~\ref{lem:sparse-cost-high-response} with \(\delta=0\), the never-demoted waiting contribution is
\[
    \frac{\mathbb P(R=0)\rho\mathbb E[(S-(S-\tau_\rho)R)^2]}
    {2(1-\rho^{\mathrm{HIGH}}(0))}
    =\Omega\!\left(\frac1{1-\rho}\right).
\]
Consequently, the mean response time is \(\Omega(1/(1-\rho))\), which exceeds the optimal size-aware benchmark in~\eqref{eq:srpt-heavy-traffic-benchmark}, of order \(1/((1-\rho)G^{-1}(\rho))\), by an unbounded factor. The comparison is uniform over all feasible deterministic trigger choices.
\end{proof}

\subsection[Repair at Arrival]{Repair at Arrival}
\label{app:sparse-error-free-proof}
\label{app:sparse-error-demotion-bound}
The following bound combines the two SOAP identities in Lemmas~\ref{lem:sparse-cost-high-response} and~\ref{lem:sparse-cost-low-response}. It translates lower bounds on the fraction of jobs demoted and their remaining service into an upper bound on mean response time. We use it below for the repair at arrival and in Appendix~\ref{app:sparse-error-candidate-proof} for the delayed candidates. Its proof is deferred to Appendix~\ref{app:sparse-error-demotion-bound-proof}.

\begin{lemma}[Response time from demoted work]
\label{lem:sparse-error-demotion-bound}

At load \(0<\rho<1\), consider an SPD policy $\pi$ with free predictions, a deterministic trigger \(\tau\ge0\), and FCFS in both queues with a prediction rule as defined in Section~\ref{sec:sparse-costly}. Let \(R\) indicate eventual demotion. If \(\mathbb P(R=1)\ge1-\rho\) and \(S-\tau\ge m>0\) whenever \(R=1\), then
\[
 \mathbb E[T^{\pi}]\le
 \frac{\mathbb E[S^2]}{(1-\rho)m}
 +\mathbb E[S]+\frac1\lambda+\frac{\tau}{\lambda m}.
\]
\end{lemma}

At \(\tau=0\), we can account for the worst allowed underprediction by lowering the residual demotion barrier to \((1-\varepsilon)b_\rho(\theta)\), while keeping the \(\theta\)-tuned query probability \(q_\rho(\theta)\).

\begin{proposition}[Robustness with free predictions]
\label{prop:sparse-error-free}
Suppose the service distribution is in the lognormal-or-lighter regime. For fixed \(\theta\in(0,\bar\theta)\) and \(\varepsilon\in[0,1)\), the pair \((\theta,0)\) is feasible for all sufficiently high loads. Under free \(\varepsilon\)-inexact predictions the policy
\[
 \mathrm{SPD}\bigl((1-\varepsilon)b_\rho(\theta),q_\rho(\theta),0\bigr)
\]
is heavy-traffic order-optimal, uniformly over the allowed prediction rules.
\end{proposition}
\begin{proof}
Lemma~\ref{lem:sparse-candidate-abundance} gives \(q_\rho(\theta)\to0\), so the pair \((\theta,0)\) is feasible for all sufficiently high loads. Every queried job with \(S>b_\rho(\theta)\) is demoted, since \(\widehat S\ge(1-\varepsilon)S>(1-\varepsilon)b_\rho(\theta)\). Conversely, a demoted job satisfies \((1+\varepsilon)S\ge\widehat S>(1-\varepsilon)b_\rho(\theta)\), so its size exceeds \(\frac{1-\varepsilon}{1+\varepsilon}b_\rho(\theta)\). If \(p_\rho^{\mathrm{dem}}\) denotes the demoted fraction, independent sampling gives \(p_\rho^{\mathrm{dem}}\ge q_\rho(\theta)\overline F(b_\rho(\theta))=1-\rho\), and high-priority slack is at least \(\lambda_\rho\frac{1-\varepsilon}{1+\varepsilon}p_\rho^{\mathrm{dem}}b_\rho(\theta)\).

Applying Lemma~\ref{lem:sparse-error-demotion-bound} with \(\tau=0\) and \(m=\frac{1-\varepsilon}{1+\varepsilon}b_\rho(\theta)\), we obtain \(\mathbb E[T]=O(1/((1-\rho)b_\rho(\theta))+1)\), uniformly over the allowed prediction rules. Since \(b_\rho(\theta)=\theta G^{-1}(\rho)\), this gives the required heavy-traffic upper bound. SRPT optimality gives \(\mathbb E[T]\ge\mathbb E[T^{\mathrm{SRPT}}]\) for every allowed prediction rule. The SRPT heavy-traffic characterization~\eqref{eq:srpt-heavy-traffic-benchmark} supplies the matching lower bound.
\end{proof}

\subsection[Free-Prediction Guarantee for the Delayed Candidates (Lemma~\ref{lem:sparse-error-free-candidates})]{Free-Prediction Guarantee for the Delayed Candidates (Lemma~\ref{lem:sparse-error-free-candidates})}
\label{app:sparse-error-candidate-proof}
We now apply Lemma~\ref{lem:sparse-error-demotion-bound} to the delayed candidates.

\begin{proof}[Proof of Lemma~\ref{lem:sparse-error-free-candidates}]
Each \(\theta_j\) lies in \((0,\bar\theta)\). The largest candidate endpoint is 
\[
 \tau_{\rho,1}+b_\rho(\theta_1)=\frac{\bar\theta}{2}G^{-1}(\rho),
\]
which is eventually below \(\bar s_\rho\) by Lemma~\ref{lem:free-bit-threshold-location}; successive endpoints are multiplied by \(\frac{1-\varepsilon}{2(1+\varepsilon)}\in(0,1)\), so all other endpoints are smaller. Thus every pair is feasible, and Lemma~\ref{lem:sparse-delayed-query-vanishing} gives vanishing ex-ante query probabilities.
Their error-adjusted residual barriers are nonnegative. The test \(\widehat S-\tau_{\rho,j}>\widetilde b_\rho(\theta_j,\tau_{\rho,j};\varepsilon)\) demotes every queried job with \(S>\tau_{\rho,j}+b_\rho(\theta_j)\), and the \((\theta_j,\tau_{\rho,j})\)-tuned query probability remains \(q_\rho(\theta_j,\tau_{\rho,j})\). Thus, if \(p_{\rho,j}^{\mathrm{dem}}\) denotes the probability that an arriving job is eventually demoted by candidate \(j\), independent sampling gives \(p_{\rho,j}^{\mathrm{dem}}\ge q_\rho(\theta_j,\tau_{\rho,j})\overline F(\tau_{\rho,j}+b_\rho(\theta_j))=1-\rho\). Conversely, a demoted job satisfies
\[
 (1+\varepsilon)S\ge\widehat S
 >(1-\varepsilon)\bigl(\tau_{\rho,j}+b_\rho(\theta_j)\bigr)
 =2(1+\varepsilon)\tau_{\rho,j}.
\]
Dividing by \(1+\varepsilon\) gives \(S>2\tau_{\rho,j}\), so at demotion the job retains \(S-\tau_{\rho,j}>\tau_{\rho,j}=\frac{1-\varepsilon}{1+3\varepsilon}b_\rho(\theta_j)\) units of service. Let \(\pi_j\) denote the policy for candidate \(j\). Lemma~\ref{lem:sparse-error-demotion-bound} with \(m=\tau_{\rho,j}=\frac{1-\varepsilon}{1+3\varepsilon}b_\rho(\theta_j)\) gives
\[
 \mathbb E[T^{\pi_j}(0)]
 =O\!\left(\frac{1}{(1-\rho)b_\rho(\theta_j)}+1+
 \frac{\tau_{\rho,j}}{b_\rho(\theta_j)}\right).
\]
Each \(\theta_j\) is a fixed positive constant and \(\tau_{\rho,j}/b_\rho(\theta_j)\) is fixed, so this is the required order. SRPT optimality gives \(\mathbb E[T^{\pi_j}(0)]\ge\mathbb E[T^{\mathrm{SRPT}}]\) for every candidate and allowed prediction rule. The SRPT heavy-traffic characterization~\eqref{eq:srpt-heavy-traffic-benchmark} supplies the matching lower bound.

There are finitely many candidates, so one can choose common constants and a common load threshold for all \(j\). The proof uses the prediction rule only through \((1-\varepsilon)S\le\widehat S\le(1+\varepsilon)S\); these same bounds therefore hold for every rule satisfying that error bound, including biased and randomized rules.
\end{proof}

\subsection[Proof of the Response-Time Bound]{Proof of the Response-Time Bound (Lemma~\ref{lem:sparse-error-demotion-bound})}
\label{app:sparse-error-demotion-bound-proof}

\begin{proof}[Proof of Lemma~\ref{lem:sparse-error-demotion-bound}]
Set \(p^{\mathrm{dem}}=\mathbb P(R=1)\). The residual-service assumption gives
\[
 1-\rho^{\mathrm{HIGH}}(0)
 =1-\rho+\lambda\mathbb E[(S-\tau)R]
 \ge\lambda m p^{\mathrm{dem}}.
\]
Apply Lemmas~\ref{lem:sparse-cost-high-response} and~\ref{lem:sparse-cost-low-response} with free predictions. Since \(0\le S-(S-\tau)R\le S\) and \(\mathbb P(R=0)\le1\), the total waiting contribution is at most
\[
\begin{aligned}
 &\frac{\lambda\mathbb E[S^2]}{2(1-\rho^{\mathrm{HIGH}}(0))}
 +\frac{p^{\mathrm{dem}}\lambda\mathbb E[S^2]}
 {2(1-\rho)(1-\rho^{\mathrm{HIGH}}(0))}\le\frac{\mathbb E[S^2]}{2m p^{\mathrm{dem}}}
 +\frac{\mathbb E[S^2]}{2(1-\rho)m}
 \le\frac{\mathbb E[S^2]}{(1-\rho)m}.
\end{aligned}
\]
The first inequality uses \(1-\rho^{\mathrm{HIGH}}(0)\ge\lambda m p^{\mathrm{dem}}\), cancelling \(p^{\mathrm{dem}}\) from the demoted-job term. The second uses \(p^{\mathrm{dem}}\ge1-\rho\). For residence, use \(\mathbb E[SR]=\mathbb E[(S-\tau)R]+\tau p^{\mathrm{dem}}\), together with the displayed slack identity and lower bound:
\[
 \mathbb E[S(1-R)]+\frac{\mathbb E[SR]}{1-\rho^{\mathrm{HIGH}}(0)}
 \le\mathbb E[S]+\frac1\lambda+\frac{\tau}{\lambda m}.
\]
Adding the waiting and residence contributions proves the bound.
\end{proof}

\section[Proofs for SRPT in the Low-Priority Queue]{Proofs for SRPT in the Low-Priority Queue (Section~\ref{sec:sparse-srpt})}
\label{app:sparse-srpt}

Appendix~\ref{app:sparse-srpt-identity} decomposes the chosen policy's mean response time into the contributions of never-demoted jobs, demoted waiting, and demoted residence. Appendix~\ref{app:sparse-srpt-intermediate} uses bounds on these three contributions to prove heavy-traffic order-optimality with inference costs (Theorem~\ref{thm:sparse-srpt-intermediate-costly}), recovering Theorem~\ref{thm:sparse-srpt-intermediate} when predictions are free. Appendix~\ref{app:sparse-srpt-helper-proofs} proves the three supporting lemmas.

\subsection[The SOAP Identity for the Chosen Policy]{The SOAP Identity for the Chosen Policy}
\label{app:sparse-srpt-identity}
With FCFS in both queues, ranks only increase at demotion, so there is no recycled relevant work. With an SRPT background, a demoted job's rank rises at demotion and then decreases with its remaining service. An old job that crosses above a fixed rank threshold can therefore later cross back below it, creating a recycled relevant-work interval. The demoted tagged job's worst future rank also changes during service. We therefore derive both the relevant-work terms and the residence integral explicitly below.

Consider \(\mathrm{SPD}(\tau_\rho,1,\tau_\rho,\mathrm{SRPT})\) with \(\tau_\rho>0\): query when \(S>\tau_\rho\), and demote when \(S>2\tau_\rho\). For ease of presentation, we refer to this policy as \(\mathrm{SRPT\text{-}low}\). Each query requires \(\delta_\rho\ge0\) units of server work, performed at high priority before any demotion; \(\delta_\rho=0\) is the case of free predictions. We write \(T^{\mathrm{SRPT\text{-}low}}(\delta_\rho)\) for response time including inference. A demoted job receives \(\tau_\rho+\delta_\rho\) units at high priority, then enters low priority with \(S-\tau_\rho\) ordinary service remaining. Assume
\[
 \rho(\delta_\rho)
 =\rho+\lambda_\rho\delta_\rho\overline F(\tau_\rho)<1.
\]
High priority has rank zero, and the low-priority rank is the true remaining service. A new job of size \(S>s\) can interrupt a demoted tagged job of size \(s\) while receiving high-priority service and inference. At demotion, however, the new job has \(S-\tau_\rho>s-\tau_\rho\) service left, whereas the tagged job has at most \(s-\tau_\rho\) left. SRPT therefore gives the tagged job priority over this larger job until it completes. To apply Proposition~\ref{prop:soap-response-time}, we next compute the relevant-work terms for jobs already present when the tagged job arrives and for jobs that arrive afterward.

\smallskip\noindent\emph{Relevant work and the SOAP denominators.}
The high-priority work and load of a generic arrival are
\[
 X_0^{\mathrm{old}}[0]
 =S+\delta_\rho\mathbf1\{S>\tau_\rho\}
   -(S-\tau_\rho)\mathbf1\{S>2\tau_\rho\},
 \qquad
 \rho^{\mathrm{HIGH}}(\delta_\rho)
 =\lambda_\rho\mathbb E[X_0^{\mathrm{old}}[0]].
\]
The loads without inference cost are \(\rho(0)=\rho\) and \(\rho^{\mathrm{HIGH}}(0)\).

For a demoted tagged job of size \(s>2\tau_\rho\), the initial worst future rank is \(s-\tau_\rho\). 
An old competitor starts at rank zero and is initially relevant. If \(S \leq s\), its rank never exceeds \(s-\tau_\rho\), so all its work remains relevant. If \(S>s\), it leaves relevance at demotion and reenters when its remaining service reaches \(s-\tau_\rho\). Its original and recycled relevant intervals are therefore
\[
\begin{aligned}
 X_0^{\mathrm{old}}[s-\tau_\rho]
 &=(S+\delta_\rho\mathbf1\{S>\tau_\rho\})\mathbf1\{S\le s\}
   +(\tau_\rho+\delta_\rho)\mathbf1\{S>s\},\\
 X_1^{\mathrm{old}}[s-\tau_\rho]
 &=(s-\tau_\rho)\mathbf1\{S>s\},
 \qquad X_i^{\mathrm{old}}[s-\tau_\rho]=0\quad(i\ge2).
\end{aligned}
\]
The second interval records a larger old job's final \(s-\tau_\rho\) units, which it may already have reached before the tagged arrival. A new job cannot reach this interval ahead of the tag. Its original relevant work agrees with \(X_0^{\mathrm{old}}\) except at \(S=s\): old jobs win equal-rank FCFS ties and new jobs lose them. Since ties have probability zero (the service distribution has a density; see Section~\ref{subsec:service-time-classes}), \(\rho_0^{\mathrm{old}}[s-\tau_\rho]=\rho^{\mathrm{new}}[s-\tau_\rho]\), and
\begin{equation}
\label{eq:sparse-srpt-relevant-slack}
 1-\rho^{\mathrm{new}}[s-\tau_\rho]
 =1-\rho(\delta_\rho)
  +\lambda_\rho\mathbb E[(S-\tau_\rho)\mathbf1\{S>s\}],
 \qquad s\ge2\tau_\rho.
\end{equation}
Here \emph{slack} means this SOAP denominator factor: total-system slack plus the load excluded from new relevant work. At the smallest demotion rank, \(\rho^{\mathrm{new}}[\tau_\rho]=\rho^{\mathrm{HIGH}}(\delta_\rho)\). By contrast, a never-demoted tag has rank zero throughout and \(\rho^{\mathrm{new}}[0]=0\), since new arrivals lose its high-priority FCFS tie.

\smallskip\noindent\emph{Residence follows the worst future rank.}
For a demoted tag of size \(s>2\tau_\rho\), let \(a\) count attained server work, including inference. Its total work is \(s+\delta_\rho\), and its worst future rank is
\[
 r(a)=
 \begin{cases}
  s-\tau_\rho,&0\le a\le\tau_\rho+\delta_\rho,\\
  s+\delta_\rho-a,&\tau_\rho+\delta_\rho<a<s+\delta_\rho.
 \end{cases}
\]
The initial block uses the rank the tag will acquire at demotion; after demotion, remaining size decreases and is its own worst future rank. Splitting the attained-work interval at \(a=\tau_\rho+\delta_\rho\) and \(a=s+\delta_\rho-\tau_\rho\) gives
\[
\begin{aligned}
 \int_0^{s+\delta_\rho}\frac{da}{1-\rho^{\mathrm{new}}[r(a)]}
 &=\int_0^{\tau_\rho+\delta_\rho}\frac{da}{1-\rho^{\mathrm{new}}[r(a)]}
 +\int_{\tau_\rho+\delta_\rho}^{s+\delta_\rho-\tau_\rho}\frac{da}{1-\rho^{\mathrm{new}}[r(a)]}+\int_{s+\delta_\rho-\tau_\rho}^{s+\delta_\rho}\frac{da}{1-\rho^{\mathrm{new}}[r(a)]}\\
 &=\frac{\tau_\rho+\delta_\rho}{1-\rho^{\mathrm{new}}[s-\tau_\rho]}
 +\int_{2\tau_\rho}^{s}
       \frac{du}{1-\rho^{\mathrm{new}}[u-\tau_\rho]}
 +\frac{\tau_\rho}{1-\rho^{\mathrm{HIGH}}(\delta_\rho)}.
\end{aligned}
\]
The first interval has constant worst future rank \(s-\tau_\rho\); in the middle interval, set \(u=s+\delta_\rho-a+\tau_\rho\); in the final interval, \(0<r(a)\le\tau_\rho\), so new arrivals contribute only their high-priority work and \(\rho^{\mathrm{new}}[r(a)]=\rho^{\mathrm{HIGH}}(\delta_\rho)\), giving \(\tau_\rho/(1-\rho^{\mathrm{HIGH}}(\delta_\rho))\). These are SOAP contributions after rank substitution, not the physical elapsed times in each queue.

\smallskip\noindent\emph{The response-time decomposition.}
A never-demoted tag sees only older high-priority work and then its own service. For a demoted tag, the equal old/new loads give the squared waiting denominator below. Averaging the residence integral over tagged sizes replaces its upper limit by the weight \(\mathbb P(S>u)=\overline F(u)\). We group the response-time identity into never-demoted response time, demoted waiting, and demoted residence. Proposition~\ref{prop:soap-response-time} gives
\begin{equation}
\label{eq:sparse-srpt-response-identity}
\begin{aligned}
 \mathbb E[T^{\mathrm{SRPT\text{-}low}}(\delta_\rho)]&={}\underbrace{F(2\tau_\rho)
 \frac{\lambda_\rho\mathbb E[(X_0^{\mathrm{old}}[0])^2]}
      {2(1-\rho^{\mathrm{HIGH}}(\delta_\rho))}
 +\mathbb E[(S+\delta_\rho\mathbf1\{S>\tau_\rho\})\mathbf1\{S\le2\tau_\rho\}]
 }_{\text{never-demoted jobs: Lemma~\ref{lem:sparse-srpt-high-negligible}}}\\[6pt]
 &\quad+\underbrace{\int_{2\tau_\rho}^{\infty}
 \frac{\lambda_\rho\sum_{i=0}^{1}\mathbb E[(X_i^{\mathrm{old}}[s-\tau_\rho])^2]}
      {2(1-\rho^{\mathrm{new}}[s-\tau_\rho])^2}f(s)\,ds
 }_{\text{demoted waiting: Lemma~\ref{lem:sparse-srpt-demoted-comparison}}}\\[6pt]
 &\quad+\underbrace{(\tau_\rho+\delta_\rho)\int_{2\tau_\rho}^{\infty}
       \frac{f(s)}{1-\rho^{\mathrm{new}}[s-\tau_\rho]}\,ds
 +\int_{2\tau_\rho}^{\infty}
       \frac{\overline F(s)}{1-\rho^{\mathrm{new}}[s-\tau_\rho]}\,ds
 +\frac{\tau_\rho\overline F(2\tau_\rho)}{1-\rho^{\mathrm{HIGH}}(\delta_\rho)}
 }_{\text{demoted residence: Lemma~\ref{lem:sparse-srpt-residence-overhead}}}.
\end{aligned}
\end{equation}
The first group is \(\mathbb E[T^{\mathrm{SRPT\text{-}low}}(\delta_\rho)\mathbf1\{S\le2\tau_\rho\}]\): the entire response-time contribution of jobs that finish at high priority, weighted by their probability. The other two groups concern only demoted jobs, \(S>2\tau_\rho\): the second is their SOAP waiting, and the third is their full SOAP residence. Its three terms follow the same order as the conditional residence formula above: the initial service and inference block, remaining service above \(\tau_\rho\), and the final \(\tau_\rho\) units.

\subsection[The SRPT Response-Time Bound]{The SRPT Response-Time Bound}
\label{app:sparse-srpt-intermediate}
\label{app:sparse-srpt-intermediate-proof}

We prove the following extension of Theorem~\ref{thm:sparse-srpt-intermediate} to the inference-cost model of Section~\ref{sec:sparse-costly}.

\begin{theorem}[Order-optimality with full-size predictions and inference costs]
\label{thm:sparse-srpt-intermediate-costly}
Suppose the service distribution is in the intermediate regime or satisfies \(H(x)\sim A(\log x)^m\), with \(A>0\) and \(m\ge2\). Fix \(\gamma\in(0,1)\) and \(\beta\in(1-\gamma,1)\). Use exact full-size predictions with \(0\le\delta_\rho\le(1-\rho)^\gamma/\lambda_\rho\), and set
\[
 \tau_\rho=\overline F^{-1}\!\left((1-\rho)^\beta\right),
 \qquad b_\rho=\tau_\rho,\qquad q_\rho=1.
\]
Then \(\mathrm{SPD}(b_\rho,q_\rho,\tau_\rho,\mathrm{SRPT})\) is eventually stable and heavy-traffic order-optimal, with ex-ante query probability \((1-\rho)^\beta\to0\).
\end{theorem}

For any \(\beta\in(0,1)\), choosing \(\gamma\in(1-\beta,1)\) and setting \(\delta_\rho=0\) recovers Theorem~\ref{thm:sparse-srpt-intermediate}. We use the tuning above throughout the proof. The three lemmas bound the three groups in~\eqref{eq:sparse-srpt-response-identity}, in order: never-demoted jobs contribute negligibly, while demoted waiting and demoted residence are each bounded at the SRPT order. We state these facts first, prove the theorem from them, and give their proofs in Appendix~\ref{app:sparse-srpt-helper-proofs}. All bounds are for sufficiently high load, with constants independent of \(\rho\); setting \(\delta_\rho=0\) also gives the free-prediction case.

Both regimes in the theorem satisfy \(H(2x)/H(x)\to1\), the tail ratio used below. For a polylogarithmic cumulative hazard, \(H(x)\sim A(\log x)^m\) with \(A>0\) and \(m\ge2\), so
\[
 \frac{H(2x)}{H(x)}
 =(1+o(1))\left(\frac{\log(2x)}{\log x}\right)^m
 \longrightarrow1.
\]
In the intermediate regime, Lemma~\ref{lem:free-bit-log-quadratic} gives \(h(x)=o(\log x/x)\), and hence
\[
 H(2x)-H(x)=\int_x^{2x}h(u)\,du=o(\log x)=o(H(x)).
\]
The tuning also gives
\[
 p_\rho^{\mathrm{qry}}=\overline F(\tau_\rho)=(1-\rho)^\beta\longrightarrow0,
 \qquad
 \lambda_\rho\delta_\rho p_\rho^{\mathrm{qry}}
 \le(1-\rho)^{\beta+\gamma}=o(1-\rho).
\]
Hence \(1-\rho(\delta_\rho)\sim1-\rho\), so the policy is eventually stable and retains at least half the ordinary-service slack. Also \(\tau_\rho\to\infty\) and \(\delta_\rho\to0\), giving \(\delta_\rho\le\tau_\rho\) eventually.

\begin{lemma}[Response time of never-demoted jobs]
\label{lem:sparse-srpt-high-negligible}
Jobs that finish at high priority contribute at most
\[
 \mathbb E[T^{\mathrm{SRPT\text{-}low}}(\delta_\rho)\mathbf1\{S\le2\tau_\rho\}]
 \le\frac{C}{\tau_\rho\overline F(2\tau_\rho)}+C
 =o\!\left(\mathbb E[T^{\mathrm{SRPT}}]\right).
\]
\end{lemma}
The choice \(\beta<1\) sends enough work to low priority to make the response-time contribution of never-demoted jobs negligible.

\begin{lemma}[Waiting of demoted jobs]
\label{lem:sparse-srpt-demoted-comparison}
The SOAP waiting contribution of demoted jobs satisfies
\[
 \int_{2\tau_\rho}^{\infty}
 \frac{\lambda_\rho\sum_{i=0}^{1}\mathbb E[(X_i^{\mathrm{old}}[s-\tau_\rho])^2]}
      {2(1-\rho^{\mathrm{new}}[s-\tau_\rho])^2}f(s)\,ds
 \le16\mathbb E[T^{\mathrm{SRPT}}].
\]
\end{lemma}
A larger competitor retains at least half its size at demotion. Our policy therefore excludes at least half the large-job work excluded by ordinary SRPT, while its relevant-work second moments increase by only a constant factor.

\begin{lemma}[Residence of demoted jobs]
\label{lem:sparse-srpt-residence-overhead}
The full SOAP residence contribution of demoted jobs satisfies
\[
\begin{aligned}
 &(\tau_\rho+\delta_\rho)\int_{2\tau_\rho}^{\infty}
       \frac{f(s)}{1-\rho^{\mathrm{new}}[s-\tau_\rho]}\,ds
 +\int_{2\tau_\rho}^{\infty}
       \frac{\overline F(s)}{1-\rho^{\mathrm{new}}[s-\tau_\rho]}\,ds
 +\frac{\tau_\rho\overline F(2\tau_\rho)}{1-\rho^{\mathrm{HIGH}}(\delta_\rho)}\\
 &\qquad\le2\mathbb E[T^{\mathrm{SRPT}}]
 +O\!\left(\log\frac1{1-\rho}\right).
\end{aligned}
\]
\end{lemma}
The initial block adds only a logarithmic cost, the middle term is at most twice ordinary SRPT's residence contribution, and the final \(\tau_\rho\) units add a bounded cost.

\begin{proof}[Proof of Theorem~\ref{thm:sparse-srpt-intermediate-costly}]
The setup above gives eventual stability and a vanishing ex-ante query probability. Lemmas~\ref{lem:sparse-srpt-high-negligible}--\ref{lem:sparse-srpt-residence-overhead} bound the three groups in the response-time identity~\eqref{eq:sparse-srpt-response-identity}, respectively, yielding
\[
 \mathbb E[T^{\mathrm{SRPT\text{-}low}}(\delta_\rho)]
 \le18\mathbb E[T^{\mathrm{SRPT}}]
      +o\!\left(\mathbb E[T^{\mathrm{SRPT}}]\right).
\]
For the lower bound, replace inference intervals by idle time. This gives an ordinary-service schedule with the same completion times, whose mean response time is at least that of SRPT. The SRPT benchmark~\eqref{eq:srpt-heavy-traffic-benchmark} now gives the claimed order.
\end{proof}

\subsection[Proofs of Auxiliary Lemmas]{Proofs of Auxiliary Lemmas (Lemmas~\ref{lem:sparse-srpt-high-negligible}--\ref{lem:sparse-srpt-residence-overhead})}
\label{app:sparse-srpt-helper-proofs}

\begin{proof}[Proof of Lemma~\ref{lem:sparse-srpt-high-negligible}]
Each demoted job removes at least \(\tau_\rho\) units from high priority. Consequently,
\[
 1-\rho^{\mathrm{HIGH}}(\delta_\rho)
 =1-\rho(\delta_\rho)
   +\lambda_\rho\mathbb E[(S-\tau_\rho)\mathbf1\{S>2\tau_\rho\}]
 \ge\lambda_\rho\tau_\rho\overline F(2\tau_\rho).
\]
Since \(X_0^{\mathrm{old}}[0]\le S+\delta_\rho\), its second moment is uniformly bounded. In the first group of~\eqref{eq:sparse-srpt-response-identity}, the waiting term is therefore at most \(C/(\tau_\rho\overline F(2\tau_\rho))\), and the service term is at most \(\mathbb E[S]+\delta_\rho=O(1)\).
It remains to compare this bound with SRPT. The ratio limit established above and the trigger definition give
\[
 \overline F(2\tau_\rho)
 =e^{-(1+o(1))H(\tau_\rho)}
 =(1-\rho)^{\beta+o(1)}.
\]
The tail-work bound in Lemma~\ref{lem:free-bit-tail-work-bound}, applied at the workload cutoff, gives
\[
 H(G^{-1}(\rho))
 =\log\frac1{1-\rho}+\log G^{-1}(\rho)+O(1).
\]
Because \(H(x)/\log x\to\infty\), this implies
\[
 \log G^{-1}(\rho)=o\!\left(\log\frac1{1-\rho}\right),
 \qquad H(G^{-1}(\rho))\sim\log\frac1{1-\rho}.
\]

Likewise \(H(\tau_\rho)=\beta\log(1/(1-\rho))\) gives
\(\log\tau_\rho=o(\log(1/(1-\rho)))\). Thus
\[
 \frac1{\tau_\rho\overline F(2\tau_\rho)}
 =(1-\rho)^{-\beta+o(1)},
 \qquad
 \mathbb E[T^{\mathrm{SRPT}}]
 =\Theta\!\left(\frac1{(1-\rho)G^{-1}(\rho)}\right)
 =(1-\rho)^{-1+o(1)}.
\]
Since \(\beta<1\), the never-demoted contribution is negligible compared with SRPT. With free predictions, total slack is exactly \(1-\rho(0)=1-\rho\), and the remaining estimates are unchanged.
\end{proof}

\begin{proof}[Proof of Lemma~\ref{lem:sparse-srpt-demoted-comparison}]
We compare the SOAP denominators and second moments with ordinary SRPT. For an SRPT tag of size \(s\), new arrivals contribute only when \(S<s\), giving relevant load \(\rho G(s)\). Our tag has worst future rank \(s-\tau_\rho\); larger arrivals also contribute their initial high-priority blocks, but their residual work is excluded. By the relevant-load identity~\eqref{eq:sparse-srpt-relevant-slack}, for \(s\ge2\tau_\rho\),
\[
\begin{aligned}
 1-\rho^{\mathrm{new}}[s-\tau_\rho]
 &=1-\rho(\delta_\rho)
   +\lambda_\rho\mathbb E[(S-\tau_\rho)\mathbf1\{S>s\}]\\
 &\ge\frac12\left(1-\rho
       +\lambda_\rho\mathbb E[S\mathbf1\{S>s\}]\right)\\
 &=\frac12\bigl(1-\rho+\rho(1-G(s))\bigr)
 =\frac12(1-\rho G(s)).
\end{aligned}
\]
Since inference load is \(o(1-\rho)\), it is at most \((1-\rho)/2\) for sufficiently high loads, leaving \(1-\rho(\delta_\rho)\ge(1-\rho)/2\). The inequality combines this bound with \(S-\tau_\rho\ge S/2\) on \(\{S>s\ge2\tau_\rho\}\). The last line uses \(G(s)=\mathbb E[S\mathbf1\{S\le s\}]/\mathbb E[S]\) and \(\rho=\lambda_\rho\mathbb E[S]\), so the load from jobs larger than \(s\) is \(\lambda_\rho\mathbb E[S\mathbf1\{S>s\}]=\rho(1-G(s))\).
Thus each SOAP denominator factor loses at most a factor of two.

For the old-work numerator, substitute the two relevant intervals from Appendix~\ref{app:sparse-srpt-identity}. A competitor with \(S\le s\) contributes all its work in one interval; a larger competitor contributes the separate intervals \(\tau_\rho+\delta_\rho\) and \(s-\tau_\rho\). Thus
\[
\begin{aligned}
 \sum_{i=0}^{1}\mathbb E[(X_i^{\mathrm{old}}[s-\tau_\rho])^2]
 &=\mathbb E[(S+\delta_\rho\mathbf1\{S>\tau_\rho\})^2\mathbf1\{S\le s\}]
 +\bigl((\tau_\rho+\delta_\rho)^2+(s-\tau_\rho)^2\bigr)\overline F(s)\\
 &\le4\mathbb E[S^2\mathbf1\{S\le s\}]+2s^2\overline F(s)
 \le4\bigl(\mathbb E[S^2\mathbf1\{S\le s\}]+s^2\overline F(s)\bigr).
\end{aligned}
\]
For the first inequality, an unqueried job contributes \(S\), while a queried job has \(\delta_\rho\le\tau_\rho<S\), so its total work is at most \(2S\). For a larger competitor, each interval is at most \(s\), since \(\delta_\rho\le\tau_\rho\) and \(s\ge2\tau_\rho\).
Combining these bounds in the demoted waiting term of~\eqref{eq:sparse-srpt-response-identity}, and extending the integral to all tagged sizes, gives
\[
 \int_{2\tau_\rho}^{\infty}
 \frac{\lambda_\rho\sum_{i=0}^{1}\mathbb E[(X_i^{\mathrm{old}}[s-\tau_\rho])^2]}
      {2(1-\rho^{\mathrm{new}}[s-\tau_\rho])^2}f(s)\,ds
 \le16\int_0^\infty
 \frac{\lambda_\rho\bigl(\mathbb E[S^2\mathbf1\{S\le s\}]+s^2\overline F(s)\bigr)}
      {2(1-\rho G(s))^2}f(s)\,ds.
\]
The integral on the right is ordinary SRPT's waiting contribution. Indeed, Proposition~\ref{prop:soap-response-time} gives its waiting and residence contributions as
\[
\begin{aligned}
 \mathbb E[T^{\mathrm{SRPT}}]
 &=\int_0^\infty
 \frac{\lambda_\rho\bigl(\mathbb E[S^2\mathbf1\{S\le s\}]+s^2\overline F(s)\bigr)}
      {2(1-\rho G(s))^2}f(s)\,ds
 +\int_0^\infty\frac{\overline F(s)}{1-\rho G(s)}\,ds\\
 &\ge\int_0^\infty
 \frac{\lambda_\rho\bigl(\mathbb E[S^2\mathbf1\{S\le s\}]+s^2\overline F(s)\bigr)}
      {2(1-\rho G(s))^2}f(s)\,ds.
\end{aligned}
\]
As a result, the demoted waiting contribution is at most \(16\mathbb E[T^{\mathrm{SRPT}}]\), proving the lemma.
\end{proof}

\begin{proof}[Proof of Lemma~\ref{lem:sparse-srpt-residence-overhead}]
We bound the three residence terms in their displayed order. For the initial block, retain the dependence on tagged size rather than using the worst denominator for every tag. Raising the size threshold through \(s\) makes an additional \(s-\tau_\rho\) units per competitor relevant, so
\[
 \frac{d}{ds}\rho^{\mathrm{new}}[s-\tau_\rho]
 =\lambda_\rho(s-\tau_\rho)f(s),
 \qquad\text{for almost every }s\ge2\tau_\rho.
\]
The same density weights the tagged sizes. Since \(s-\tau_\rho\ge\tau_\rho\), changing from size to relevant load gives
\[
\begin{aligned}
 (\tau_\rho+\delta_\rho)\int_{2\tau_\rho}^{\infty}
       \frac{f(s)}{1-\rho^{\mathrm{new}}[s-\tau_\rho]}\,ds
 &\le\frac{\tau_\rho+\delta_\rho}{\lambda_\rho\tau_\rho}
       \int_{2\tau_\rho}^{\infty}
       \frac{\frac{d}{ds}\rho^{\mathrm{new}}[s-\tau_\rho]}
            {1-\rho^{\mathrm{new}}[s-\tau_\rho]}\,ds=\frac{\tau_\rho+\delta_\rho}{\lambda_\rho\tau_\rho}
       \log\frac{1-\rho^{\mathrm{HIGH}}(\delta_\rho)}{1-\rho(\delta_\rho)}.
\end{aligned}
\]
At the lower endpoint, new arrivals contribute only high-priority work; as \(s\to\infty\), all their work becomes relevant. These give the logarithm's two endpoints. Since \(\delta_\rho\le\tau_\rho\), \(\lambda_\rho=\Theta(1)\), and \(1-\rho(\delta_\rho)\ge(1-\rho)/2\), the prefactor is bounded and the logarithm is at most \(\log(2/(1-\rho))\).

For the middle term, the denominator comparison in the proof of Lemma~\ref{lem:sparse-srpt-demoted-comparison} gives
\[
 \int_{2\tau_\rho}^{\infty}\frac{\overline F(s)}{1-\rho^{\mathrm{new}}[s-\tau_\rho]}\,ds
 \le2\int_0^{\infty}\frac{\overline F(s)}{1-\rho G(s)}\,ds
 \le2\mathbb E[T^{\mathrm{SRPT}}].
\]
The tail-probability weight is the same as in ordinary SRPT residence, so only the denominator comparison is needed.

For the final \(\tau_\rho\) units of demoted jobs, the load bound \(1-\rho^{\mathrm{HIGH}}(\delta_\rho)\ge\lambda_\rho\tau_\rho\overline F(2\tau_\rho)\), derived in the proof of Lemma~\ref{lem:sparse-srpt-high-negligible}, gives
\[
 \frac{\tau_\rho\overline F(2\tau_\rho)}{1-\rho^{\mathrm{HIGH}}(\delta_\rho)}
 \le\frac1{\lambda_\rho}=O(1).
\]
Adding the initial, middle, and final residence bounds proves the lemma. The SRPT estimate \(\mathbb E[T^{\mathrm{SRPT}}]=(1-\rho)^{-1+o(1)}\), established in the proof of Lemma~\ref{lem:sparse-srpt-high-negligible}, also shows that the logarithmic overhead is negligible.
\end{proof}

\end{document}